\documentclass[aip,pof,reprint,letterpaper,superscriptaddress,twocolumn,10pt,amsfonts,amssymb,amsmath,footinbib]{revtex4-1}

\usepackage{graphicx}
\usepackage{dcolumn}
\usepackage{bm}
\usepackage[utf8]{inputenc}
\usepackage[T1]{fontenc}
\usepackage{mathptmx}
\usepackage{etoolbox}

\makeatletter
\def\@email#1#2{%
 \endgroup
 \patchcmd{\titleblock@produce}
  {\frontmatter@RRAPformat}
  {\frontmatter@RRAPformat{\produce@RRAP{*#1\href{mailto:#2}{#2}}}\frontmatter@RRAPformat}
  {}{}
}%
\makeatother

\usepackage[hidelinks]{hyperref}
\usepackage{graphics}
\usepackage{color}

\usepackage{ifthen}

\usepackage[light,condensed,math]{anttor}
\newcommand{\anttlc}{\fontfamily{anttlc}\selectfont}

\def\lt{<}

\newcount\fontnumber
\newif\ifoneaccent\oneaccenttrue
\def\ifnextchar#1#2#3{\let\tempe #1\def\tempa{#2}\def\tempb{#3}\futurelet
  \tempc\ifnch}
\def\ifnch{\ifx\tempc\tempe\let\tempd\tempa\else\let\tempd\tempb\fi\tempd}
\def\gobble#1{}
\def\greekmode{%
\catcode`\<=13%
\catcode`\>=13%
\catcode`\'=11%
\catcode`\`=13%
\catcode`\~=11%
\catcode`\"=13%
\lccode`\<=`\<%
\lccode`\>=`\>%
\lccode`\'=`\'%
\lccode`\~=`\~%
\lccode`\"=`\"%
\def\rg{\fontnumber=1\tengr}%
\def\sl{\fontnumber=2\tengrsl}%
\def\it{\fontnumber=3\tengrit}%
\def\bf{\fontnumber=4\tengrbf}%
\def\smc{\fontnumber=5\tengrsmc}%
\def\I##1{\setbox0\hbox{##1}\ifdim\ht0=1ex\accent'174 ##1%
  \else{\ooalign{\hidewidth\char'174\hidewidth\crcr\unhbox0}}\fi}}%
\newcount\vwl
\newcount\acct
{
  \greekmode
  \gdef>{\ifnextchar ~{\expandafter\smoothcircumflex\gobble}{\char\lq\>}}
  \gdef<{\ifnextchar ~{\expandafter\roughcircumflex\gobble}{\char\lq\<}}
  \gdef\smoothcircumflex#1{\acct=\rq134 \vwl=\lq#1 \dobreathingcircumflex}
  \gdef\roughcircumflex#1{\acct=\rq100 \vwl=\lq#1 \dobreathingcircumflex}
  \gdef\dobreathingcircumflex{\ifnum\vwl\lt\rq140 
    \char\the\acct\kern -.2em\char\the\vwl\else
    \doaccent\fi}
  \gdef\doaccent{\accent\the\acct \char\the\vwl\relax}
  \gdef"{\ifnextchar '{\expandafter\diaeresisacute\gobble}{\accent\lq\"}}
  \gdef\diaeresisacute#1{\accent\rq043 #1}
  \gdef`{'}
}
 
\newif\ifgreek\greekfalse
 
\def\begingreek{\bgroup\font\tengr=rgrrg10\font\tengrsl=rgrsl10%
\font\tengrbf=rgrbf10\font\tengrit=rgrti10\font\tengrsmc=rgrsc10%
\greektrue\greekmode\rg}
\def\beginmgreek{\bgroup\font\tengr=mrgrrg10\font\tengrsl=mrgrsl10%
\font\tengrbf=mrgrbf10\font\tengrit=mrgrti10\font\tengrsmc=rgrsc10%
\greektrue\greekmode\rg}
\def\endgreek{\egroup}
\def\endmgreek{\egroup}
\def\monotoniko{%
\font\tengr=mrgrrg10\font\tengrsl=mrgrsl10%
\font\tengrbf=mrgrbf10\font\tengrit=mrgrti10%
\ifnum\fontnumber=5\smc%
  \else\ifnum\fontnumber=4\bf%
    \else\ifnum\fontnumber=3\it%
      \else\ifnum\fontnumber=2\sl%
        \else\rg%
      \fi%
    \fi%
  \fi%
\fi%
}
\def\polutoniko{%
\font\tengr=rgrrg10\font\tengrsl=rgrsl10%
\font\tengrbf=rgrbf10\font\tengrit=rgrti10%
\ifnum\fontnumber=5\smc%
  \else\ifnum\fontnumber=4\bf%
    \else\ifnum\fontnumber=3\it%
      \else\ifnum\fontnumber=2\sl%
        \else\rg%
      \fi%
    \fi%
  \fi%
\fi%
}
\let\math=$%
{\catcode`\$=13%
}
\def\grave#1{{\edef\next{\the\font}\smc\accent\rq001\next#1}}
\def\roughgrave#1{{\edef\next{\the\font}\smc\accent\rq002\next#1}}
\def\smoothgrave#1{{\edef\next{\the\font}\smc\accent\rq003\next#1}}
\def\diaeresisgrave#1{{\edef\next{\the\font}\smc\accent\rq004\next#1}}
\def\diaeresiscircumflex#1{{\edef\next{\the\font}\smc\accent\rq005\next#1}}
\def\breve#1{{\edef\next{\the\font}\smc\accent\rq006\next#1}}
\def\macron#1{{\edef\next{\the\font}\smc\accent\rq007\next#1}}
\def\rhorough{{\tengrsmc
\ifnum\fontnumber=5\char\rq162
  \else\ifnum\fontnumber=4\char\rq016
    \else\ifnum\fontnumber=3\char\rq014
      \else\ifnum\fontnumber=2\char\rq012
        \else\char\rq010
      \fi
    \fi
  \fi
\fi
}}
\def\rhosmooth{{\tengrsmc
\ifnum\fontnumber=5\char\rq162
  \else\ifnum\fontnumber=4\char\rq017
    \else\ifnum\fontnumber=3\char\rq015
      \else\ifnum\fontnumber=2\char\rq013
        \else\char\rq011
      \fi
    \fi
  \fi
\fi
}}
\def\digamma{{\smc\char\rq135}}

\def\Digamma{{\tengrsmc
\ifnum\fontnumber=5\char\rq021
  \else\ifnum\fontnumber=4\char\rq027
    \else\ifnum\fontnumber=3\char\rq025
      \else\ifnum\fontnumber=2\char\rq023
        \else\char\rq021
      \fi
    \fi
  \fi
\fi
}}
\def\vardigamma{{\tengrsmc
\ifnum\fontnumber=5\char\rq020
  \else\ifnum\fontnumber=4\char\rq026
    \else\ifnum\fontnumber=3\char\rq024
      \else\ifnum\fontnumber=2\char\rq022
        \else\char\rq020
      \fi
    \fi
  \fi
\fi
}}

\newcommand{\greeky} [1]{\mbox{\begingreek#1\endgreek}}    

\newcommand{\sgreeky}[2]{\scalebox{#1}{#2}} 
\newcommand{\greeksizeincaption}{0.80}
\newcommand{\etasizeincaption}{0.918}
\newcommand{\thetasizeincaption}{0.875}
\newcommand{\greeksizeinfootnote}{0.75}  
\newcommand{\betasizeinfootnote}{0.70}
\newcommand{\greeksizeinsubscript}{0.66} 
\newcommand{\greeksizeinfootnotesubscript}{0.5725} 

\newcommand{\alphay}{\greeky{a}}
  
  \newcommand{\alphayc}{\sgreeky{\greeksizeincaption}{\protect\alphay}}    

\newcommand{\betay}{\greeky{b}}
  
  \newcommand{\betayc}{\sgreeky{\greeksizeincaption}{\protect\betay}}       
  \newcommand{\betayf}{\sgreeky{\betasizeinfootnote}{\betay}}              

\newcommand{\zetay}{\greeky{z}}

\newcommand{\etay}{\greeky{h}}
  \newcommand{\etays}{\sgreeky{\greeksizeinsubscript}{\etay}} 
  \newcommand{\etayss}{\sgreeky{\greeksizeinfootnotesubscript}{\etay}} 
  \newcommand{\etayc}{\sgreeky{\etasizeincaption}{\protect\etay}}    

\newcommand{\thetay}{\greeky{j}}
  \newcommand{\thetays}{\sgreeky{\greeksizeinsubscript}{\thetay}} 
  \newcommand{\thetayf}{\sgreeky{\greeksizeinfootnote}{\thetay}}              
  \newcommand{\thetayc}{\sgreeky{\thetasizeincaption}{\thetay}}    

\newcommand{\lambday}{\greeky{l}}
  \newcommand{\lambdayc}{\sgreeky{\thetasizeincaption}{\lambday}}    

\newcommand{\xiy}{\greeky{x}}
  \newcommand{\xiyc}{\sgreeky{\greeksizeincaption}{\xiy}}
  \newcommand{\xiyf}{\sgreeky{\greeksizeinfootnote}{\xiy}}
  \newcommand{\xiys}{\sgreeky{\greeksizeinsubscript}{\xiy}} 

\newcommand{\piy}{\greeky{p}}
  \newcommand{\piys}{\sgreeky{\greeksizeinsubscript}{\protect\piy}}    

\newcommand{\rhoy}{\greeky{r}}

\newcommand{\chiy}{\greeky{q}}
  \newcommand{\chiys}{\sgreeky{\greeksizeinsubscript}{\protect\chiy}}    

\newcommand{\omegay}{\greeky{w}}

\usepackage{dg-latexcmds}
\usepackage{submath}

\newcommand{\gdir}{./}
\newcommand{\wbox}{false} 

\newcommand{\pg}{PG-m}
\newcommand{\pge}{PG-sge}
\newcommand{\vdw}{vdWG-m}

  \newcommand{\Refmb}{\oref{dg2024pof}}

  \newcommand{\ma}{\onlinecite{dg2019ejmb}}
  \newcommand{\subma}{$_{\ssub{0.65}{\mbox{[\ma]}}}$}
  \newcommand{\Reqma}[1]{\mbox{Eq.\;(#1)}\subma}
  \newcommand{\Reqqma}[1]{\mbox{Eqs.\;(#1)}\subma}
  \newcommand{\Reqsma}[2]{\mbox{\mbox{Eqs.\;(#1)}\subma\,and\;(#2)\subma}}
  \newcommand{\Rsema}[1]{\mbox{Sec.\;#1}\subma}
  \newcommand{\Rfima}[1]{\mbox{Fig.\;#1}\subma}
  \newcommand{\Rfisma}[2]{\mbox{\mbox{Figs.\;#1}\subma\,and\;#2\subma}}
  \newcommand{\Refsma}[1]{\mbox{\mbox{Refs.\;#1}\subma}}

  \newcommand{\mb}{\onlinecite{dg2024pof}}
  \newcommand{\submb}{$_{\ssub{0.65}{\mbox{[\mb]}}}$}
  \newcommand{\Reqmb}[1]{\mbox{Eq.\;(#1)}\submb}
  \newcommand{\Reqqmb}[1]{\mbox{Eqs.\;(#1)}\submb}
  \newcommand{\Reqsmb}[2]{\mbox{\mbox{Eqs.\;(#1)}\submb\,and\;(#2)\submb}}
  \newcommand{\Rsemb}[1]{\mbox{Sec.\;#1}\submb}
  \newcommand{\Rfimb}[1]{\mbox{Fig.\;#1}\submb}
  \newcommand{\Rfismb}[2]{\mbox{\mbox{Figs.\;#1}\submb\,and\;#2\submb}}
  \newcommand{\Rtamb}[1]{\mbox{Table\;#1}\submb}

  \newcommand{\tcb}{\mbox{TC\submb}}

  \newcommand{\uw}{u_{t}}
  \newcommand{\tw}{\thetay_{t}}
  \newcommand{\xw}{\xiy_{t}}
  \newcommand{\fomt}{1oM$_{2}$} 
  \newcommand{\Xs}{\ssub{0.55}{X}}

\begin{document}


\title{{\anttlc Does the fluid-static equilibrium of a self-gravitating isothermal sphere of van der Waals' gas present multiple solutions?}}

\author{\anttlc Domenico Giordano} \email{dg.esa.retired@gmail.com}  \affiliation{European Space Agency - ESTEC (retired), Noordwijk, The Netherlands}
\author{\anttlc Pierluigi Amodio}  \author{\anttlc Felice Iavernaro} \affiliation{Dipartimento di Matematica, Universit\`a di Bari Aldo Moro, Bari, Italy}
\author{\anttlc Francesca Mazzia}                                    \affiliation{Dipartimento di Informatica, Universit\`a di Bari Aldo Moro, Bari, Italy}

\date{\today}
\pacs{}

\begin{abstract}  We take up the investigation {\color{black} we left in the future-work stack in Giordano \textit{et al.} [``Fluid statics of a self-gravitational isothermal sphere of van der Waals' gas,'' Phys. Fluids \textbf{36}, 056127 (2024)]}, in which we pointed out the obvious necessity to inquire about the existence or absence of values of the characteristic numbers \itm{\alphay} and \itm{\betay} in correspondence to which the perfect-gas model's self gravitational effects, namely, upper boundedness of the gravitational number, spiraling behavior of peripheral density, oscillating behavior of central density, and the existence of multiple solutions corresponding to the same value of the gravitational number, appear also for the van der Waals' model.
The development of our investigation brings to the conversion of our M$_{2}$ scheme based on a second-order differential equation into an equivalent system of two first-order differential equations that incorporates Milne's homology invariant variables. The converted scheme \fomt\ turns out to be much more efficacious than the M$_{2}$ scheme in terms of numerical calculations' easiness and richness of results.
We use the perfect-gas model as benchmark to test the \fomt\ scheme; we re-derive familiar results and put them in a more general and rational perspective that paves the way to deal with the van der Waals' gas model.
We introduce variable transformations that turn out to be the key to study (almost) analytically the monotonicity of the peripheral density with respect to variations of the gravitational number.
The study brings to the proof that the gravitational number is not constrained by upper boundedness, the peripheral density does not spiral, and the central density does not oscillate for any couple of values assumed by the characteristic numbers \itm{\alphay} and \itm{\betay}; however, multiple solutions corresponding to the same value of the gravitational number can exist but their genesis is completely different from that of the perfect-gas model's multiple solutions. We provide the boundary in the \itm{\alphay,\betay} plane between the two regions of solution's uniqueness and multiplicity.
Finally, by the application of the \fomt\ scheme, we resolve the mystery of the {\color{black} steep curve corresponding to the ``60km'' case considered by Aronson and Hansen [``Thermal equilibrium states of a classical system with gravitation'', Astrophys. J. \textbf{177}, 145 (1972)]} and we even detect another solution that had somehow escaped the attention of those authors.

  \end{abstract}
\maketitle

\section{{\protect\anttlc Introduction}\label{intro}}

\textcolor{black}{Self-gravitating fluids, that is, fluids whose statics and dynamics are influenced by the action of their own gravity, constitute a very fascinating category of fluid dynamics.
Their queenly collocation is certainly in astrophysics, where the applications stretch from the familiar context of Newtonian gravity to the extreme context of general relativity, but they can also attract interest for purposes of more general fluid-dynamics nature.
We came in contact with them by following a trail originating from gravitomagnetism.
For familiarization with detailed description and relevant literature, we refer the reader to the introduction of \ocite{dg2019ejmb}, which started our series of studies and was based on the perfect gas model (\pg), a model that notoriously lacks accountability of molecular attraction.
Expectedly, we re-obtained the same physically questionable results that had been known in the astrophysical literature since long time, although prompted by different motivations.
Our lack of confidence with respect to the meaningfulness of those results pushed us to engage in a similar study, described in \ocite{dg2024pof}, based on the van der Waals' gas model (\vdw) with the hope that the terms accounting for molecular size and attraction in that model's equation of state could lead to results freed from the conceptual uncertainties affecting those of the \pg.
We scored success to a notable extent but did not fulfill completely our task because we left unexplored two important corners: a sensitivity analysis of characteristic numbers and the use of general relativity.}
{\color{black} Here, we take up the \textcolor{black}{former} investigation, {\color{black} left} in the future-work stack at the end of \mbox{Sec. V B 2} of \ocite{dg2024pof}}.
For obvious reasons of consistency, we use the same notation of and definitions introduced in \textcolor{black}{our previous studies}.
Also, we maintain the convention to subscript cross-references to \ocites{dg2024pof,dg2019ejmb} with the labels [\ma] and [\mb], respectively; for example, \Reqma{3} refers to \mbox{Eq. (3)} in \ocite{dg2019ejmb} and \Rfimb{3} refers to \mbox{Fig. 3} in \ocite{dg2024pof}.
We consider only the situation in which the fluid-sphere temperature is above the critical temperature [\Reqmb{51} top]; therefore, phase equilibria are excluded from consideration.

In \Rsemb{V B}, we described and discussed results relative to a test case characterized by the couple of values \itm{\alphay=0.1053,\,\betay=0.0595} of the characteristic numbers [\Reqsmb{31b}{31c}] appearing in the nondimensional \vdw's equation of state [\Reqmb{32}]; in the sequel, we will refer to this case with the label \tcb.
We provided undeniable evidence in \Rfismb{10}{11} that the physically questionable self-gravitational effects accompanying the \pg, namely, upper boundedness $(N\leq2.5176)$ of the gravitational number defined in \Reqmb{31a}, spiraling behavior of peripheral density, and oscillating behavior of the central density, do not trouble the \vdw.
That outcome led us to conjecture\footnote{Admittedly, we did not prove it.} also the absence of the last uncomfortable effect featured by the \pg: the existence of multiple equilibria in correspondence to a specified gravitational number, mainly evidenced by the spiraling curve in \Rfimb{10} and explicitly exemplified in \Rfisma{7}{8} for \itm{N=2.4}.
Hereinafter, for brevity, we convene to group the mentioned self-gravitational effects related to the \pg\ under the label \pge.
We concluded \Rsemb{V B 2} with the admission that the results presented therein proved the disappearance of the \pge\ only for the assumed values of \itm{\alphay} and \itm{\betay} and with the recognition that extrapolation of such an occurrence to arbitrary values may be hazardous.
{\color{black} We also pointed out the obvious necessity of an investigation regarding the existence or absence of values of \itm{\alphay} and \itm{\betay} in correspondence to which the \pge\ appear also for the \vdw.
The contents of this communication describe our efforts to go to the bottom of this matter}.

\section{\protect\anttlc Analysis\label{an}}

\subsection{\anttlc Preliminary considerations\label{pc}}
An appropriate entry point for our investigation is the M$_{2}$ scheme described in nondimensional form at the end of \Rsemb{II B}.
At its core, there is a second-order differential equation [\Reqmb{43}], expanded here to the more convenient form
\begin{equation}\label{sode.ss.vdw.nd}
    \phi_{\etays\etays} + \frac{2}{\etay}\,\phi_{\etays} + 3N\xiy = 0 \, ,
\end{equation}
which requires the assignment of the thermodynamic model by means of the reduced chemical potential
\begin{equation}\label{rcp.nd}
  \phi = \phi(\xiy)
\end{equation}
whose natural variable is, in general, the specific volume $\zetay$ but that we reformulate here better in terms of the density \itm{\xiy=1/\zetay};
for the \vdw, it takes the explicit expression [\Reqmb{42}]
  \begin{equation}\label{rcp.vdw.nd}
     \phi(\xiy) = \ln\xiy - \ln ( 1 - \betay\xiy ) + \frac{1}{1 - \betay\xiy} - 2\alphay\xiy 
  \end{equation}
\REqb{sode.ss.vdw.nd} is accompanied by the ``gravitational'' boundary conditions [\Reqsmb{44a}{44b}]
\begin{subequations} \label{bc.rcp.nd}
  \begin{align}
    \phi_{\etays}(0) & =  0   \label{bc.rcp.r=0.nd} \\[.5\baselineskip]
    \phi_{\etays}(1) & =  -N  \label{bc.rcp.r=a.nd} \,.
  \end{align}
\end{subequations}
The former enforces the vanishing of the gravitational field in the sphere's center \itm{(\etay=0)}; the latter enforces the Gaussian value of the gravitational field at the container's internal wall \itm{(\etay=1)}.
We ought to point out that the sphere's center is a singular point for the central term in \REq{sode.ss.vdw.nd} but the discontinuity is removed by the gravitational-field vanishing [\REq{bc.rcp.r=0.nd}]; for the purpose of computational and algebraical operations, the application of de l'H\^{o}pital's theorem treats the indeterminate form of the central term when \itm{\etay\rightarrow0} and yields the cured form
\begin{equation}\label{sode.ss.vdw.nd.0}
    \phi_{\etays\etays}(0) + N\xiy(0) = 0 \,.
\end{equation}
We will need \REq{sode.ss.vdw.nd.0} later.

We have pursued several lines of attack to develop the investigation outlined at the end of \Rse{intro} in the context of the M$_{2}$ scheme but, after spending a substantial amount of time and innumerable unsuccessful efforts, we had to surrender to an unavoidable conclusion: to the best of our skills, the M$_{2}$ scheme constitutes an impenetrable barrier to the investigation we tasked ourselves with.
So, it became evident that we had to explore other avenues and, out of that necessity, we obviously sought help and inspiration from the literature.
We took particular advantage from the teachings of Chandrasekhar\cite{sc1957} and the treatments provided by Padmanabhan\cite{tp1989ajss,tp1990pr} and Horedt;\cite{gh2004} they indicate that, at least relatively to the \pg, the multiple-solution subject is best dealt with by resorting to the famous homology invariant variables \itm{u,v} introduced by Milne\cite{em1932mnras} in 1932, as we learned from the very useful bibliographical notes provided by Chandrasekhar at page 176 of \ocite{sc1957}.\footnote{More precisely, item 3 of Sec. VI at page 182.}
The core of that method is represented by a universal first-order differential equation involving the Milne's variables whose solution turns out to be a curve in the \itm{u,v} plane spiraling around the center \itm{u=1,v=2}.\footnote{For specific details, we refer the reader to Eq. (453) and Fig. 20 in \mbox{Sec. 27} at \mbox{page 168} of \ocite{sc1957}
or Eq. (18) and Fig. 1 in \mbox{Sec. II} at \mbox{page 651} of \ocite{tp1989ajss} or Eq. (4.26) and Fig. 4.1 in \mbox{Sec. 4.3} at \mbox{page 313} of \ocite{tp1990pr}}
Results and conclusions achievable by following that approach are undoubtedly remarkable.
Yet the differential equation in question is systematically looked at and treated as a {standalone} pillar of the method; such an interpretation, we found out, restricts somehow a more insightful view with respect to the method's more general possibilities and, above all, its potential ability to go beyond the \pg.

We start from a different stand and describe in \Rse{rm2s} how our departure point leads to a broader picture in which the familiar first-order differential equation found in the literature is accompanied by other both supporting and necessary equations, differential as well as algebraic, and the whole mathematical structure is generalized to the \vdw\ or, as a matter of fact, to any thermodynamic model with two intensive degrees of freedom.

\subsection{\anttlc Conversion of the M$_{2}$ scheme from second to first order\label{rm2s}}
Our colleagues the mathematicians teach that any second-order differential equation can be converted into a system of two equivalent first-order differential equations.
Let us see what this prescript entails in connection with \REq{sode.ss.vdw.nd}.

In general, the conversion procedure requires the introduction of a variable
\begin{equation}\label{v.gen}
  v = f(\etay,\phi,\phi_{\etays})
\end{equation}
defined through a chosen function $f$.
The derivation of \REq{v.gen} with respect to $\etay$ yields
\begin{equation}\label{dv.gen}
  v_{\etays} = f_{\etays} + f_{\phi}\,\phi_{\etays} + f_{\phi_{\etayss}}\,\phi_{\etays\etays}  \; ;
\end{equation}
the second derivative can be extracted from \REq{sode.ss.vdw.nd} and substituted into \REq{dv.gen} to obtain the final form
\begin{equation}\label{dv.gen.f}
  v_{\etays} = f_{\etays} + \left( f_{\phi} - \frac{2}{\etay}f_{\phi_{\etayss}} \right) \,\phi_{\etays} - 3 N \xiy \,f_{\phi_{\etayss}} \,.
\end{equation}
\REqdb{v.gen}{dv.gen.f} compose the sought system of the first-order differential equations for the unknown functions \itm{\phi,\xiy,v} of the independent variable $\etay$; they require the auxiliary knowledge of the reduced chemical potential [\REq{rcp.nd}], given explicitly by \REq{rcp.vdw.nd} for the \vdw\ but that needs not necessarily be such, and the eventual adaptation of the boundary conditions [\REqq{bc.rcp.nd}].

We follow now Milne's footsteps \footnote{Perhaps we should better say Chandrasekhar's. As a matter of fact, Milne did not consider isothermal spheres in \ocite{em1932mnras}; he introduced the \itm{u,v} variables [Eqs. (42) at page 622] for a non-isothermal sphere composed by a zone of perfect-gas polytrope of index \itm{n=3} in contact with a zone of degenerate-gas polytrope of index \itm{n=3/2}. To the best of the literature we have consulted, it seems to us that Chandrasekhar was the first to introduce the \itm{u,v} variables for isothermal spheres of perfect gas in the first edition of \ocite{sc1957} that appeared in 1939; our conjecture seems to be corroborated by the brief note\cite{sc1949aj} that Chandrasekhar and Wares published in the Astrophysical Journal in 1949 \textbf{}and by Horedt's citation in the beginning of Sec. 2.2.2 at page 34 of \ocite{gh2004}. Besides, we reassure the reader that we have thoroughly cross-checked our definitions [\REqd{v.is}{u.is}] with those of Chandrasekhar [Eqs. (400) at page 160 and the auxiliary Eqs. (373) at page 155 of \ocite{sc1957}] through a patient exercise of notation conversion and definition verification; of course, Chandrasekhar's definitions apply only to the \pg.} and set
\begin{equation}\label{v.is}
  v = - \etay\,\phi_{\etays} \, \ge 0 \,\; .
\end{equation}
Then \REq{dv.gen.f} becomes
\begin{subequations}\label{dv.is.tmp}
    \begin{equation}\label{dv.is.tmp1}
      v_{\etays}  = \phi_{\etays} + 3 N \etay\,\xiy 
    \end{equation}
    which, after elimination of \itm{\phi_{\etays}} with the aid of \REq{v.is}, can be rewritten as
    \begin{equation}\label{dv.is.tmp2}
      v_{\etays} = - \frac{v}{\etay} \left( 1 - \frac{3N\etay^{2}\xiy}{v}  \right) \; ;
    \end{equation}
\end{subequations}
with a view to the right-hand side of \REq{dv.is.tmp2}, we introduce the convenience variable
\begin{equation}\label{u.is}
  u = \frac{3N\etay^{2}\xiy}{v} = - \frac{3N\etay\,\xiy}{\phi_{\etays}} > 0
\end{equation}
and bring \REq{dv.is.tmp2} into its final form
\begin{equation}\label{dv.is}
  v_{\etays} = - \frac{v}{\etay} \left( 1 - u \right) \,.
\end{equation}
So, our conversion of the M$_{2}$ scheme has well progressed:
we have at our disposal four equations, two of which are algebraic [\REqd{rcp.nd}{u.is}] and two of which are first-order differential [\REqd{v.is}{dv.is}],
that compose a system to simultaneously determine the four unknown functions \itm{\phi,\xiy,v,u}, of the independent variable $\etay$.
The final touch consists of the elaboration of the boundary conditions to associate with this first-order system starting from \REqq{bc.rcp.nd}, a process that requires a good understanding of what happens to the variables \itm{u,v} at the sphere's center \itm{(\etay=0)} and at the container's internal wall \itm{(\etay=1)}. 
\Rseb{bc} is devoted to this matter.

\subsection{\anttlc Boundary conditions\label{bc}}
Let us begin with the sphere's center and keep in mind that $\etay$ is the independent variable. 
The variable $u$ requires special attention because \REq{u.is} becomes an indeterminate form at \itm{\etay=0} and needs to be treated with de l'H\^{o}pital's theorem whose application gives
\begin{align}\label{u.is.0}
  u(0) = &  - \lim_{\etays\rightarrow0}\frac{3N\etay\xiy}{\phi_{\etays}} = -  3N\xiy(0)\lim_{\etays\rightarrow0}\frac{1}{\phi_{\etays\etays}} \nonumber \\[.5\baselineskip]
       = & -  \frac{3N\xiy(0)}{\phi_{\etays\etays}(0)} = -  \frac{3N\xiy(0)}{-N\xiy(0)} = 3 \;,
\end{align}
a well known result\cite{sc1957,tp1989ajss,tp1990pr} whose extent of applicability, however, goes well beyond the \pg\ according to the way we obtained it; in the bottom line of \REq{u.is.0}, we have replaced the second derivative in accordance with \REq{sode.ss.vdw.nd.0}.

Moving on to the variable $v$, obviously from \REq{v.is}, we obtain
\begin{equation}\label{v.is.0}
  v(0) = 0
\end{equation}
but we should refrain to look hurriedly at \REq{v.is.0} as a boundary condition if derived in this way because the vanishing of $v(0)$ is caused by the presence of $\etay$ in \REq{v.is} regardless of the value of $\phi_{\etays}(0)$; therefore, it may be objected, and with reason, that \REq{v.is.0} does not necessarily imply the vanishing of the gravitational field in the sphere's center [\REq{bc.rcp.r=0.nd}].
Nevertheless, the objection can be circumvented with the following stratagem.
Our task is to prove that \REq{bc.rcp.r=0.nd} follows from \REq{v.is.0}.
First, we rearrange \REq{v.is} to the equivalent form
\begin{equation}\label{v.is.r}
  \phi_{\etays} = - \frac{v}{\etay} \; ;
\end{equation}
\REqb{dv.is} is already well adapted.
We can now invoke the same argument we used to justify \REq{bc.rcp.r=0.nd} in order to shield \REq{sode.ss.vdw.nd} against a possible discontinuity at \itm{\etay=0}: we must assume \REq{v.is.0} to avoid analogous discontinuities in \REqd{dv.is}{v.is.r}.
Of course, the right-hand sides of the latter equations turn into indeterminate forms
\begin{equation}\label{v-dv.is.0}
  \left\{
    \begin{aligned}
      v_{\etays}(0)    & = - \lim_{\etays\rightarrow0}\frac{v}{\etay} \left( 1 - u \right) =   2\lim_{\etays\rightarrow0}\frac{v}{\etay}   \\[.5\baselineskip]
      \phi_{\etays}(0) & = - \lim_{\etays\rightarrow0}\frac{v}{\etay}
    \end{aligned}
  \right.
\end{equation}
but we can always resort to de l'H\^{o}pital's theorem, our reliable and effective mathematical tool, to remove the indeterminateness
\begin{equation}\label{lhop}
  \lim_{\etays\rightarrow0}\frac{v}{\etay} = \lim_{\etays\rightarrow0}\frac{v_{\etays}}{1} = v_{\etays}(0)
\end{equation}
and convert \REqq{v-dv.is.0} into determinate forms
\begin{equation}\label{v-dv.is.0.c}
  \left\{
    \begin{aligned}
      v_{\etays}(0)    & = 0   \\[.5\baselineskip]
      \phi_{\etays}(0) & = 0 \; .
    \end{aligned}
  \right.
\end{equation}
At the top of \REqq{v-dv.is.0.c}, we obtain a condition for the first derivative;\footnote{The attentive reader may have noticed that \REq{v-dv.is.0.c} (top) follows also as direct consequence of \REq{dv.is.tmp1} but this fits the logic in which \REq{bc.rcp.r=0.nd} is pre-assigned. Here we are following the reverse logical path: \REq{v.is.0} is pre-assigned with the purpose in mind to prove \REq{bc.rcp.r=0.nd}; therefore, \REq{dv.is.tmp1} must be looked at posteriorly as a sort of consistency indicator in connection with \REq{v-dv.is.0.c} (top).} at the bottom, we rediscover \REq{bc.rcp.r=0.nd}, understand that it is a consequence of \REq{v.is.0}, and rest assured that the latter qualifies as genuine boundary condition that goes to accompany \REq{dv.is}.

The situation at the container's internal wall is more relaxed.
By taking into account \REq{bc.rcp.r=a.nd}, \REq{v.is} establishes the condition 
\begin{equation}\label{v.is.1}
  v(1) = N \; .
\end{equation}
\REqb{v.is.1}, however, must be transformed into a condition involving either density or reduced chemical potential in order to be a valid boundary condition to support \REq{v.is}.
The transformation comes from the evaluation of \REq{u.is} at \itm{\etay=1} which establishes a connection between the peripheral density and the terminal value of the variable $u$
\begin{equation}\label{u.is.1}
  3\,\xiy(1) - u(1) = 0
\end{equation}
and, to all effects, represents the boundary condition required for \REq{v.is}.

With the boundary conditions [\REqd{v.is.0}{u.is.1}] in hand, the conversion of the M$_{2}$ scheme to a first-order system of differential equations can be considered completed.

\subsection{\anttlc Change of independent variable \label{civ}}

Until now, we have conferred to $\etay$ the role of independent variable but there is really no mathematical necessity or obligation to do so; we may decide to take the variable $u$ as independent and work out the consequences of such a decision on the first-order system of equations derived in \Rse{rm2s} and on the boundary/auxiliary conditions discussed in \Rse{bc}.

For convenience, let us collect here the first-order system of governing equations, slightly adapted for the purpose of switching independent variable
\begin{subequations}\label{fos}
    \begin{align}
        & \phi = \phi(\xiy)                                          \label{fos.rcp} \\[1ex]
        & \phi_{\etays} + \frac{v}{\etay}  = 0                       \label{fos.v}   \\[1ex]
        & v_{\etays}    + \frac{v}{\etay} \left( 1 - u \right)  = 0  \label{fos.dv}  \\[1ex]
        & u    = \dfrac{3N\etay^{2}\xiy}{v}  \; .                    \label{fos.u}
    \end{align}
\end{subequations}
We left the reduced chemical potential unspecified in \REq{fos.rcp} because we wish to show that what we are going to deduce in the sequel is applicable to any generic thermodynamic model with two thermodynamic intensive degrees of freedom and for which the separation indicated in \Reqmb{23} is possible.\footnote{This would be the case, for example, for the thermodynamic model proposed by Stahl et al.\cite{bs1995pss} that we considered in \Rsemb{V B 4}; see \Reqmb{77a}.}
The first step consists of reformulating \REqq{fos} in terms of the variables' differentials.
\REqb{fos.rcp} expands as
\begin{equation}\label{fos.rcp.d.tmp}
  d\phi = \phi_{\ssub{0.7}\xiys}\,d\xiy = (\xiy\,\phi_{\ssub{0.7}\xiys}) \frac{d\xiy}{\xiy}
\end{equation}
and, for brevity, we introduce the purely thermodynamic function \itm{\hnd = (\xiy\,\phi_{\ssub{0.7}\xiys})^{-1} = \hnd(\xiy)}; its explicit expression for the \vdw
\begin{equation}\label{h.vdw}
  \hnd(\xiy) = \frac{\left(1 - \betay\xiy\right)^{2}}{1-2\alphay\xiy\left(1 - \betay\xiy\right)^{2}}
\end{equation}
is obtained from \REq{rcp.vdw.nd}.
\begin{subequations}\label{fos.d}
Thus,
    \begin{equation}\label{fos.rcp.d}
      d\phi = \frac{1}{\hnd} \frac{d\xiy}{\xiy} \; .
    \end{equation}
    \REqdb{fos.v}{fos.dv} are straightforward
    \begin{align}
        & d\phi        + v \frac{d\etay}{\etay} = 0                         \label{fos.v.d}   \\[1ex]
        & \frac{dv}{v} + \left( 1 - u \right) \frac{d\etay}{\etay} = 0 \; . \label{fos.dv.d} 
    \end{align}
    \REqb{fos.u} is better handled with the logarithmic differentiation and yields
    \begin{equation}\label{fos.u.d}
      \frac{du}{u} + \frac{dv}{v} = 2 \frac{d\etay}{\etay} + \frac{d\xiy}{\xiy}
    \end{equation}
\end{subequations}
The next step involves the resolution in terms of \itm{du} of all the other differentials appearing in \REqq{fos.d} and, therefrom, the extraction of the derivatives with respect to $u$ of all the other variables
\begin{subequations}\label{fos-u}
    \begin{align}
        & \xiy_{u} =   \frac{\hnd v}{u + \hnd v - 3} \frac{\xiy}{u}     \label{fos.xi.u}  \\[1ex]
        & \phi_{u} =   \frac{v}     {u + \hnd v - 3} \frac{1}{u}        \label{fos.rcp.u} \\[1ex]
        & v_{u}    = - \frac{u-1}   {u + \hnd v - 3} \frac{v}{u}        \label{fos.v.u}   \\[1ex]
        & \etay_{u} = - \frac{1}{u + \hnd v - 3} \frac{\etay}{u}  \; .  \label{fos.eta}
    \end{align}
\end{subequations}
\REqqb{fos-u} constitute the first-order system of governing equations with $u$ as independent variable.
The one among them that immediately catches the eye is \REq{fos.v.u}: it is the clear and unambiguous generalization of the differential equation standardly considered and discussed in the literature\cite{sc1957,tp1989ajss,tp1990pr} in the case of the \pg\ to any thermodynamic model with the characteristics specified just after \REq{fos.u}.
In conformity with Chandrasekhar's terminology, the vertical line \itm{u=1} is still ``the locus of points at which the curves have horizontal tangent'' but ``the locus of points at which the curves have vertical tangent'' is definitely not the universal straight line 
\begin{equation}\label{vden.pg}
  u + v - 3 = 0
\end{equation}
anymore but could be the curve
\begin{equation}\label{vden.vdw}
  u + \hnd v - 3 = 0
\end{equation}
whose shape, however, turns out to be solution-dependent due to the intervention of the thermodynamic function \itm{\hnd}.
So, we receive straightaway a first forewarning of distancing from \pg: the existence of intersections between the solution \itm{v(u)} obtained from \REqq{fos-u} and the curve defined by \REq{vden.vdw} and, in case they do exist, how they affect the shape of \itm{v(u)} remains to be seen.
Moreover, \REq{fos.v.u} does not have a standalone role because the thermodynamic function \itm{\hnd} couples it to \REq{fos.xi.u}; indeed, the two equations must be inevitably solved simultaneously.
After, when the solutions \itm{v(u)} and \itm{\xiy(u)} have been determined, we can proceed to integrate \REqd{fos.rcp.u}{fos.eta} to obtain the functions \itm{\phi(u)} and \itm{\etay(u)}, formally; practically, we can shortcut the integration by obtaining the reduced chemical potential and the radial coordinate algebraically from, respectively, \REqd{fos.rcp}{fos.u}, the latter being resolved accordingly as
\begin{equation}\label{fos.u.rev}
   \etay = \sqrt{\dfrac{u v}{3N\xiy}}  \; .
\end{equation}

We turn now to the elaboration of the boundary conditions needed for \REqd{fos.xi.u}{fos.v.u}.
{\color{black} There is a very important aspect to keep in mind with regard to the switch of independent variable from $\etay$ to $u$.
The variable $\etay$ ranges in the well defined interval [0,1] but that is not the case for the variable $u$.
We know the end point corresponding to the sphere's center (\itm{u=3}) from \REq{u.is.0}, which reverses as 
\begin{equation}\label{eta.is.3}
  \etay(3) = 0 \; ,
\end{equation}
but the other end point, that is, the terminal value \itm{u(1)} appearing in \Req{u.is.1} 
corresponding to the container's internal wall is unknown and needs to be determined.
Let us set formally
\begin{equation}\label{uw.is.eta}
  u(1) = \uw
\end{equation}
and keep in mind that the subscript $t$ does not imply partial derivation but represents just a mnemonic label referring to the word ``terminus.'' This convention applies to all future occurrences of that subscript.}
By account of \REq{eta.is.3}, \REq{v.is.0} yields the boundary condition
\begin{equation}\label{v.is.3}
  v(3) = 0
\end{equation}
and that goes with \REq{fos.v.u}.
\REqb{uw.is.eta} reverses as
\begin{equation}\label{eta.is.uw}
  \etay(\uw) = 1
\end{equation}
and this inversion of variables turns \REq{u.is.1} into the boundary condition
\begin{equation}\label{xi.is.uw}
  \xiy(\uw) = \frac{\uw}{3}
\end{equation}
that accompanies \REq{fos.xi.u}. 
The strong similarity of their mathematical structure notwithstanding, there is a profound conceptual difference that distances \REq{xi.is.uw} from \REq{u.is.1}: the latter constitutes a condition that involves two,  $\xiy$ and $u$, of the four unknowns on the fixed boundary \itm{\etay=1} while the former represents the prescription of the unknown $\xiy$ at the floating boundary \itm{u=\uw} whose determination is also part of the mathematical problem.
That is why we are entitled, with the help of \REq{eta.is.uw}, to reinterpret \REq{v.is.1} as
\begin{equation}\label{v.is.uw}
  v(\uw) = N
\end{equation}
and exploit it as an auxiliary condition to fix the terminus $\uw$.
It comes to no surprise that the responsibility of settling the integration boundary falls on a parameter of physical significance such as the gravitational number $N$.

There is one last aspect that we must address to complete this section: we need to find out what happens to \REqd{fos.xi.u}{fos.v.u} in the sphere's center because there, according to \REq{v.is.3}, their right-hand sides become indeterminate.
So, we subject them to the limit \itm{u\rightarrow3}
\begin{subequations}\label{fos-u.3}
    \begin{align}
        & \xiy_{u}(3) = \hnd(3) \left(\lim_{u\rightarrow3}\frac{v}{u + \hnd v - 3}\right) \frac{\xiy(3)}{3}     \label{fos.xi.u.3}  \\[1ex]
        & v_{u}(3)    = - \frac{2}{3} \left(\lim_{u\rightarrow3}\frac{v}{u + \hnd v - 3}\right)        \label{fos.v.u.3}   
    \end{align}
\end{subequations}
and treat the limit in parentheses with de L'H\^{o}pital's theorem
\begin{align}\label{lim.u=3}
  \lim_{u\rightarrow3}\frac{v}{u + \hnd v - 3} & = \lim_{u\rightarrow3}\frac{v_{u}}{1 + \hnd_{u} v + \hnd v_{u}} \nonumber \\
                                               & = \frac{v_{u}(3)}{1 + \hnd(3)\,v_{u}(3)} \; .
\end{align}
The substitution of \REq{lim.u=3} into \REqq{fos-u.3} yields
\begin{subequations}\label{fos-u.3.f}
    \begin{align}
        & \xiy_{u}(3) =   \frac{5}{6}\,\xiy(3)                      \label{fos.xi.u.3.f}  \\[1ex]
        & v_{u}(3)    = - \frac{5}{3}  \frac{1}{\hnd(3)} \; .       \label{fos.v.u.3.f}   
    \end{align}
\end{subequations}
\REqqb{fos-u.3.f} are essential for the numerical integration of \REqd{fos.xi.u}{fos.v.u}. 
\REqb{fos.v.u.3.f} is the generalization of a familiar result relative to the \pg;\footnote{For the case of the \pg, see Chandrasekhar's derivation that leads to his Eq. (466) at page 169 of \ocite{sc1957}. Padmanabhan also mentions the same result at page 653 of \ocite{tp1989ajss}, just below Eq. (18), and at page 314 of \ocite{tp1990pr}, just below Eq. (4.26), but without giving a derivation.} its substitution into \REq{lim.u=3} provides
\begin{equation}\label{lim.u=3.f}
  \lim_{u\rightarrow3}\frac{v}{u + \hnd v - 3} = \frac{5}{2} \frac{1}{\hnd(3)} \; ,
\end{equation}
a result that will prove useful later.

The change of independent variable is now complete: \REq{fos.rcp}, \REq{fos.xi.u}, \REq{fos.v.u}, \REq{fos.u.rev} compose a first-order system of governing equations with $u$ as the independent variable, accompanied by \REqd{v.is.3}{xi.is.uw} as boundary conditions and by \REq{v.is.uw} as the auxiliary condition to fix the terminus $\uw$; the determinate forms given by \REqq{fos-u.3.f} also belong to the system.
The latter becomes closed in the mathematical sense upon the explicit assignment of the reduced chemical potential, such as \REq{rcp.vdw.nd}, for example.
In the sequel, we will refer to this converted scheme with the label \fomt.

\subsection{\anttlc Perfect gas\label{pg}}
\subsubsection{\anttlc Preliminary remarks\label{pr}}
The \pg\ is characterized by  two basic assumptions: absence of (any kind of) molecular forces and of molecular size.
Experience teaches that real gases, those that actually exist in nature, may {behave} according to the \pg\ under specific experimental circumstances.
Is self-gravitation included in them?
Some colleagues opt for a positive answer and consider a self-gravitating perfect gas as an admissible reasonable approximation.\footnote{We conjecture that this attitude hinges on historical reasons tracing back to the pioneers in astronomy and astrophysics. See \Refsma{45-52}.}
Others\footnote{We belong to this camp.} are more inclined toward a negative answer motivated by the physical observation that within a microscopic perspective, whether classical or quantum mechanical, the hamiltonian of the \pg\ includes only kinetic-energy terms; therefore, such a model is only compatible with a vanishing gravitational constant \itm{(G\rightarrow0)} in order to have switched off the potential-energy terms in the hamiltonian.
There is no amount of conceptual funambulism that, springing from such a microscopic baseline, can make self-gravitational effects appear at the macroscopic level.
Thus, they look at a self-gravitating perfect gas as a physically inconsistent association of concepts rather than as a reasonable approximation.

This difference of opinions notwithstanding, we perpetrate now a mathematical funambulism: on the one hand, we dismiss molecular gravitational forces and molecular size
\begin{equation}\label{pga}
   \alphay = \betay = 0
\end{equation}
in the \vdw\ but, on the other hand, we retain a non-vanishing gravitational constant \itm{(G\neq0)} in the gravitational-field governing equations [\Reqsma{1a}{12}].
The latter move is the reason why the term \itm{3N\xiy} {survives} in \REq{sode.ss.vdw.nd} and the \pg\ acquires the capability to self-gravitate.\footnote{The physical fallacy hidden behind this move is tremendously brought to light by the outstanding analysis carried out by Saslaw\cite{wc1968mnras} in his first paper of the series addressing gravithermodynamics. In Sec. 3, he considered a \vdw\ as an example of ``imperfect self-gravitating gas''. In Sec. 3.2, he evaluated the van der Waals' constant $a$, notoriously connected to the molecular forces, and showed beyond any doubt in \mbox{Eq. (5)} that $a$ is linearly proportional to $G$; thus, the removal of molecular gravitational forces \itm{(a=0)} called for by the \pg\ requires {a fortiori} the inescapable vanishing of the gravitational constant \itm{(G=0)}. There is no way around it.}

\subsubsection{\anttlc Simplification of the governing equations\label{ge.pg}}
In compliance with the simplification enforced by \REq{pga}, the reduced chemical potential
  \begin{equation}\label{rcp.pg.nd}
     \phi(\xiy) = \ln\xiy 
  \end{equation}
follows from \REq{rcp.vdw.nd} and the thermodynamic function $\hnd$ [\REq{h.vdw}] undergoes a drastic simplification that reduces it to a constant
  \begin{equation}\label{h.pg}
     \hnd(\xiy) = 1 \;  . 
  \end{equation}
In turn, \REq{h.pg} allows the simplification of \REqd{fos.xi.u}{fos.v.u} to the forms
\begin{subequations}\label{fos-u.pg}
    \begin{align}
        & \xiy_{u} =   \frac{v}  {u + v - 3} \frac{\xiy}{u}     \label{fos.xi.u.pg}  \\[1ex]
        & v_{u}    = - \frac{u-1}{u + v - 3} \frac{v}{u}        \label{fos.v.u.pg}   
    \end{align}
\end{subequations}
the second of which is the differential equation we all became acquainted with from Chandrasekhar's textbook.\cite{sc1957} 
Something really remarkable\footnote{One of those strokes of magic that mathematics is surprisingly capable of sometimes.} happened mathematically in the transition from \REqd{fos.xi.u}{fos.v.u} to \REqq{fos-u.pg}:
\REq{fos.v.u.pg} decouples from \REq{fos.xi.u.pg} as a consequence of \REq{h.pg}, and becomes a standalone first-order differential equation whose integration with the boundary condition enforced by \REq{v.is.3} provides a universal solution represented by the spiraling curve in the \itm{u,v} plane whose geometrical characteristics, well described in the literature, we all are familiar with.
All isothermal spheres belong to this curve; the one corresponding to a given gravitational number is picked out by \REq{v.is.uw}, the selector\footnote{Padmanabhan\cite{tp1989ajss,tp1990pr} used the total energy as selector but the two approaches are consistent. To prove the equivalence, we start from \Reqma{121}, substitute $\lambda$ for its left-hand side in accordance to Padmanabhan's notation, $\uw/3$ for $\protect\xiyf(1,N)\equiv\protect\xiyf(\uw)$ [\REq{xi.is.uw}], and $v(\uw)$ for $N$ [\REq{v.is.uw}]. Thus, the total energy reads $\lambda = \left(\uw - {3}/{2}\right)/v(\uw)$
that compares with \mbox{Eq. (22)} of \ocite{tp1989ajss} or \mbox{Eq. (4.29)} of \ocite{tp1990pr}. The latter expression can be rewritten as $v(\uw) = \left(\uw - {3}/{2}\right)/\lambda$ which reveals the terminus $\uw$ being fixed by the intersection between the universal solution $v(u)$ and the straight line $ v = \left(u - {3}/{2}\right)/\lambda $ shown in Fig. 2 of \ocite{tp1989ajss} or Fig. 4.2 of \ocite{tp1990pr}. However, it must be kept in mind that multiple solutions with same total energy (Padmanabhan's selector) correspond to different gravitational numbers and that multiple solutions with same gravitational number (our selector) correspond to different total energies.} tasked to fix the terminus $\uw$.
Afterwards, with the universal solution $v(u)$ and the terminus $\uw$ in hand, the density profile follows from the integration of \REq{fos.xi.u.pg} with its accompanying boundary condition [\REq{xi.is.uw}]
\begin{equation}\label{xi-u.pg}
   \xiy(u) = \frac{\uw}{3} \exp\left[ \int_{\uw}^{u} \frac{v(s)}{s + v(s) - 3} \frac{1}{s}\,ds \right]
\end{equation}
in which $s$ is a dummy integration variable; the integral clearly requires numerical treatment.
The evaluation of the central density
\begin{equation}\label{xi-u.pg.0}
   \xiy(3) = \frac{\uw}{3} \exp\left[ \int_{\uw}^{3} \frac{v(s)}{s + v(s) - 3} \frac{1}{s}\,ds \right]
\end{equation}
does not present any difficulty because, due to \REqd{lim.u=3.f}{h.pg}, the integrand stays finite
\begin{equation}\label{lim.u=3.icd}
  \lim_{u\rightarrow3}\frac{v}{u + v - 3}\frac{1}{u} = \frac{5}{6} \; .
\end{equation}
Finally, the radial-coordinate profile follows from \REq{fos.u.rev}.
The universal solution $v(u)$ is unique and, given its spiraling behavior, it takes little effort to imagine that \REq{v.is.uw} can generate multiple values of the terminus $\uw$ for a specified gravitational number; for each of those values, we can run \REqd{xi-u.pg}{fos.u.rev} and generate the corresponding profiles.
Here we have the simple and transparent explanation about the existence of the multiple solutions we stumbled upon in \ocite{dg2019ejmb} by somewhat blind calculations with the isothermal Lane-Emden equation [\Reqma{42a}].

It is evident that the \fomt\ scheme, customized here for the \pg, is very appealing from a computational point of view.
Just to mention an example, all the calculations we carried out during the study described in \ocite{dg2019ejmb} were based on the M$_{2}$ scheme [\REqd{sode.ss.vdw.nd}{bc.rcp.nd}] with the reduced chemical potential of the \pg\ [\REq{rcp.pg.nd}], which corresponds to the isothermal Lane-Emden equation [\Reqma{42a}]; within that approach, we had to find out through experience that calculation success with a gravitational number presupposing multiple solutions\footnote{See for example \Rfisma{7}{8} which illustrate the two solutions corresponding to \itm{N=2.4}.} is hindered and plagued by the necessity to start from adequate initial radial profiles of density whose identification, we humbly admit, was a real form of art requiring sharp mathematical intuition.
The path that goes through the numerical machinery required by the \fomt\ scheme turns out to be smoother in this regard;
the mentioned hindrance disappears because there is no need of devising any initial profile.
\begin{figure}[b!] 
  \vspace{0.5\baselineskip}
  \gfbox{\wbox}{\includegraphics[keepaspectratio=true, trim = 9ex  8ex 7ex 12ex , clip , width=.97\columnwidth]{\gdir/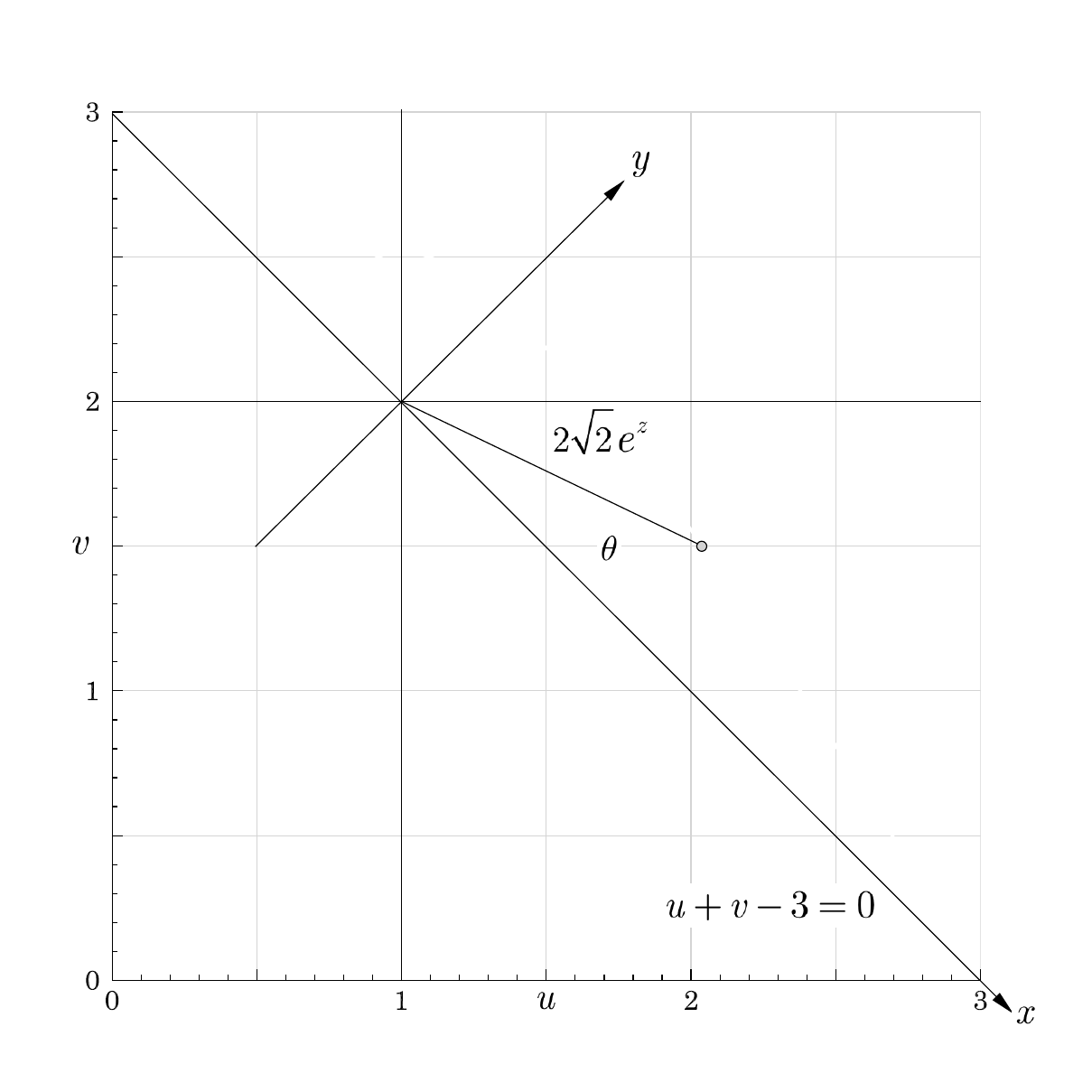}}\\  [-0.5\baselineskip]
  \caption{Coordinate transformations for the integration of \REq{fos.v.u.pg}.\label{ct}}
\end{figure}

The numerical integration of \REq{fos.v.u.pg} is definitely the fundamental step and it must be taken for the obvious necessity to respect self-includedness.
Surprisingly, to the best of the literature we consulted, we did not find any author that engaged in such a task.
Chandrasekhar and Wares\cite{sc1949aj} tabulated the universal solution $v(u)$ but, according to what they declared in their brief note published in the Astrophysical Journal in 1949, they did not integrate directly \REq{fos.v.u.pg} but rather their isothermal Lane-Emden equation [Eq. (374) in \ocite{sc1957}] and calculated {a posteriori} the \itm{u,v} variables from their definitions [Eq. (400) in \ocite{sc1957}].
So, we rose to the challenge and decided to have a go at it.

\subsubsection{\anttlc Integration of \REq{fos.v.u.pg}\label{dvdu.int}}

Admittedly, we start from the vantage position of knowing how the solution looks like and that helps to realize that it is not a good idea to integrate \REq{fos.v.u.pg} in the \itm{u,v} plane due to the expected numerical chicanery required to deal with the points of the curve at which the slope $v_{u}$ becomes vertical.
So, as illustrated in \Rfi{ct}, first we shift the axis origin \itm{(u=0, v=0)} to the center of the spiral \itm{(u=1, v=2)} and rotate the axes 45$^{\circ}$ clockwise
\begin{subequations}\label{cts}
\begin{equation}\label{ct.xy}
   \left\{ \begin{aligned}  u & = \left(y+x+\sqrt{2}\right)/\sqrt{2} \\[.1\baselineskip]   v & = \left(y-x+2\sqrt{2}\right)/\sqrt{2}  \end{aligned} \color{white}\right\}\color{black} \; ;
\end{equation}
then we introduce normalized polar coordinates
\begin{equation}\label{ct.tz}
   \left\{ \begin{aligned}  x & = 2 \sqrt{2} e^{z} \cos\protect\thetay \\[.4\baselineskip] y & = 2 \sqrt{2} e^{z} \sin\protect\thetay  \end{aligned} \color{white}\right\}\color{black}
\end{equation}
in order to generate the final transformation
    \begin{equation}\label{ct.uv}
      \left\{   \begin{aligned} u & = 1+2 e^{z}(\sin\thetay+\cos\thetay) \\[.5\baselineskip] v & = 2 \left[ 1 + e^{z}(\sin\thetay-\cos\thetay)\right] \end{aligned}    \right.
    \end{equation}
\end{subequations}
that converts \REq{fos.v.u.pg} into
\begin{widetext}
\begin{equation}\label{fos.v.u.pg.zt}
  z_{\thetays} =
  - \frac{2 \sin\thetay \left[ e^{z} ( \sin\thetay + \cos\thetay ) + \dfrac{1}{2} \right] ( \sin\thetay + \cos\thetay ) - \left[ e^{z} ( \sin\thetay - \cos\thetay ) + 1 \right] ( \sin^{2}\thetay - \cos^{2}\thetay )}
         {2 \sin\thetay \left[ e^{z} ( \sin\thetay + \cos\thetay ) + \dfrac{1}{2} \right] ( \sin\thetay - \cos\thetay ) + \left[ e^{z} ( \sin\thetay - \cos\thetay ) + 1 \right] ( \sin\thetay + \cos\thetay )^{2}    }
\end{equation}
\end{widetext}
and \REq{v.is.3} into the boundary condition
\begin{equation}\label{z.is.0}
  z(0) = 0 \; .
\end{equation}
It may seem, at first sight and maybe with reason, that, by looking at the differential equation in polar coordinates [\REq{fos.v.u.pg.zt}], the series of coordinate transformations has injected mathematical cumbersomeness; but that fee turns out to be irrelevant because the new first-order differential equation is not as thorny as \REq{fos.v.u.pg} is for the purpose of numerical operations.
Indeed, the denominator on the right-hand side of \REq{fos.v.u.pg.zt} never vanishes for \itm{\thetay>0}.
It does vanish for \itm{\thetay=0} but so does the numerator and the first derivative \itm{z_{\thetays}} becomes an indeterminate form.
This is the only preoccupation, really, but it is easily disposed of because the indeterminateness can be treated with a linearization technique based on the smallness of a vanishing $\thetay$ whose application leads to
\begin{equation}\label{fos.v.u.pg.zt.0}
  z_{\thetays}(0) = - 4
\end{equation}
which we provide but omit the lengthy details.\footnote{\label{en.hints}\textcolor{black}{Some hints for the interested reader. Assume \itm{\protect\thetayf\ll1} and linearize numerator and denominator of \REq{fos.v.u.pg.zt}, including the exponentials; then passing the resulting expression to the limit \itm{\protect\thetayf\rightarrow0} yields a quadratic equation for \itm{z_{\theta}(0)}. Compatibility with \REq{fos.xi.u.3.f} selects the applicable root.}}
The numerical integration of \REq{fos.v.u.pg.zt} with the boundary condition and particular form given, respectively, by \REqd{z.is.0}{fos.v.u.pg.zt.0} is straightforward; the universal solution $z(\thetay)$ and its first derivative are illustrated in \Rfi{z} in the interval [0\,,\,6\,\piy].
\begin{figure}[h]
  \vspace{0.5\baselineskip}
  \gfbox{\wbox}{\includegraphics[keepaspectratio=true, trim = 8ex  2ex 3ex 12ex , clip , width=.97\columnwidth]{\gdir/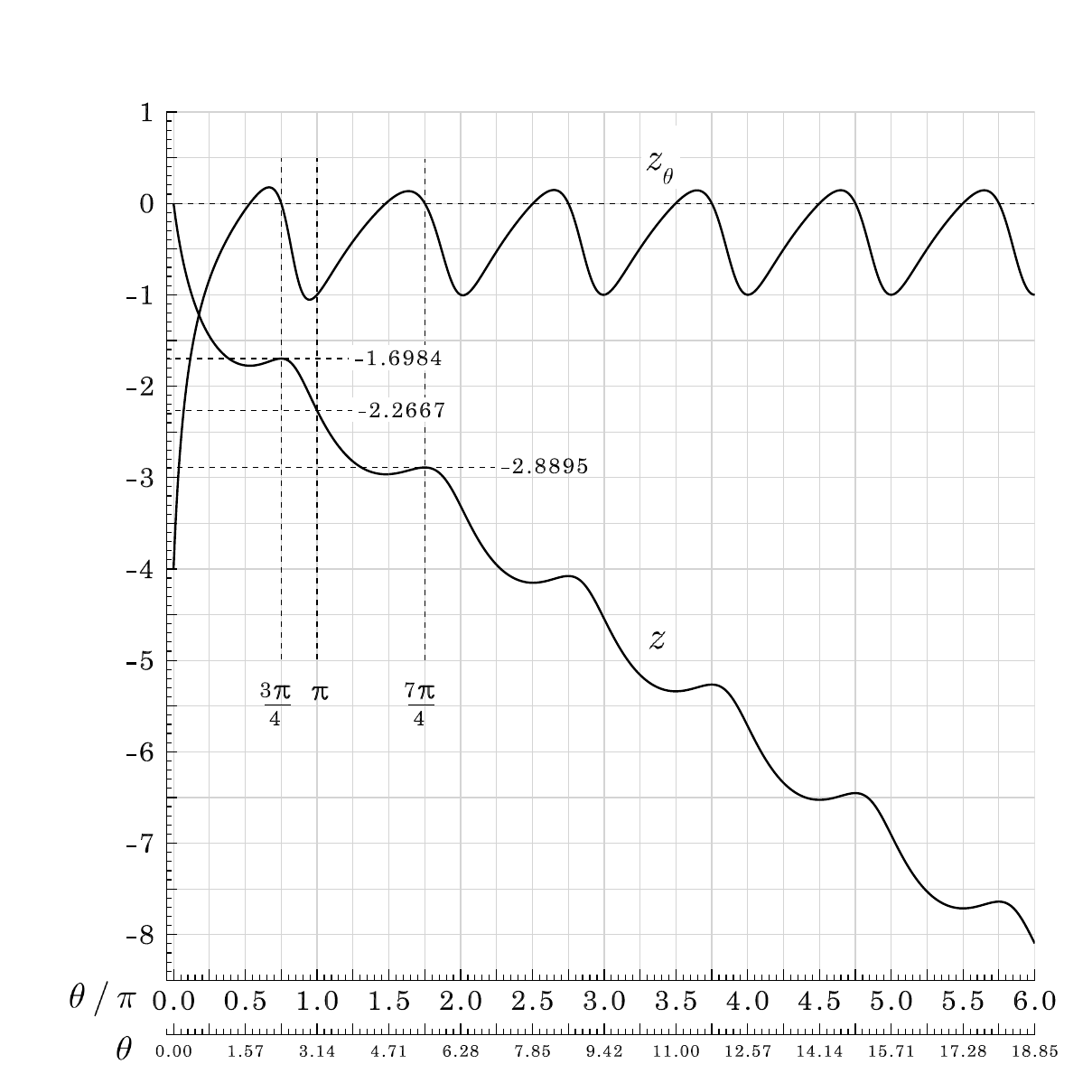}}\\  [-0.5\baselineskip]
  \caption{The universal solution $z(\protect\thetayc)$ and its first derivative.}\label{z}
\end{figure}
Afterward, \REqq{ct.uv} generate the graph of the \itm{u,v} variables in a parametric form shown in \Rfi{uv-t}
\begin{figure}[h]
  \vspace{0.5\baselineskip}
  \gfbox{\wbox}{\includegraphics[keepaspectratio=true, trim = 8ex  2ex 3ex 12ex , clip , width=.97\columnwidth]{\gdir/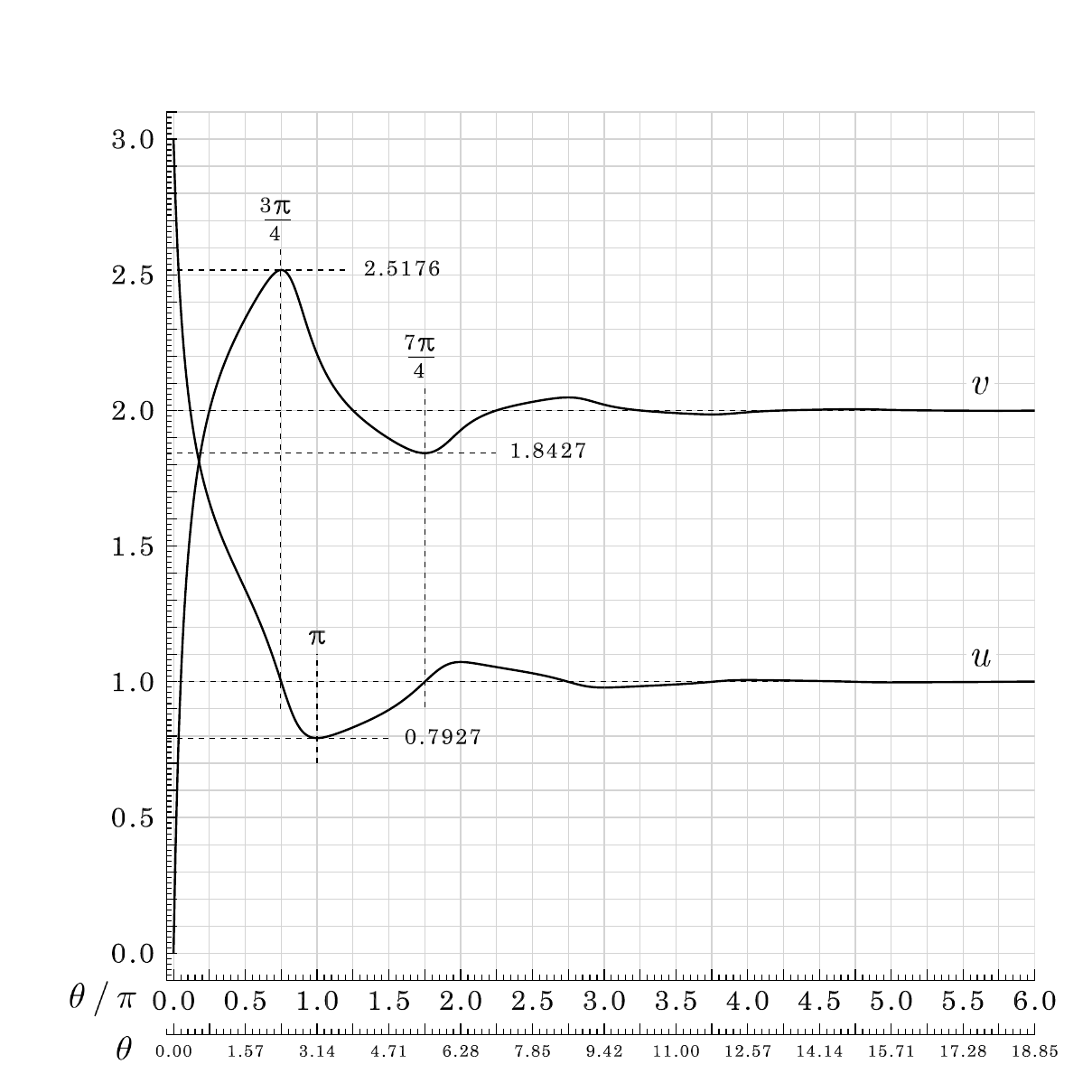}}\\  [-0.5\baselineskip]
  \caption{The universal solutions $u(\protect\thetayc)$ and $v(\protect\thetayc)$ in parametric form.}\label{uv-t}
\end{figure}
and the elimination between them of the parameter $\thetay$ leads to the universal solution $v(u)$ represented by the well known curve illustrated in \Rfi{uv};
\begin{figure}[h]
  \vspace{0.5\baselineskip}
  \gfbox{\wbox}{\includegraphics[keepaspectratio=true, trim = 9ex  8ex 12ex 12ex , clip , width=.97\columnwidth]{\gdir/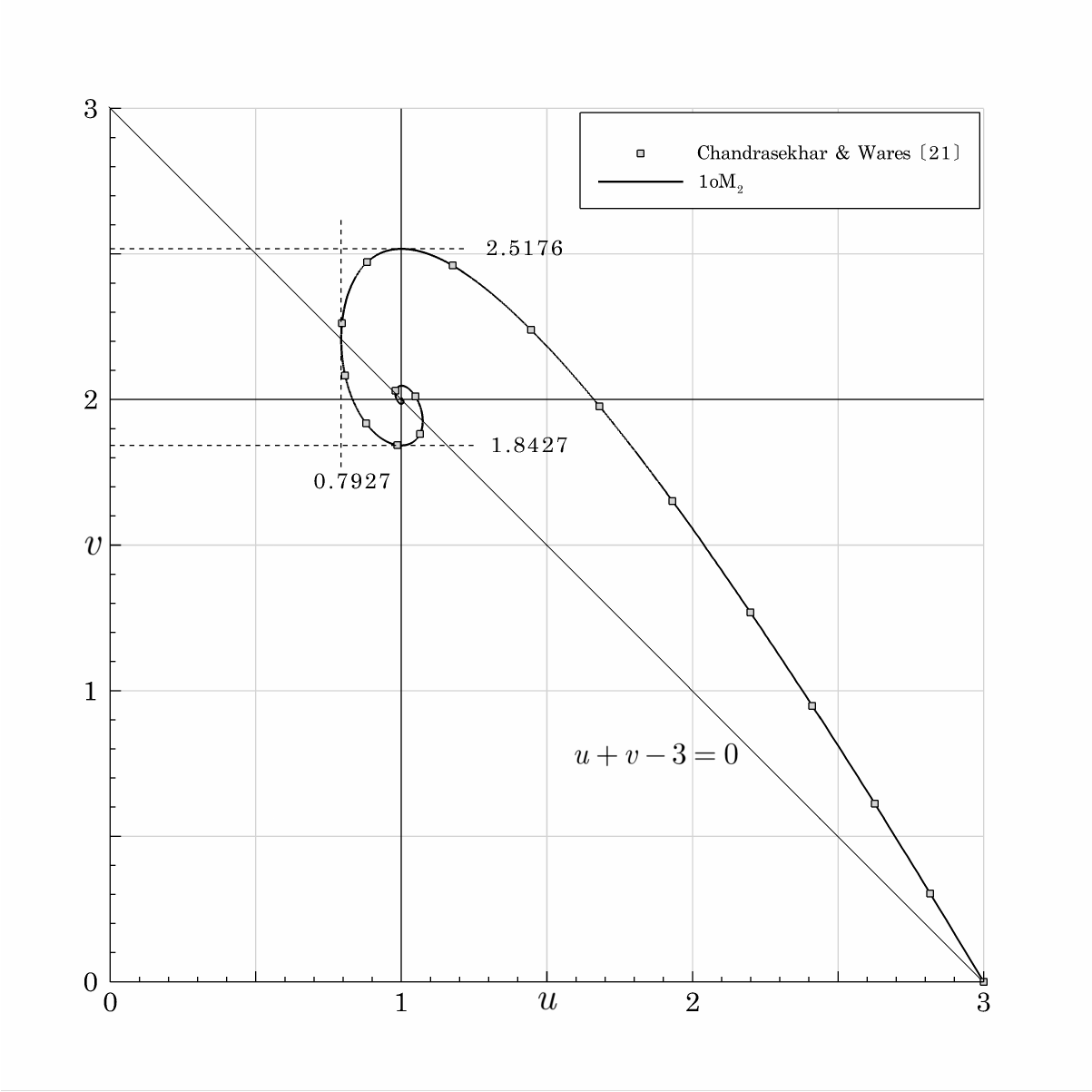}}\\  [-0.5\baselineskip]
  \caption{The universal solution $v(u)$.}\label{uv}
\end{figure}
we have superposed the values published by Chandrasekhar and Wares\cite{sc1949aj} as evidence of validation of our calculations.

With reference to \Rfi{z}, the angular positions of the extrema of $z(\thetay)$ are determined by the zeros of \REq{fos.v.u.pg.zt}.
Those of the maxima arise from
\begin{equation}\label{fos.v.u.pg.zt.0.max}
  \sin\thetay + \cos\thetay = 0
\end{equation}
and follow the progression
\begin{equation}\label{zms}
  \subeqn{\thetay^{\ssub{0.65}{M,k}} = \left(k+\frac{3}{4}\right)\piy}{k=0,1,2,\ldots}{1}
\end{equation}
\REqb{ct.uv} (top) confirms analytically that all maxima of the function $z(\thetay)$ fall on the vertical line \itm{u=1}, as reflected in \Rfid{uv-t}{uv}.
The same progression makes vanish also the first derivative of $v(\thetay)$
\begin{equation}\label{dvdt}
  v_{\thetays}  = 2 e^{z}\left[ z_{\thetays}(\sin\thetay-\cos\thetay) + (\sin\thetay+\cos\thetay)\right]
\end{equation}
but \REq{ct.uv} (bottom) converts the maxima of $z$ into the maxima of $v$ for even values \itm{(k=0,2,\ldots)} of \REq{zms} and flips maxima into minima for odd values \itm{(k=1,3,\ldots)}, as shown in \Rfi{uv-t}.
The first two values \itm{\thetay^{\ssub{0.65}{M,0}}=3\piy/4}, \itm{\thetay^{\ssub{0.65}{M,1}}=7\piy/4} and their maxima $z(3\piy/4)=-1.6984$, $z(7\piy/4)=-2.8895$ are important because they identify the endpoints of the variable $v$, as well as of the gravitational number according to the dictamen of \REq{v.is.uw}, which fix the boundaries delimiting the zone within which multiple solutions exist.
These endpoints {can be calculated} from \REq{ct.uv} (bottom) and turn out to be \itm{v(3\piy/4)=2.5176}, the disquieting upper bound of the gravitational number, and \itm{v(7\piy/4)=1.8427}, the somewhat less worrisome lower bound; %
\footnote{We consider the possibility to calculate these endpoints as another mark of superiority of the \fomt\ scheme with respect to the M$_{2}$ scheme. In the latter scheme, we basically treated the isothermal Lane-Emden equation as a boundary-value problem in which the gravitational number is prescribed [\Reqqma{42}]; in that calculational context, upper and lower bounds of the gravitational number can only be obtained through a strategy of tedious trial and error to gain accurate decimal digits which require repetitive and time consuming running of the numerical algorithm, the tediousness being particularly acute for the upper bound.
In connection with \REq{ct.uv} (bottom), Darwin's statement
``... I am unable to find any analytical relationship by which the minimum value of \itm{\frac{1}{3}\protect\betayf^{2}} can be deduced.''
at page 19 of \ocite{gd1889ptrs}, comes to mind again; in Darwin's notation, \itm{\frac{1}{3}\protect\betayf^{2}} corresponds to our \itm{1/N}.
We quoted it in \Rsema{4}, at the bottom of the right column of page 85, where we also conceded the same incapability.
We are pleased to have moved forward to a position in which we are not obliged to admit the same defeat this time.}
they and their associated boundaries are indicated in \Rfid{uv-t}{uv} from whose visual inspection we recognize the well known possibilities
\begin{equation}\label{Ncases}
   \begin{cases}
      N < 1.8427                & \text{one solution}       \\[1ex]
      1.8427 \leq N \leq 2.5176 & \text{multiple solutions} \\[1ex]
      2.5176 < N                & \text{no solution}
   \end{cases}
\end{equation}
conveniently formulated in terms of the gravitational number.
The minima of the function $z(\thetay)$ do not seem to have any particular relevance; for the sake of completeness, we mention briefly that their angular positions arise from
\begin{equation}\label{fos.v.u.pg.zt.0.min}
  \left[2\sin(2\thetay) - \cos(2\thetay)\right]e^{z(\thetays)} + \cos\thetay = 0
\end{equation}
and follow a progression very nearly to
\begin{equation}\label{zmins}
  \subeqn{\thetay^{\ssub{0.65}{m,k}} \simeq \left(k+\frac{1}{2}\right)\piy}{k=0,1,2,\ldots}{1}
\end{equation}
due to the negligibility of the first term on the left-hand side of \REq{fos.v.u.pg.zt.0.min} for increasing $k$, in practice \itm{k>2} as it can be appreciated graphically from \Rfi{z}.
According to \REq{zmins}, the minima fall very nearly onto the $y$ axis of \Rfi{ct}, which we did not draw in \Rfi{uv} to avoid cluttering the figure.

The vanishing of the first derivative
\begin{equation}\label{dudt}
  u_{\thetays}  = 2 e^{z}\left[ z_{\thetays}(\sin\thetay+\cos\thetay) - (\sin\thetay-\cos\thetay)\right]
\end{equation}
of \REq{ct.uv} (top) identifies the angular positions of the extrema of the function $u(\thetay)$; with reference to \Rfi{uv-t}, they follow the progression
\begin{equation}\label{ums}
  \subeqn{\thetay^{\ssub{0.65}{E,k}} = \left(k+1\right)\piy}{k=0,1,2,\ldots}{1}
\end{equation}
The first value \itm{\thetay^{\ssub{0.65}{E,0}}=\piy} and the corresponding value $z(\piy)=-2.2667$ are important because they identify the first and deepest minimum of $u$ which is also the left end point in the \itm{u,v} plane that sets the boundary at which the curve of the universal function $v(u)$ begins to spiral.
This end point can be calculated from \REq{ct.uv} (top) and turns out to be \itm{u(\piy)=0.7927}; at the same angular position, \REq{ct.uv} (bottom) yields \itm{v(\piy)=2.2073}.
It is straightforward to check that the point with those coordinates belongs also to the  oblique line [\REq{vden.pg}] and, therefore, it is also the leftmost point of vertical slope on the curve in \Rfi{uv}.

\subsubsection{\anttlc A validation exercise for the \fomt\ scheme with the \pg\label{ve}}

Now that we are in possession of the universal functions $z(\thetay)$, $u(\thetay)$, $v(\thetay)$, and  $v(u)$, we can concentrate on the completion of the calculational program of the \fomt\ scheme, which consists of the application of \REq{v.is.uw} to determine the terminus $\uw$, of \REq{xi-u.pg} to determine the density profile, and of \REq{fos.u.rev} to determine the radial-coordinate profile.
In that regard, we thought it is appropriate to reconsider the case \itm{N=2.4} as validation exercise for the \fomt\ scheme because we had already calculated it with the isothermal Lane-Emden equation [\Reqma{42a}] in \ocite{dg2019ejmb}.

Regarding the solution of the auxiliary condition [\REq{v.is.uw}], we can operate either directly with the function $v(u)$ or with the latter's parametric form
  \begin{equation}\label{ct.uv.N}
    \left\{   \begin{aligned} \uw & = 1+2 e^{z(\thetays_{\ssub{0.5}{t}})}(\sin\tw+\cos\tw) \\[.5\baselineskip]
                              v(\uw) & = 2 \left[ 1 + e^{z(\thetays_{\ssub{0.5}{t}})}(\sin\tw-\cos\tw)\right] = N \; .\end{aligned}    \right.
  \end{equation}
extractable from \REq{ct.uv} after proper adaptation.
The bottom equation gives the terminus $\tw$; the top equation gives the terminus $\uw$.
For example, in the noteworthy case of \itm{N=2}, the bottom equation becomes
\begin{equation}\label{ct.uv.N=2.b}
  \sin\tw-\cos\tw=0
\end{equation}
and singles out the angular progression \itm{\tw=\left(k+1/4\right) \piy} \itm{(k=0,1,2,\ldots)}
from which the top equation yields the infinite termini
\begin{equation}\label{ct.uv.N=2.t}
  \uw = 1 + 2 \sqrt{2} (-1)^{k} e^{z\left[\left(k+\frac{1}{4}\right)\piys\right]}
\end{equation}
In the case of \itm{N=2.4}, \REq{v.is.uw} or \REq{ct.uv.N} yields two termini
\begin{equation}\label{v.is.uw.N=2.4}
  \uw = \begin{cases} 1.2652 & \quad\text{a-solution} \\[1ex] 0.8309 & \quad\text{b-solution} \end{cases}
\end{equation}
\begin{figure}[h]
  \vspace{0.5\baselineskip}
  \gfbox{\wbox}{\includegraphics[keepaspectratio=true, trim = 9ex  8ex 7ex 12ex , clip , width=.97\columnwidth]{\gdir/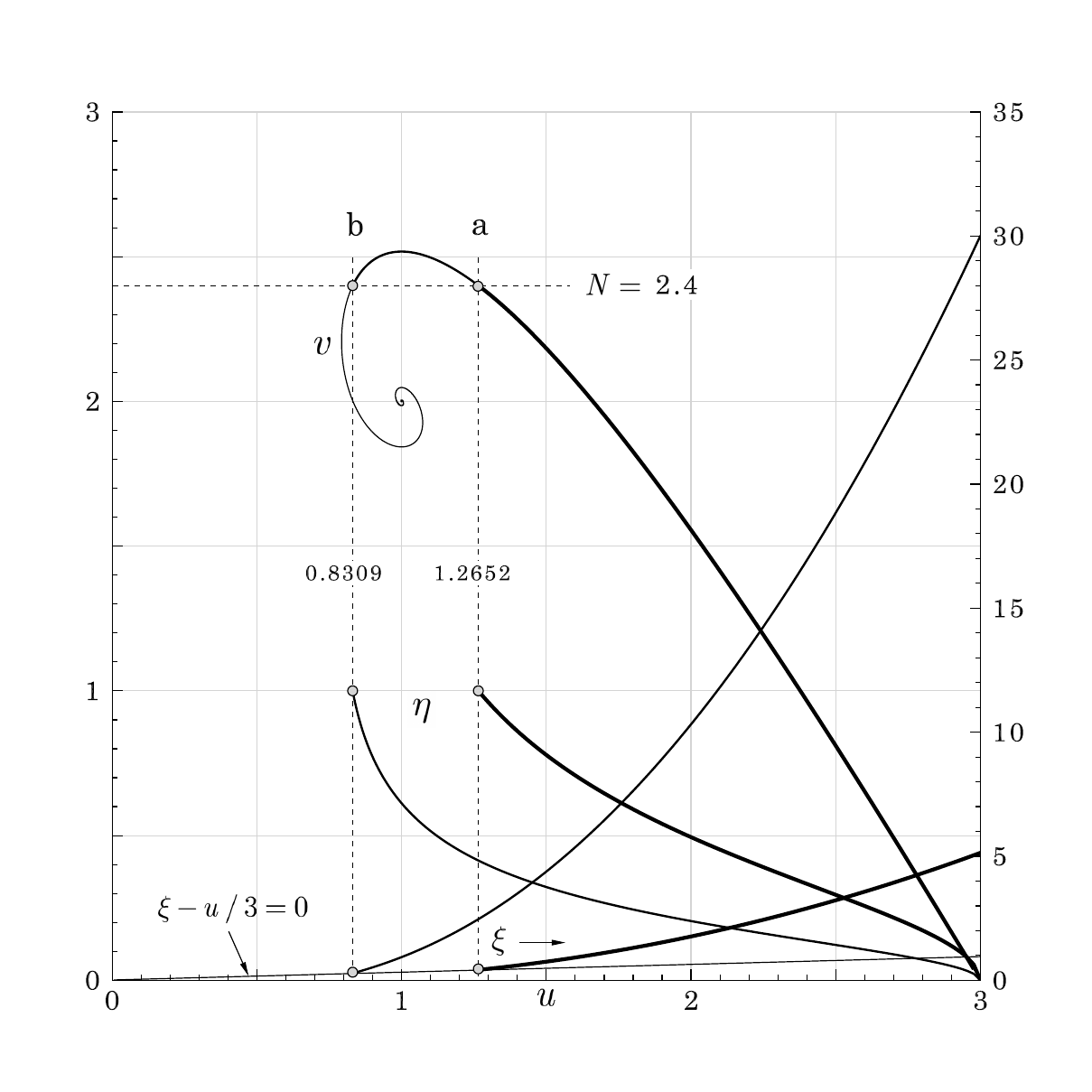}}\\  [-0.5\baselineskip]
  \caption{The two solutions at $N=2.4$ with $u$ as independent variable.}\label{uv2.4}
\end{figure}
The corresponding solutions are overlaid on the universal curve $v(u)$ in \Rfi{uv2.4}; the a-solution (thicker line) extends, and covers the b-solution, down to the terminus \itm{\uw=1.2652}; beyond that, the b-solution emerges and extends further down to the terminus \itm{\uw=0.8309}.
The density and radial-coordinate profiles generated by \REqd{xi-u.pg}{fos.u.rev} for each of the termini are also shown in \Rfi{uv2.4}.
Then, in \Rfi{xieta2.4}, we have reconstructed \Rfima{7} by plotting density vs radial coordinate and validated the \fomt\ scheme by superposing the results we obtained in \ocite{dg2019ejmb}.
\begin{figure}[h]
  \vspace{0.5\baselineskip}
  \gfbox{\wbox}{\includegraphics[keepaspectratio=true, trim = 14ex  6ex 4ex 12ex , clip , width=.97\columnwidth]{\gdir/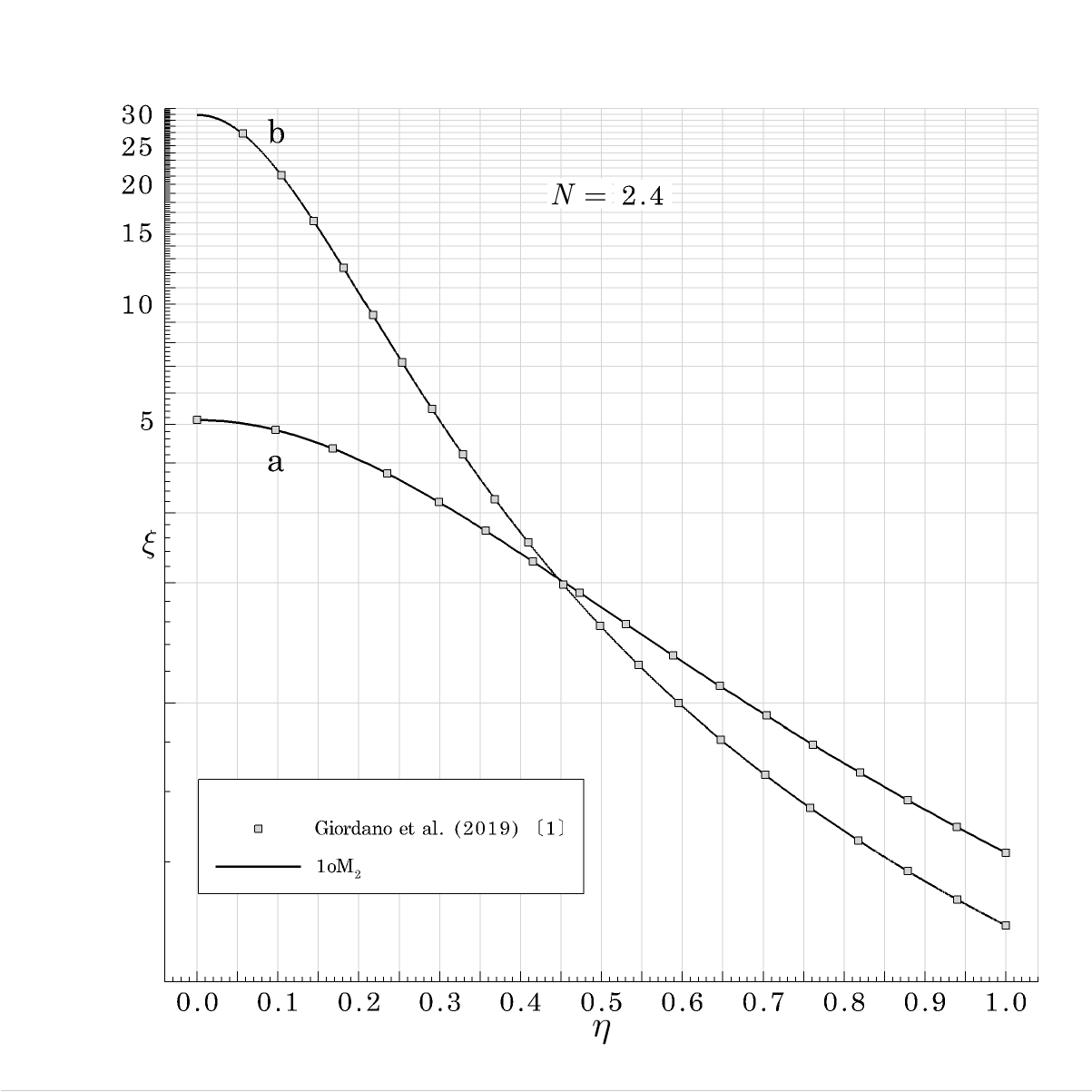}}\\  [-0.5\baselineskip]
  \caption{The two density profiles at $N=2.4$; reconstruction of \Rfima{7}.}\label{xieta2.4}
\end{figure}

\subsubsection{\anttlc Postliminary remarks\label{por}}
At the cost of somewhat minimal repetition, we have revisited in \Rses{pr}{ve} the familiar results relative to the \pg\ with the main purpose in mind to put them in a more general and rational perspective;
that has also permitted us to acquire a broader and deeper understanding of the corresponding phenomenology and has served as an introductory preface to the \vdw.
The results are impeccable from a mathematical point of view but, from a physical one, they are inevitably subjected to the Damocles' sword represented by the mathematical funambulism  we have perpetrated in the paragraph closing \Rse{pr}, the one containing \REq{pga}.

\subsection{\anttlc Van der Waals' gas\label{vdwg}}

\subsubsection{\anttlc Preliminary remarks\label{pr.vdw}}

Let us recall from \Rse{intro} the main driving motivation of this study: we wish to investigate the existence/absence of values of \itm{\alphay} and \itm{\betay} in correspondence to which the \pge\ could appear also for the \vdw.
Now, on the basis of what we have learned in \Rse{pg}, a preliminary, and admittedly intuitive and naive, reflection could come hurriedly to mind: the \pge\ are a direct consequence of the absence of molecular gravitational forces and molecular size [\REq{pga}] which, in turn, relegates the thermodynamic function $\hnd$ [\REq{h.vdw}] to the role of a unit constant [\REq{h.pg}]; therefore, the \pge\ cannot occur for \vdw\ because \REqd{pga}{h.pg} do not hold and $\hnd$ retains its full status of mathematical bond between \REqd{fos.xi.u}{fos.v.u} compelling their simultaneous solution.
And, on the basis of such a reflection, the proposed investigation could be dismissed.
However a perspicacious reader may rightly object: first, one may not exclude \mbox{a priori} the existence of situations in which the interplay of $\alphay, \betay$ and $\xiy$ make negligible with respect to 1 the terms in \REq{h.vdw}; second, the \vdw\ may bring about unphysical results of its own and that have nothing to do with the \pge.
We concede the legitimacy of the objection and agree that the investigation retains its worthiness; however, before plunging into it, we deem it important to prefix two considerations.

The main goal of \Refmb\ was to show that the adoption of a thermodynamic model which consistently, although qualitatively, accounts for molecular gravitational attraction leads to the removal of the \pge, at least for some specific values of the physical parameters intervening in the model; the results described and discussed in \Refmb\ undeniably prove the achievement of that goal.
Moreover, experience teaches that every physical model has limits.
Thus, if there exist values of $\alphay$ and $\betay$ for which the \vdw\ presents unphysical results, well so be it; such an occurrence certainly does not detract the merit indicated in the former consideration.
As a matter of fact, we are already aware of some physical limitations of \vdw; just to mention two of them, the isotherms may trespass into the negative-pressure plan [\Reqmb{47} and \Reqqmb{48}] and the compressibility ratio [\Reqmb{50e}] is fixed to 3/8=0.375, a rigid constraint that affects the applicability of \vdw\ to real gases.\footnote{For example, the \vdw\ cannot describe water vapor because the latter's compressibility ratio is $\sim 0.23$.}
So, it should be hardly a surprise if it turns out that there exist values of $\alphay$ and $\betay$ producing physically inconsistent results for the \vdw\ in self-gravitating circumstances.

With these considerations preliminarily expressed, we can now proceed to the required investigation.
We upgrade \REq{pga} to
\begin{equation}\label{vdwa}
   \alphay \neq 0 \qquad \betay \neq 0
\end{equation}
and the consequent logical step is to have a look in depth at the characteristics of the thermodynamic function $\hnd$.

\subsubsection{\anttlc The thermodynamic function \scalebox{1.1}{$\hnd$} for the \vdw.\label{vdw.h}}
The non-negligibility of molecular size [\REq{vdwa} (right)] permits a slight adaptation of \REq{h.vdw} to an interesting form
\begin{equation}\label{h.vdw.s}
   \hnd(\chiy) = \frac{\left(1 - \chiy\right)^{2}}{1-2\omegay\chiy\left(1 - \chiy\right)^{2}}
\end{equation}
that reveals how $\hnd$ effectively depends only on the ratio
\begin{equation}\label{ab.r}
  \omegay = \frac{\alphay}{\betay} = \frac{A}{BR\overline{T}}
\end{equation}
[\Reqsmb{31b}{31c}] and on the scaled density
\begin{equation}\label{chi}
  \chiy = \betay\,\xiy
\end{equation}
that ranges in the interval [0,1].
For all intents and purposes, the ratio $\omegay$ defined in \REq{ab.r} can be considered itself a legitimate independent characteristic number that we can decide to use instead of $\alphay$.
The convenience in this swap resides in the fact that $\omegay$ does not depend on the average density as $\alphay$ does [\Reqmb{31b}].
Therefore, by working with $\omegay$ and $\betay$, we keep separate the effects of changing temperature [\REq{ab.r}] and average density [\Reqmb{31c}];
for example, we can experimentally control $\betay$ by changing the average density without the need to change the temperature, as it would instead be the case had we decided to operate with $\alphay$.

The characteristics of \REq{h.vdw.s} are shown in the graph of \Rfi{hvdw}.
\begin{figure}[]
  \vspace{0.5\baselineskip}
  \gfbox{\wbox}{\includegraphics[keepaspectratio=true, trim = 14ex  6ex 4ex 12ex , clip , width=.97\columnwidth]{\gdir/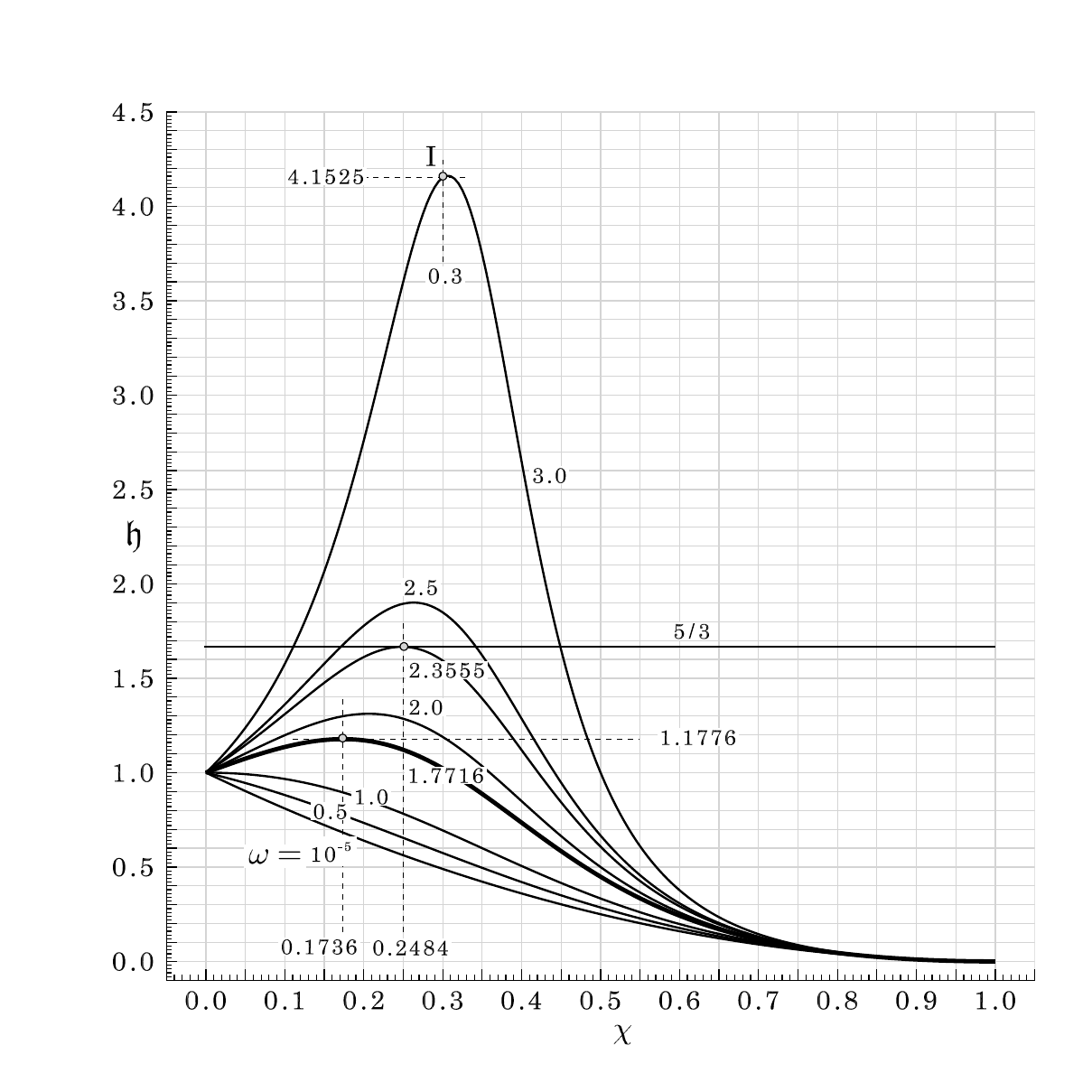}}\\  [-0.5\baselineskip]
  \caption{The thermodynamics function $\hnd$ for the \vdw; the thicker curve labeled $\protect\omegay=1.7716$ corresponds to \tcb. See caption of \Rfi{uindN1.45} for information related to point I.}\label{hvdw}
\end{figure}
The first derivative
\begin{equation}\label{hx.vdw.s}
   \hnd_{\chiys} = - 2 \frac{\left(1-\chiy\right)\left[1 - \omegay \left(1-\chiy\right)^{3}\right]}{\left[1-2\omegay\chiy\left(1 - \chiy\right)^{2}\right]^{2}}
\end{equation}
vanishes at \itm{\chiy=1}; thereat, the second derivative
\begin{equation}\label{hxx.vdw.s}
   \hnd_{\chiys\chiys} = 2 \frac{ 1+2\omegay\left(1-\chiy\right)^{2} \left[3-7\left(1-\chiy\right)+2\omegay\left(1-\chiy\right)^{4}\right]  }{\left[1-2\omegay\chiy\left(1 - \chiy\right)^{2}\right]^{3}}
\end{equation}
takes on the positive value \itm{\hnd_{\chiys\chiys}(1) = 2} and identifies the minimum \itm{\hnd^{\ssub{0.65}{m}}(1)=0}, which is shared by all the curves.
The first derivative vanishes also at
\begin{equation}\label{chiofmax}
  \chiy = 1 - \frac{1}{\omegay^{1/3}} \; ;
\end{equation}
thereat, the second derivative becomes negative
\begin{equation}\label{sdofmax}
  \hnd_{\chiys\chiys}\left(1 - \frac{1}{\omegay^{1/3}}\right) = - \frac{6}{\left(3-2\omegay^{1/3}\right)^{2}}
\end{equation}
and identifies the maximum
\begin{equation}\label{hmax}
  \hnd^{\ssub{0.65}{M}}\left(1 - \frac{1}{\omegay^{1/3}}\right) = \frac{1}{3\omegay^{2/3} - 2\omegay}
\end{equation}
The maximum is not visible if \itm{\omegay<1} because the corresponding value returned by \REq{chiofmax} falls to the left of the interval [0,1] (first two curves from bottom in \Rfi{hvdw}).
The maximum enters the interval at \itm{\chiy=0} if \itm{\omegay=1} and takes on the value \itm{\hnd^{\ssub{0.65}{M}}(0)=1} (third curve from bottom).
After that, the maximum penetrates into the interval, increases in magnitude, and its position shifts rightward monotonically.
We stopped at \itm{\omegay=3} in order not to trespass the phase-equilibrium threshold [\Reqmb{51} (top)] in compliance with our initially opted for restriction to deal only with gas phases (\Rse{intro}).

\Rfib{hvdw} gives ample evidence that the deviations of the \vdw's thermodynamic function $\hnd$ from the \pg's constant level [\REq{h.pg}] are abundantly beyond any hope of negligibility and forewarn about the strong coupling to be expected between \REqd{fos.xi.u}{fos.v.u}.

\subsubsection{\anttlc Integration of the governing equations\label{ge.vdw}}
Within the more general situation enforced by \REq{vdwa}, the thermodynamic function $\hnd$ applies in all its entirety  [\REq{h.vdw}] and we must deal with the full set of equations described in \Rse{civ}.
In principle, there is no impediment to resort again to the coordinate transformation introduced for the \pg\ that brings forth the polar coordinates \itm{z,\thetay} [\REqq{ct.uv} and \Rfi{ct}];
furthermore, this time we even make more efficacious its application by noticing that we can rearrange \REqq{ct.uv} as
\begin{equation}\label{ct.uv,inv}
  \left\{ \begin{aligned} e^{z}(\sin\thetay+\cos\thetay) & = \frac{u - 1}{2} \\[.5\baselineskip] e^{z}(\sin\thetay-\cos\thetay) & = \frac{v - 2}{2} \end{aligned}    \right.
\end{equation}
and rephrase the first derivatives given in \REqd{dvdt}{dudt} into the more convenient, although equivalent, forms
\begin{equation}\label{dvudt}
  \left\{ \begin{aligned} v_{\thetays} & = z_{\thetays} \left( v - 2 \right) + \left(u - 1\right)  \\[.5\baselineskip] u_{\thetays} & = z_{\thetays} \left( u - 1 \right) - \left(v - 2\right) \; .\end{aligned}   \right.
\end{equation}
Moreover, the way in which $\xiy_{u}$ and $\xiy$ appear in \REq{fos.xi.u} suggests introducing the convenience variable
\begin{equation}\label{lnxi}
  w = \ln\xiy \; .
\end{equation}
Then \REqd{fos.xi.u}{fos.v.u} become\footnote{It is easy to verify that setting \itm{\hnd=1} in \REq{fos.w.z.vdw.z} retrieves the differential equation of the \pg\ [\REq{fos.v.u.pg.zt}] and decouples it from \REq{fos.w.z.vdw.w}.}
\begin{subequations}\label{fos.w.z.vdw}   
\begin{align} \negthickspace \negthickspace \negthickspace
   w_{\thetays} & =  \hnd \frac{v}{u+\hnd v - 3} \frac{z_{\thetays} \left( u - 1 \right) - \left(v - 2\right)}{u} \label{fos.w.z.vdw.w}
      \\[\baselineskip]
   z_{\thetays} & = - \frac{\left( u - 1 \right) - \dfrac{v}{u + \hnd v - 3} \dfrac{u-1}{u} \left(v - 2\right)}
                           {\left( v - 2 \right) + \dfrac{v}{u + \hnd v - 3} \dfrac{u-1}{u} \left(u - 1\right)} \label{fos.w.z.vdw.z}
\end{align}
\end{subequations}
in which the variables $u$ and $v$ must obviously be considered functions of $z$ and $\thetay$ in compliance with \REqq{ct.uv}.
The sphere's center \itm{(\thetay=0)} requires again a bit of attention but we have already removed the indeterminateness of the term \itm{v/(u+\hnd v - 3)} in \REq{lim.u=3.f} [\itm{\hnd(3)\rightarrow\hnd(0)}]; so, with additional account of \REqd{u.is.0}{v.is.0}, we obtain the determinate forms\footnote{Another quick verification: setting \itm{\hnd(0)=1} in \REq{fos.w.z.vdw.z.0} retrieves \REq{fos.v.u.pg.zt.0}.}
\begin{subequations}\label{fos.w.z.vdw.0}   
\begin{align} \negthickspace \negthickspace \negthickspace
   w_{\thetays}(0) & = - \frac{10\,    \hnd(0)}{5 - 3\, \hnd(0)} \label{fos.w.z.vdw.w.0} \\[.5\baselineskip]
   z_{\thetays}(0) & = - \frac{5 + 3\, \hnd(0)}{5 - 3\, \hnd(0)} \label{fos.w.z.vdw.z.0}
\end{align}
\end{subequations}
The central boundary condition [\REq{v.is.3}] turns again into the initial condition [\REq{z.is.0}] for the second differential equation [\REq{fos.w.z.vdw.z}].
The peripheral boundary condition [\REq{xi.is.uw}] must now be interpreted in terms of the terminus $\tw$ as
\begin{equation}\label{xi.is.uw.t}
  \xiy\left[\uw(\thetay_{t})\right] = \frac{u_{t}(\thetay_{t})}{3}
\end{equation}
and goes with the first differential equation [\REq{fos.w.z.vdw.w}]. 
Finally, for the auxiliary condition [\REq{v.is.uw}], we can recycle the parametric equations [\REq{ct.uv.N}] we have already used for the \pg\ because they are thermodynamic-model independent.
Here, once more, we reassure the reader that the mathematical cumbersomeness of \REqq{fos.w.z.vdw} is only apparent and of no hindrance at all for their numerical solution.
The differential equations turned out to be surprisingly docile when fed into the first-order finite-difference forward-marching scheme and the associated ``shooting'' technique we adopted for their numerical solution.
The shooting parameter is the central density; the numerical machinery is set in motion from an initial guess $\xiy_{c}$ which, from \REq{lnxi}, provides the initial value \itm{w(0)=\ln\xiy_{c}} necessary to get \REq{fos.w.z.vdw.w} started;\footnote{We wish to stress that this is not the same as imposing the central density as boundary condition; we have already expressed our opinion about this matter in \orefs{dg2019ejmb,dg2024pof}.} the initial condition [\REq{z.is.0}] does the same with \REq{fos.w.z.vdw.z}.
The algorithm marches forward with $\thetay$ until the ``target'' peripheral boundary condition [\REq{xi.is.uw.t}] is met and the terminus $\tw$ is detected; then \REq{ct.uv.N} (top) yields the terminus $\uw$ and \REq{ct.uv.N} (bottom) gives the gravitational number $N$.
The procedure is iterated until convergence to the prescribed gravitational number is achieved.
A very simple and efficient algorithm that, however, is affected by a geometrical limitation: with reference to \Rfi{ct}, forward marching with $\thetay$ is possible only if the slope of the \itm{v(u)} curve at the sphere's center [\REq{fos.v.u.3.f}] is steeper than that of the $x$ axis
\begin{equation}\label{vu3}
   v_{u}(3) < - 1
\end{equation}
so that the curve lies initially in the half plane \itm{y>0}.
The geometrical constraint enforced by \REq{vu3} sets a bound on the initial value
\begin{equation}\label{h.0}
  \hnd(0) < \frac{5}{3}
\end{equation}
of the thermodynamic function $\hnd$ as a consequence of \REq{fos.v.u.3.f}; 
as a matter of fact, the denominator in \REqq{fos.w.z.vdw.0} gave already an early warning of the mathematical danger lurking behind the occurrence \itm{\hnd(0)=5/3}.
This geometrical aspect remained invisible for the \pg\ because, according to \REq{h.pg}, the inequalities expressed in \REqd{vu3}{h.0} are always satisfied; the former is even evidenced geometrically in \Rfi{uv}.
\REqb{h.0} constitutes a rigid barrier for the applicability of the previously described algorithm, visualized by the horizontal line at the level 5/3 in \Rfi{hvdw}: if the central scaled density \itm{\chiy_{c} = \betay\xiy_{c}} makes \itm{\hnd(0)} fall above that line in defiance of \REq{h.0}, as exemplified by point I, then the forward-marching algorithm cannot be started.
This shortcoming becomes of concern if \itm{\omegay\ge2.3555}.
If \REq{h.0} is infringed then there is a workaround to overcome the unsuitability of the described algorithm but we will consider that situation in \Rse{i.nohu3}.
In \Rse{i.hu3}, we will describe and discuss results for situations in which \REq{h.0} is always satisfied.

\subsubsection{\anttlc Another validation exercise for the \fomt\ scheme with TC$_{\ssub{0.65}{\mbox{[2]}}}$\label{i.hu3}} 

We thought appropriate to reexamine \tcb\  to describe the performance of the algorithm outlined in \Rse{ge.vdw} and to validate it with results obtained by the M$_2$ scheme.
\begin{figure}[h]
  \vspace{0.5\baselineskip}
  \gfbox{\wbox}{\includegraphics[keepaspectratio=true, trim = 1.5ex  2ex 6.5ex 12ex , clip , width=.97\columnwidth]{\gdir/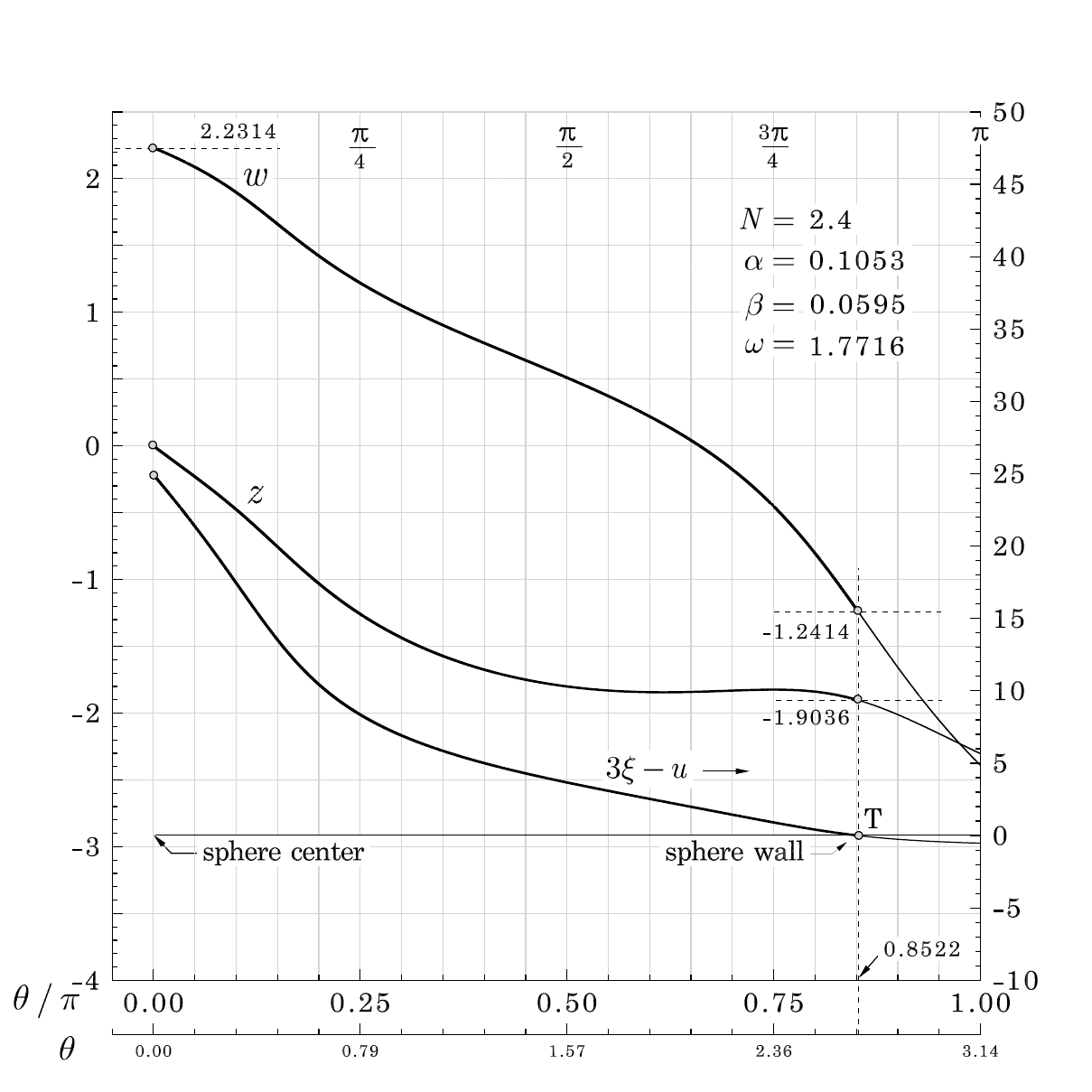}}\\  [-0.5\baselineskip]
  \caption{The curves $w(\protect\thetayf)$ and $z(\protect\thetayf)$ at \itm{N=2.4} relative to \protect\tcb; the ``target'' peripheral boundary condition [\REq{xi.is.uw.t}] is met in the point T corresponding to the terminus \itm{\protect\tw/\protect\piy=0.8522}.}\label{zw2.4-vdw}
\end{figure}
We begin with the case \itm{N=2.4}.
\begin{figure}[t]
  \vspace{0.5\baselineskip}
  \gfbox{\wbox}{\includegraphics[keepaspectratio=true, trim = 10ex  2ex 3ex 12ex , clip , width=.97\columnwidth]{\gdir/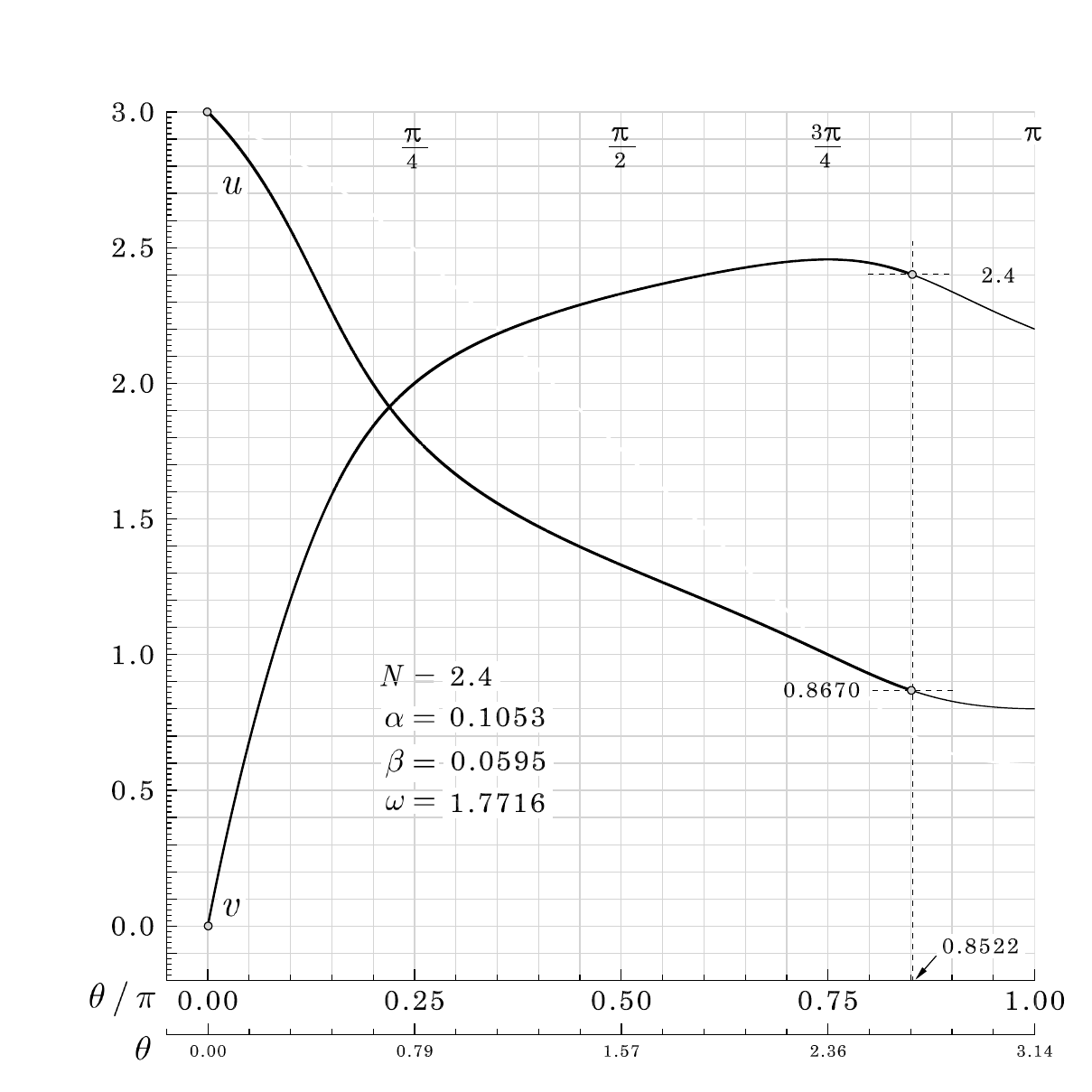}}\\  [-0.5\baselineskip]
  \caption{The curves $u(\protect\thetayf)$ and $v(\protect\thetayf)$ at \itm{N=2.4} relative to \protect\tcb.}\label{uv2.4-vdw}
\end{figure}
The thicker lines in \Rfi{zw2.4-vdw} show the curves \itm{w(\thetay), z(\thetay)}, and the evolution of the function \itm{(3\,\xiy - u)} during the forward march with the angle $\thetay$ until the target condition [\REq{xi.is.uw.t}] is met in the point T in correspondence to which the algorithm detects the first terminus \itm{\thetay_{t}=0.8522}; thereat, \itm{w(0.8522)=-1.2414} and \itm{z(0.8522)=-1.9036}.
Of course, we continued the forward march sufficiently further (thinner lines), typically to \itm{\thetay=2\piy} although it is possible to go as far as wished, to check the existence of other intersections with the level 0 of the right vertical axis but there were not and we considered it a proof of the solution's uniqueness.
\begin{figure}[t]
  \vspace{0.5\baselineskip}
  \gfbox{\wbox}{\includegraphics[keepaspectratio=true, trim = 16ex  8ex  0ex 12ex , clip , width=.97\columnwidth]{\gdir/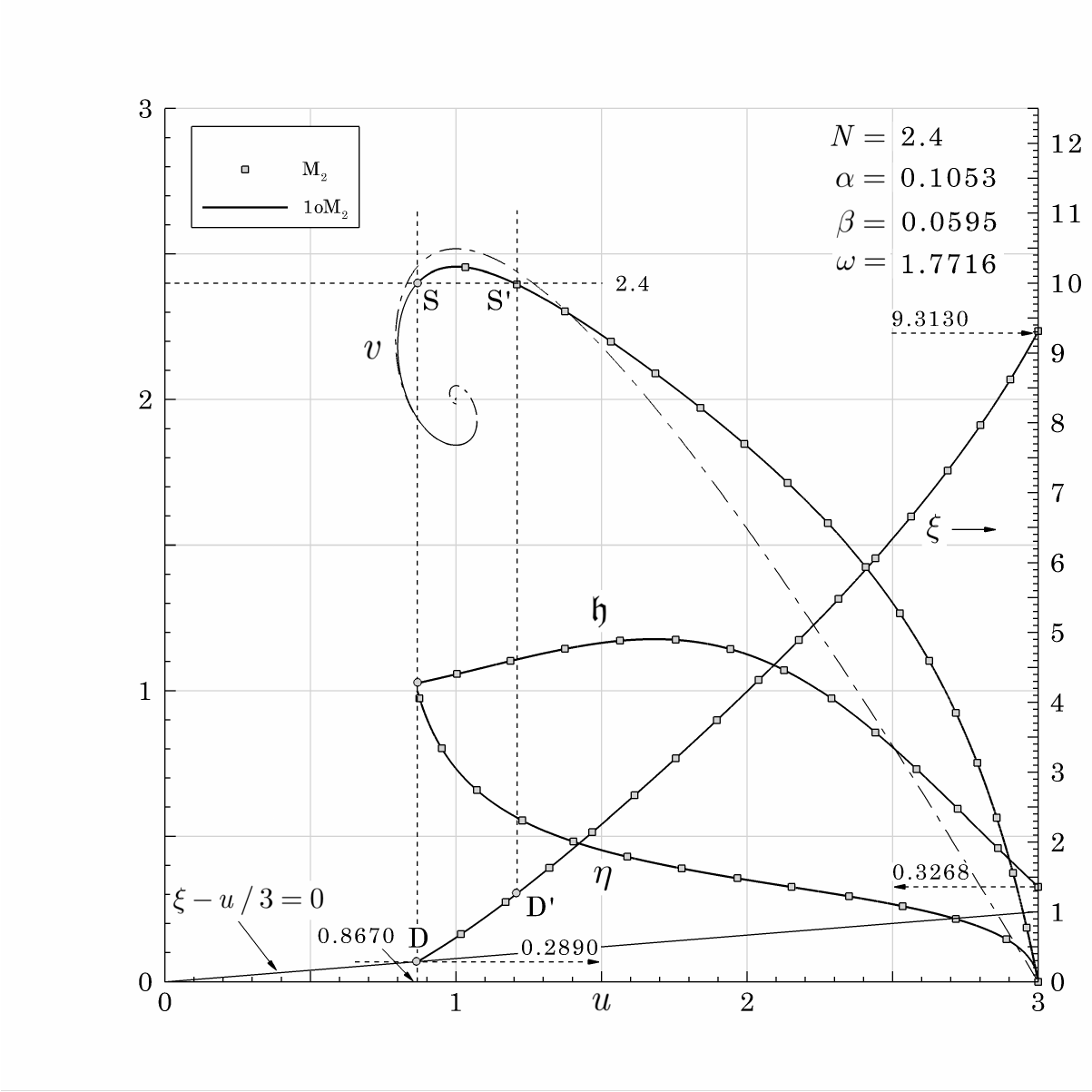}}\\  [-0.5\baselineskip]
  \caption{The curves of density $\protect\xiyc$, variable $v$, radial coordinate $\etayc$ forming the unique solution at \itm{N=2.4} relative to \protect\tcb\ and thermodynamic function $\hnd$ with $u$ as independent variable; results from the M$_{2}$ scheme are superposed as validation evidence. The \pg's curve (dash-dot line) from \Rfi{uv2.4} is show for comparison.}\label{uindN2.4}
\end{figure}
\begin{figure}[h!]
  \vspace{0.5\baselineskip}
  \gfbox{\wbox}{\includegraphics[keepaspectratio=true, trim = 15ex  4ex  1ex 12ex , clip , width=.97\columnwidth]{\gdir/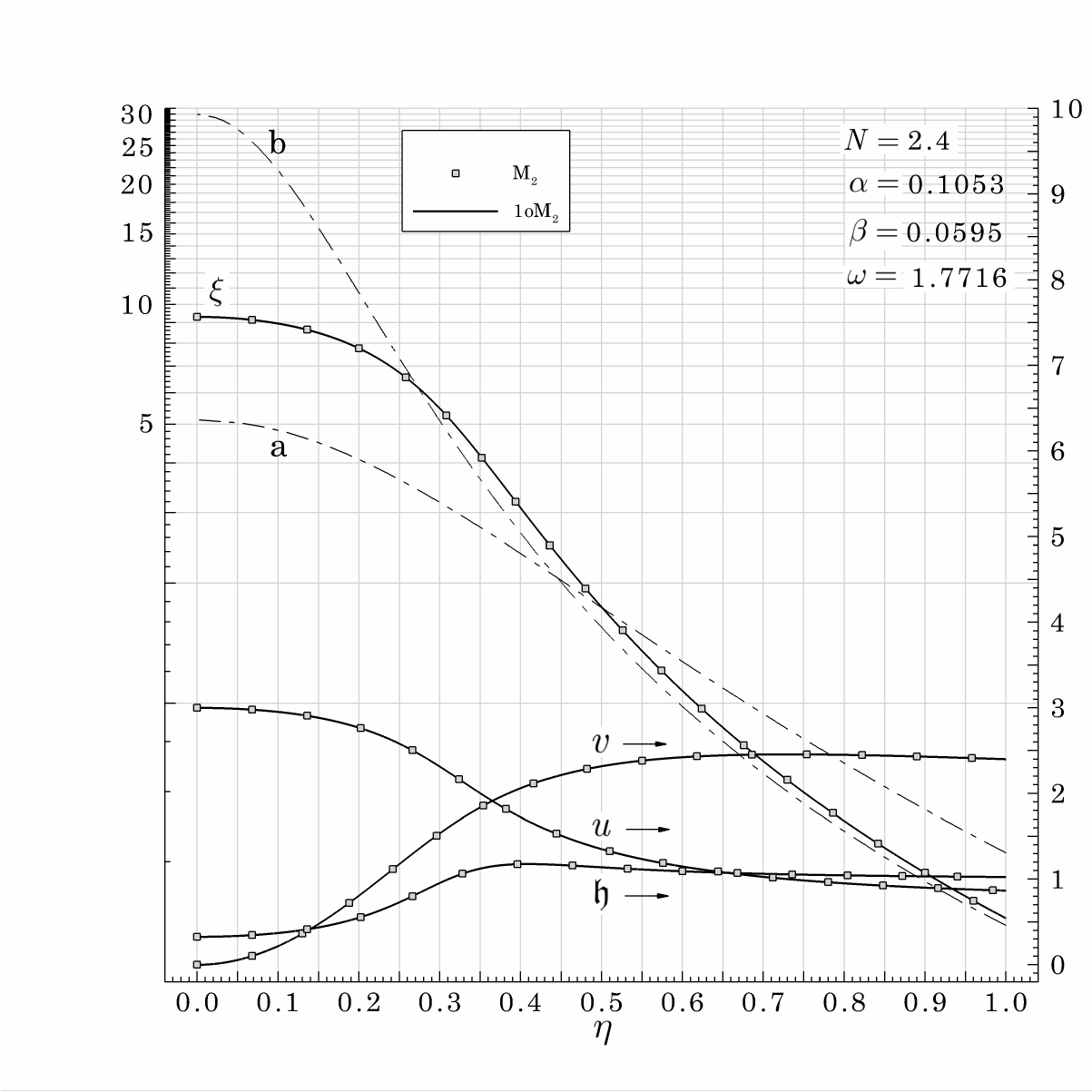}}\\  [-0.5\baselineskip]
  \caption{The graph of \Rfi{uindN2.4} reworked with $\etayc$ as independent variable. The \pg's density profiles (dash-dot lines) from \Rfi{xieta2.4} are shown for comparison.}\label{etaindN2.4}
\end{figure}
The curves \itm{u(\thetay),v(\thetay)} in parametric form are shown in \Rfi{uv2.4-vdw}; the vertical line in correspondence to the terminus \itm{\protect\tw/\piy=0.8522} identifies the terminus \itm{\protect\uw=0.8670} [\REq{ct.uv.N} (top)] and its intersection with the curve $v(\thetay)$ proves the convergence to the prescribed gravitational number \itm{N=2.4} [\REq{ct.uv.N} (bottom)].
The achievement of the converged situation illustrated in \Rfid{zw2.4-vdw}{uv2.4-vdw} required only a few ``shooting'' attempts with the central density.

The curves (solid lines) of density $\xiy$ (right vertical axis), variable $v$ and radial coordinate $\etay$ vs the independent variable $u$ are shown in \Rfi{uindN2.4} with analogous results obtained from the M$_{2}$ scheme [\REqd{sode.ss.vdw.nd}{bc.rcp.nd}, solid squares] superposed as validation evidence.
The curve of the thermodynamic function $\hnd$ is also shown, mainly to emphasize its substantial departure from the \pg's constant unit value [\REq{h.pg}] and the consequent strong coupling effect between \REqd{fos.w.z.vdw.w}{fos.w.z.vdw.z}.
The \pg's universal curve $v(u)$ from \Rfi{uv2.4} is overlaid (dash-dot line) for comparison.
The \vdw's curve \itm{v(u)} also spirals around the center \itm{u=1,v=2} and presents two intersections with the horizontal line at the level 2.4; on the contrary of the \pg's case however, the right intersection S' does not qualify as a solution point because its corresponding density, at point D', does not fall on the target line \itm{\xiy-u/3=0} [\REqd{xi.is.uw}{xi.is.uw.t}].
The graph of \Rfi{uindN2.4} can be reworked into the more familiar graph with $\etay$ as the independent variable, as shown \Rfi{etaindN2.4}.
The \pg's radial density profiles from \Rfi{xieta2.4} are overlaid (dash-dot lines) for comparison with the \vdw's radial density profile.
There are strong density differences in a central core whose thickness is more or less 25\% of the radius, after which the \vdw's profile seems to agree more with the profile corresponding to the b-solution of the \pg.
The curve of the thermodynamic function $\hnd$ settles very nearly to the constant level \itm{\hnd \simeq 1} [\REq{h.pg}] in a peripheral zone whose extension is about 40\% of the radius, a clear indication that therein the \vdw\ behaves essentially as the \pg\ because the density is sufficiently low; however, the similarity does not hold upwards of that zone.
This result is in agreement with the conclusion we drew\footnote{Paragraph just above \Rfimb{10} at page 11.} in connection with the radial profiles of pressure and density illustrated in \Rfimb{9}.

The numerical algorithm can be applied for any desired value of the gravitational number.
Curves and solution points shift accordingly but the uniqueness of solution is preserved; \Rfi{uindN2.2-2.4} illustrates, for example, the effect of a change from \itm{N=2.2} to \itm{N=2.4}. 
\begin{figure}[h]
  \vspace{0.5\baselineskip}
  \gfbox{\wbox}{\includegraphics[keepaspectratio=true, trim = 16ex  8ex  0ex 12ex , clip , width=.97\columnwidth]{\gdir/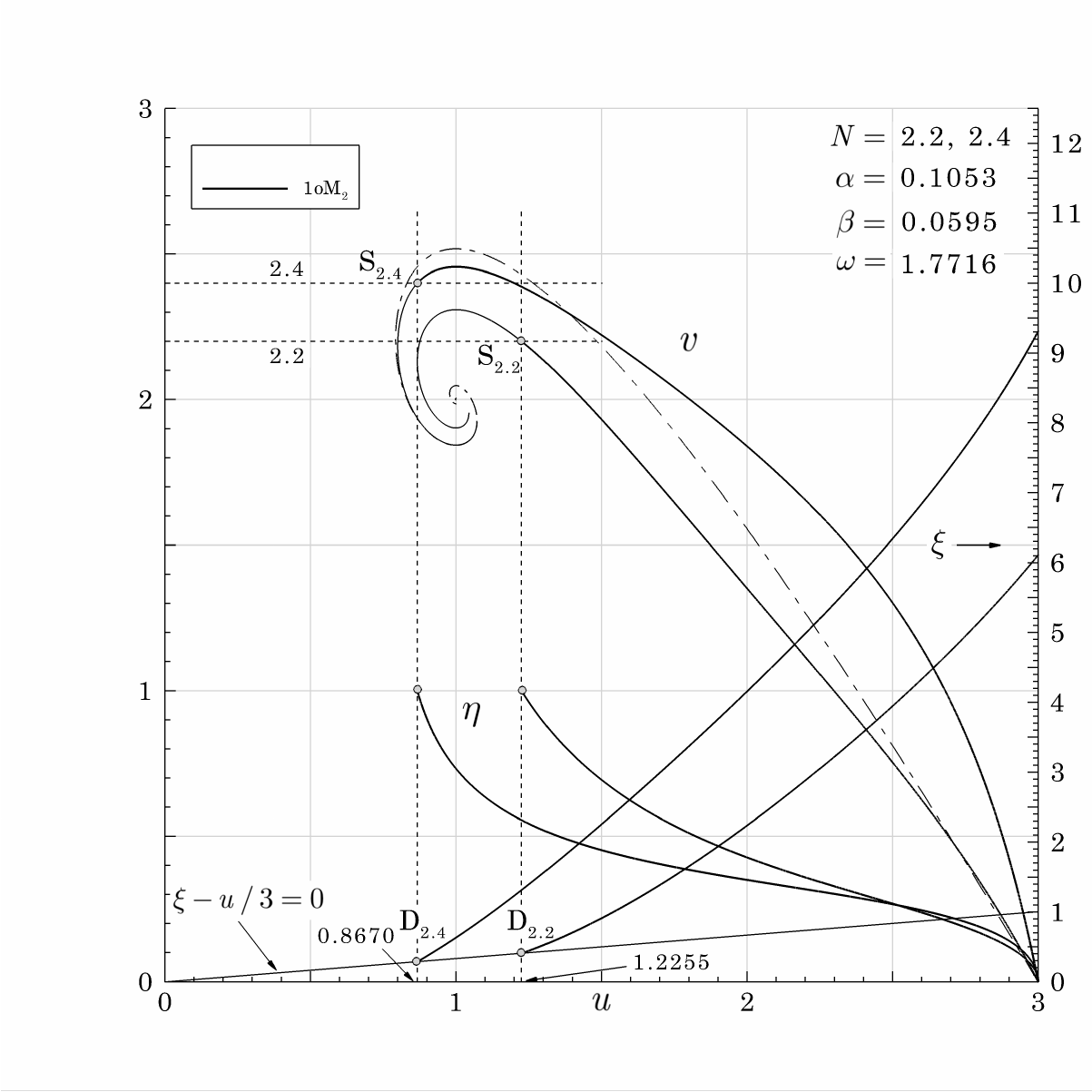}}\\  [-0.5\baselineskip]
  \caption{The effect of a change in gravitational number from \itm{N=2.2} to \itm{N=2.4} on the solutions relative to \protect\tcb. The unaffected \pg's \itm{v(u)} curve (dash-dot line) is shown for comparison.}\label{uindN2.2-2.4}
\end{figure}
In \Rfi{xicxittc2}, we have reproduced the curves of peripheral \itm{(\etay=1)} and central \itm{(\etay=0)} densities vs gravitational number presented, respectively, in \Rfismb{10}{11} (labeled \itm{\protect\alphay=0.1053}); the agreement between results from M$_{2}$ and \fomt\ schemes is very satisfactory.
\begin{figure}[h]
  \vspace{0.5\baselineskip}
  \gfbox{\wbox}{\includegraphics[keepaspectratio=true, trim = 6ex  6ex  6ex 12ex , clip , width=.97\columnwidth]{\gdir/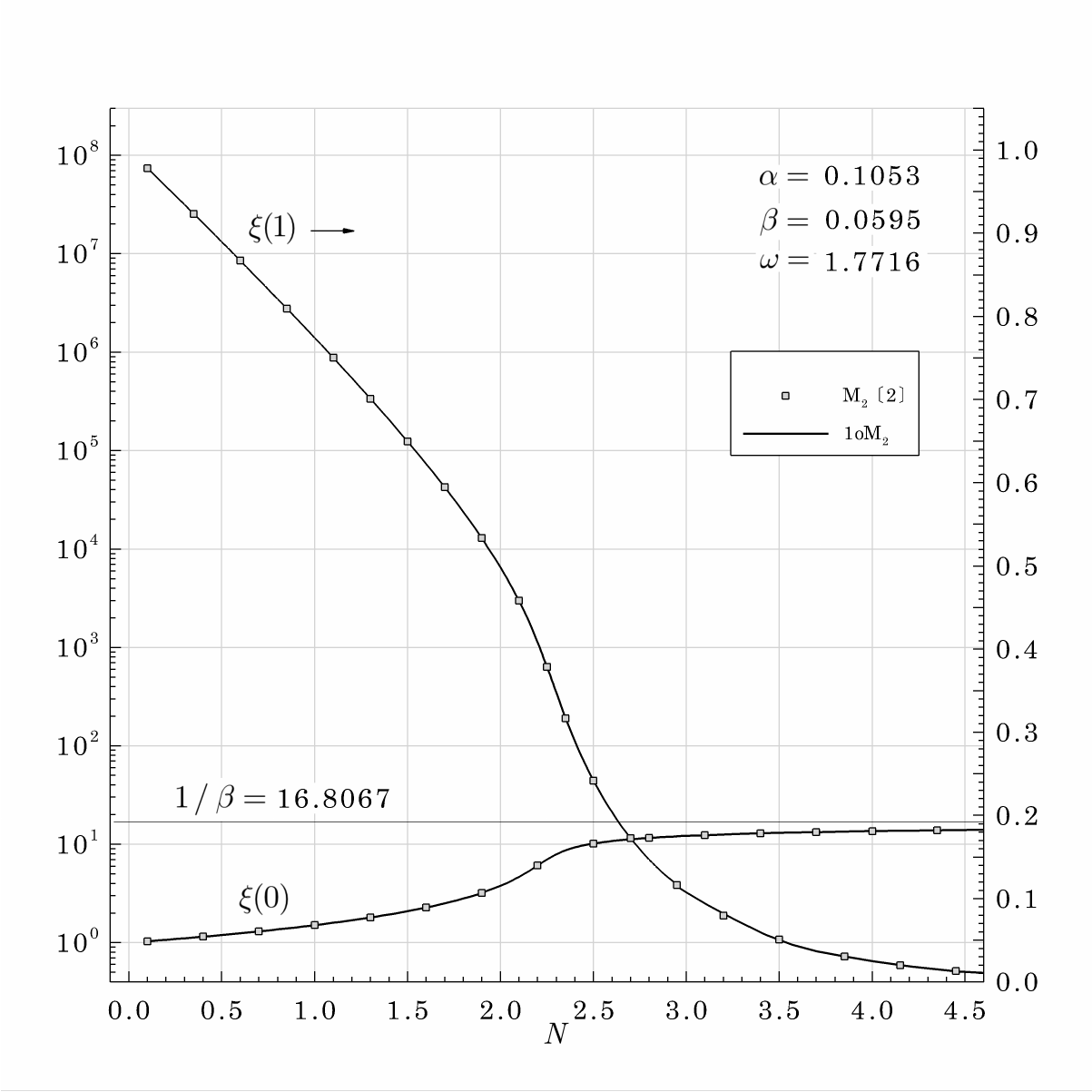}}\\  [-0.5\baselineskip]
  \caption{Peripheral and central densities versus gravitational number relative to \protect\tcb. The curves from \Rfismb{10}{11} labeled with \itm{\protect\alphay=0.1053} are superposed (here as solid squares) as validation evidence.}\label{xicxittc2}
\end{figure}

\subsubsection{\anttlc Integration if \REq{h.0} is not satisfied\label{i.nohu3}}

\REqb{h.0} is always respected for \tcb\ with any value of the gravitational number but there can be combinations of the characteristic numbers for which the limitation breaks down.
The infringement of \REq{h.0} implies, in turn, the violation of \REq{vu3}: the slope of the \itm{v(u)} curve at the sphere's center is less steep than that of the $x$ axis (\Rfi{ct}), the curve lies initially in the half plane \itm{y<0}, and the forward march with the angle $\thetay$ is blocked.
The artifice to bypass this setback consists of the application of the Cartesian-coordinate transformation [\REq{ct.xy}] rather than the polar-coordinate one [\REq{ct.uv}], together with the convenient reversal of the $x$ axis
\begin{equation}\label{xrev}
  x = 2\sqrt{2} - X
\end{equation}
meant to permit the start of the integration again from the sphere's center \itm{(X=0)}.
We also recycle the variable $w$ defined by \REq{lnxi}.
Then \REqd{fos.xi.u}{fos.v.u} become
\begin{subequations}\label{fos.w.y.vdw}   
\begin{align} 
   w_{\Xs} & = - \frac{\hnd }{\sqrt{2}}\frac{v}{u + \hnd v - 3} \frac{1 - y_{\Xs} }{u}  \label{fos.w.y.vdw.w}
      \\[\baselineskip]
   y_{\Xs} & = - \frac{1 - \dfrac{v}{u + \hnd v - 3} \dfrac{u-1}{u} }{1 + \dfrac{v}{u + \hnd v - 3} \dfrac{u-1}{u} } \label{fos.w.y.vdw.y}
\end{align}
\end{subequations}
in which the variables $u$ and $v$ must be considered functions of $y$ and $X$ in compliance with \REqd{ct.xy}{xrev}.
\REqb{lim.u=3.f} helps one more time to obtain the determinate forms of \REqq{fos.w.y.vdw}
at the sphere's center \itm{(X=0)}
\begin{subequations}\label{fos.w.y.vdw.0}   
\begin{align} \negthickspace \negthickspace \negthickspace
   w_{\Xs}(0) & = - \frac{5}{\sqrt{2}} \frac{\hnd(0)        }{5 + 3\, \hnd(0)} \label{fos.w.y.vdw.w.0} \\[.5\baselineskip]
   y_{\Xs}(0) & =                      \frac{5 - 3\, \hnd(0)}{5 + 3\, \hnd(0)} \label{fos.w.y.vdw.y.0}
\end{align}
\end{subequations}
The denominator of \REqq{fos.w.y.vdw.0} is immune to the danger carried by \itm{\hnd(0)=5/3} for the case of the polar coordinates (\Rse{ge.vdw}); if that occurrence prevails, the numerator of \REq{fos.w.y.vdw.y.0} reveals that the curve $v(u)$ is tangent \itm{[y_{\Xs}(0) = 0]} to the $x$ axis in the sphere's center.
The central boundary condition [\REq{v.is.3}] turns into the initial condition
\begin{equation}\label{y.is.0}
  y(0) = 0
\end{equation}
for the second differential equation [\REq{fos.w.y.vdw.y}].
The peripheral boundary condition [\REq{xi.is.uw}] must be interpreted in terms of the terminus $X_{t}$ as
\begin{equation}\label{xi.is.uw.t.tmp}
  \xiy\left[u_{t}(X_{t})\right] = \frac{u_{t}(X_{t})}{3}
\end{equation}
and goes with the first differential equation [\REq{fos.w.y.vdw.w}]. 
In Cartesian coordinates, the auxiliary condition [\REq{v.is.uw}] translates into the parametric equations
\begin{equation}\label{ct.yX.N}
  \left\{   \begin{aligned} \uw    & = \left(y(X_{t}) - X_{t} + 3\sqrt{2}                                 \right)/\sqrt{2} \\[.5\baselineskip]
                            v(\uw) & = \left(y(X_{t}) + X_{t} \parbox{0em}{\textcolor{white}{+ 3$\sqrt{2}$}}\right)/\sqrt{2} = N \end{aligned}    \right.
\end{equation}
deducible from \REqd{ct.xy}{xrev} after proper adaptation.
The central-density based shooting method applied in the case of the polar coordinates can be exploited also for the numerical solution of \REqq{fos.w.y.vdw}.
The initial value \itm{w(0)=\ln\xiy_{c}} and \REq{y.is.0} get started \REqd{fos.w.y.vdw.w}{fos.w.y.vdw.y}, respectively.
The algorithm marches forward with $X$ until the peripheral boundary condition [\REq{xi.is.uw.t.tmp}] is met and the terminus $X_{t}$ is detected.
Then, \REq{ct.yX.N} (top) and (bottom) gives the terminus $\uw$ and the gravitational number $N$.
The procedure is iterated until convergence to the prescribed gravitational number is achieved.
A numerical example whose conditions in the sphere's center (\itm{\xiy=5.0453, \chiy=\betay\,\xiy=0.3, \hnd=4.1525}) correspond to point I in \Rfi{hvdw} is provided in \Rfid{yw1.45-vdw}{uindN1.45}.
The former figure reveals that the target point T falls on the terminus \itm{X_{t}/2\sqrt{2}=0.6864} to which, as indicated in the latter figure, correspond the terminus \itm{\uw=1.7020} and the solution point S at \itm{v(1.7020)=N=1.45}.
The $v(u)$ curve presents a rather unusual shape; it starts below the $x$ axis with a slope calculable from \REq{fos.v.u.3.f} as \itm{v_{u}(3)=-5/3/4.1525=-0.4014>-1}, moves away in the half plane \itm{y<0} until it reaches an inversion point, the minimum \itm{y=-0.1501} at \itm{X/2\sqrt{2}=0.2194} in \Rfi{yw1.45-vdw} or the point \itm{u=2.4551, v=0.3326} in \Rfi{uindN1.45}, after which it returns into the half plane \itm{y>0}, and, finally, goes into the familiar spiraling.
\begin{figure}[h]
  \vspace{0.5\baselineskip}
  \gfbox{\wbox}{\includegraphics[keepaspectratio=true, trim = 9ex  5ex 6.5ex 12ex , clip , width=.97\columnwidth]{\gdir/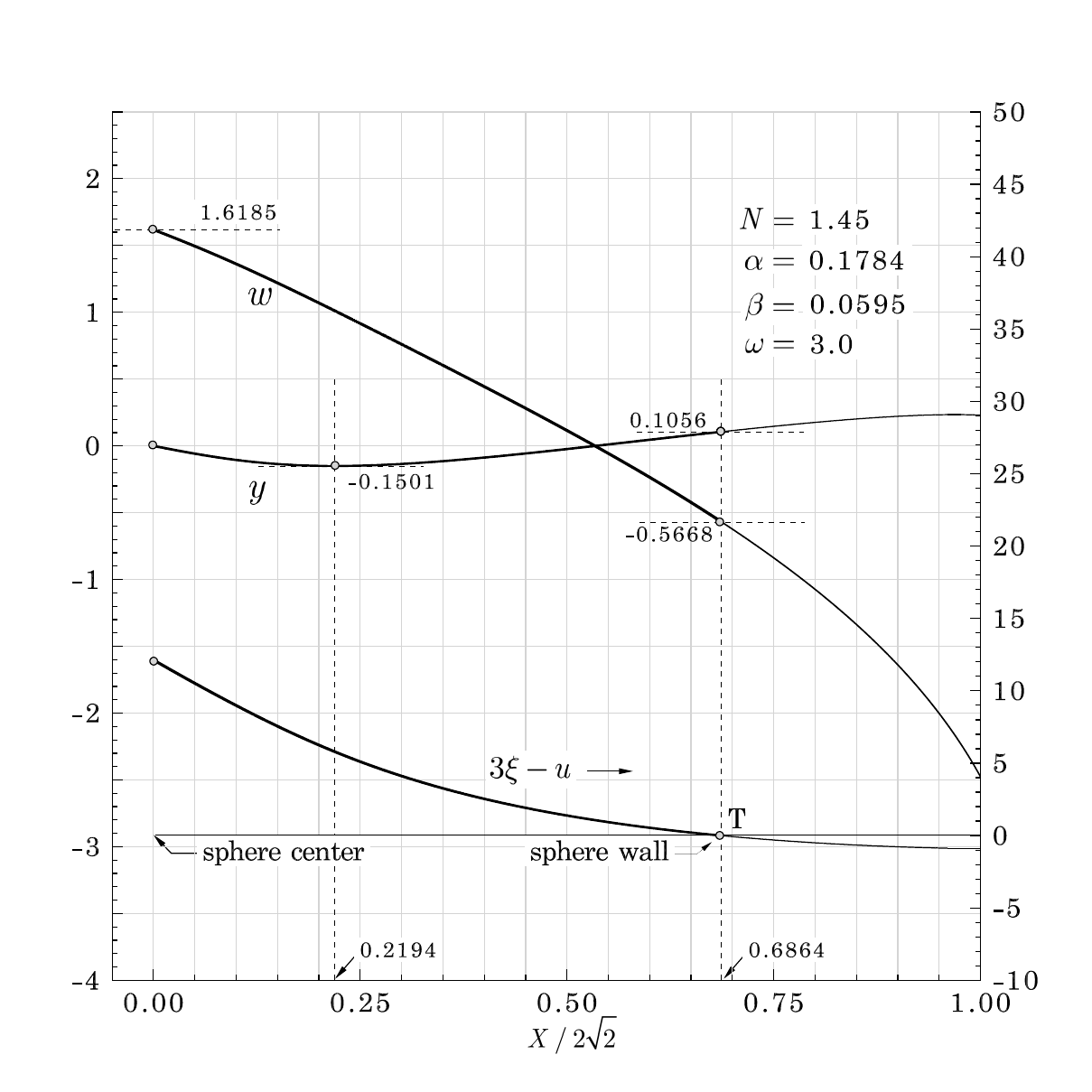}}\\  [-0.5\baselineskip]
  \caption{The curves $w(X)$ and $y(X)$ at \itm{N=1.45} relative to characteristic numbers for which \REq{h.0} is infringed; the ``target'' peripheral boundary condition [\REq{xi.is.uw.t.tmp}] is met in the point T corresponding to the terminus \itm{X_{t}/2\sqrt{2}=0.6864}.}\label{yw1.45-vdw}
\end{figure}
\begin{figure}[h]
  \vspace{0.5\baselineskip}
  \gfbox{\wbox}{\includegraphics[keepaspectratio=true, trim = 16ex  7ex  0ex 12ex , clip , width=.97\columnwidth]{\gdir/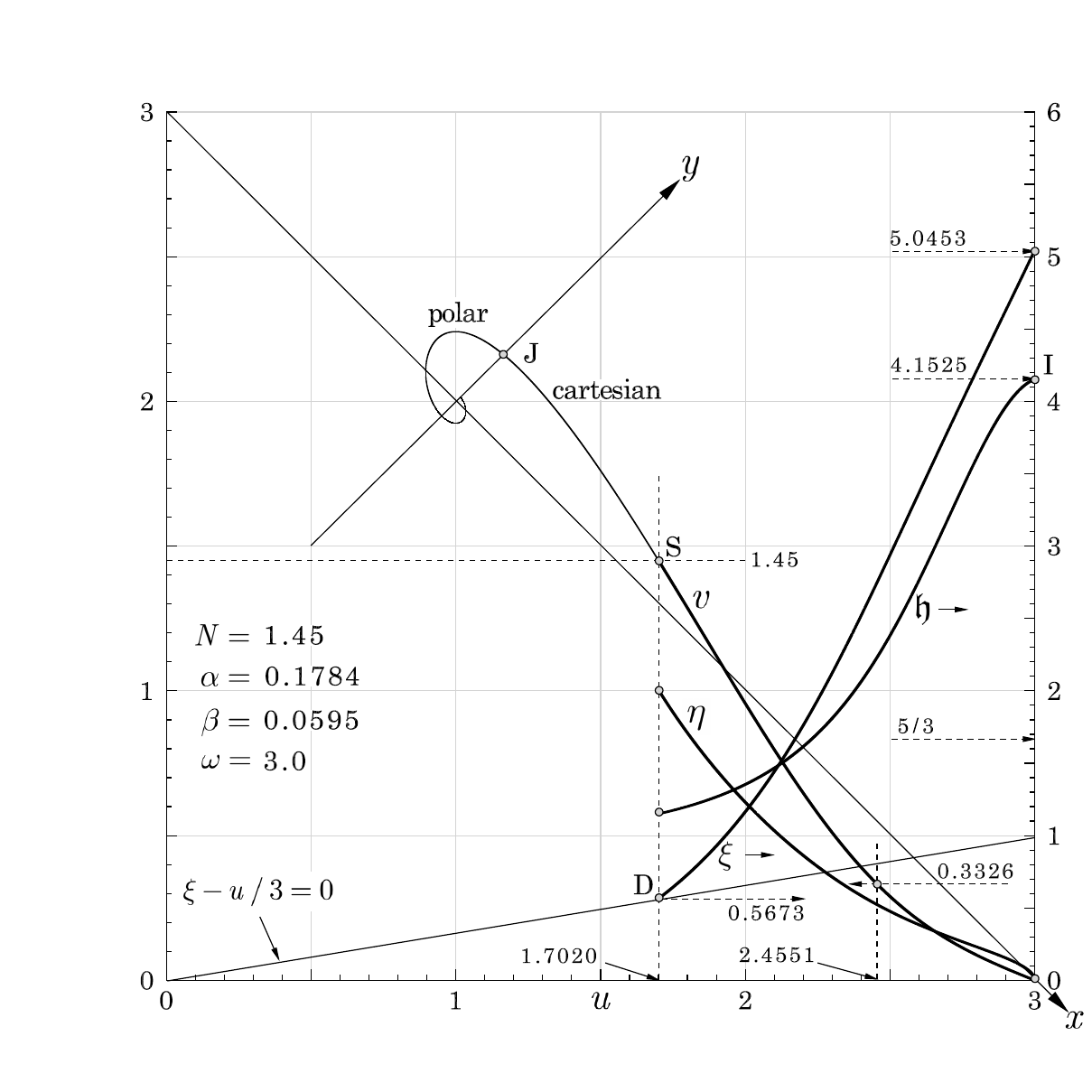}}\\  [-0.5\baselineskip]
  \caption{The curves of density $\protect\xiyc$, variable $v$, radial coordinate $\etayc$ forming the unique solution at \itm{N=1.45} relative to characteristic numbers for which \REq{h.0} is infringed and thermodynamic function $\hnd$ with $u$ as independent variable; the conditions  at the sphere's center \itm{(\scalebox{0.85}{$\protect\chiy_{c}$}=\scalebox{0.85}{$\protect\betay\protect\xiy_{c}$}=0.3)} correspond to point I in \Rfi{hvdw}.}\label{uindN1.45}
\end{figure}

It is not a wise idea to push the integration with the Cartesian coordinates beyond the spiral's center \itm{(X=2\sqrt{2})} because the location of the solution point S is not known in the beginning; there is a risk that it may fall past the point of the $v(u)$ curve at which the tangent becomes parallel (\itm{y_{X}\rightarrow\infty}) to the $y$ axis, in which case the algorithm breaks down.
A safe approach consists of building the $v(u)$ curve piecewise: the Cartesian-coordinate based algorithm integrates from the sphere's center \itm{(X=0)} to the junction point J \itm{(X=2\sqrt{2})} in which we extract the conditions \itm{y(2\sqrt{2}), w(2\sqrt{2})}, convert them to
\begin{subequations}\label{v.conv}   
\begin{align} 
   w\left(\frac{\piy}{2}\right) & = w\left(2\sqrt{2}\right) \label{v.conv.w} \\[\baselineskip]
   z\left(\frac{\piy}{2}\right) & = \ln\frac{y(2\sqrt{2})}{2\sqrt{2}} \label{v.conv.z}
\end{align}
\end{subequations}
and feed \REqq{v.conv} as initial conditions into the polar-coordinate based algorithm to continue the integration as far as wished.
This is the way we built the $v(u)$ curve shown in \Rfi{uindN1.45}; although there was no necessity because the solution point S falls before the junction point J; nevertheless we did it for the purpose of demonstration.

\subsubsection{\anttlc Monotonicity study\label{ms}}

Let us go back to \Rfi{xicxittc2}.
From the point of view of physical consistency, the merit of the peripheral density's curve \itm{\xiy(1)} resides in its decreasing monotonicity with respect to the gravitational number's increase, and particularly in its asymptotic vanishing.
The monotonicity is a consequence of the presence of molecular attraction and size in the \vdw's equation of state [\Reqmb{32}] via the characteristic numbers $\alphay$ and $\betay$.
The model fits well for \tcb\ but the existence of values of those characteristic numbers determining effects similar to \pge\ cannot be excluded a priori only on the basis of just one successful test case.
It seems reasonable to expect that a good insight into this problem can be inferred by studying the influence of those characteristic numbers on the slope of the peripheral density's curve.
For this purpose, an opportunity is offered by the combination of the peripheral boundary condition [\REq{xi.is.uw}] with the parametric equations [\REq{ct.uv.N}] for the terminus $\uw$ generated by the auxiliary condition [\REq{v.is.uw}]; we remind the reader that \REqq{ct.uv.N}, although introduced within the \pg's context, are thermodynamic-model independent. 
Setting for brevity \itm{\xiy(\uw) = \xw}, a rapid manipulation of \REqd{xi.is.uw}{ct.uv.N} leads to
\begin{equation}\label{ct.uv.N.s}
  \left\{   \begin{aligned} \xw & = \frac{1}{3}\left[1+2 e^{z(\thetays_{\ssub{0.5}{t}})}(\sin\tw+\cos\tw)\right] \\[.5\baselineskip]
                            N   & = 2 \left[ 1 + e^{z(\thetays_{\ssub{0.5}{t}})}(\sin\tw-\cos\tw)\right] \; .\end{aligned}    \right.
\end{equation}
from which we obtain the derivative
\begin{equation}\label{dxidN}
  \tds{}{\xw}{N} = - \dfrac{1}{3} \frac{\sin\tw-\cos\tw - \tds{}{z(\tw)}{\tw}(\sin\tw+\cos\tw)}{\sin\tw+\cos\tw + \tds{}{z(\tw)}{\tw}(\sin\tw-\cos\tw)}  
\end{equation}
The derivative of \itm{z(\thetay_{t})} on the right-hand side of \REq{dxidN} can be expressed in terms of its slope \itm{\lambday(\thetay_{t})}
\begin{equation}\label{zt.slo}
  \tds{}{z(\tw)}{\tw} = \tan\,\lambday(\thetay_{t})
\end{equation}
which, by the way, should not be confused with the slope corresponding to \itm{z_{\thetays}(\tw)}, as in \Rfi{zw2.4-vdw}, for example.
The substitution of \REq{zt.slo} into \Req{dxidN} produces the final plain expression
\begin{equation}\label{dxidN.1}
  \tds{}{\xw}{N} = - \dfrac{1}{3} \frac{\sin(\lambday(\tw)-\tw) + \cos(\lambday(\tw)-\tw)}{\sin(\lambday(\tw)-\tw) - \cos(\lambday(\tw)-\tw)}
\end{equation}
from which we deduce that the responsibility for the monotonicity of the peripheral density versus gravitational number falls on the denominator \itm{\sin(\lambday(\tw)-\tw) - \cos(\lambday(\tw)-\tw)}; if the latter vanishes then the monotonicity breaks down and this happens if
\begin{equation}\label{den=0}
  \subeqn{\lambday(\tw) = \tw + \frac{\piy}{4} ( 1 + 4k)}{k=0,\pm1,\pm2,\ldots}{1.5} \; \,
\end{equation}
that is, if the function \itm{\lambday(\thetay_{t})} has intersections with at least one of the straight lines defined on the right-hand side of \REq{den=0}.
Obviously, the influence of the characteristic numbers \itm{\alphay, \betay} or, better, \itm{\omegay, \betay} is embedded within \REq{dxidN.1}.

\begin{figure}[h]
  \vspace{0.5\baselineskip}
  \gfbox{\wbox}{\includegraphics[keepaspectratio=true, trim = 6ex  2ex 2ex 12ex , clip , width=.97\columnwidth]{\gdir/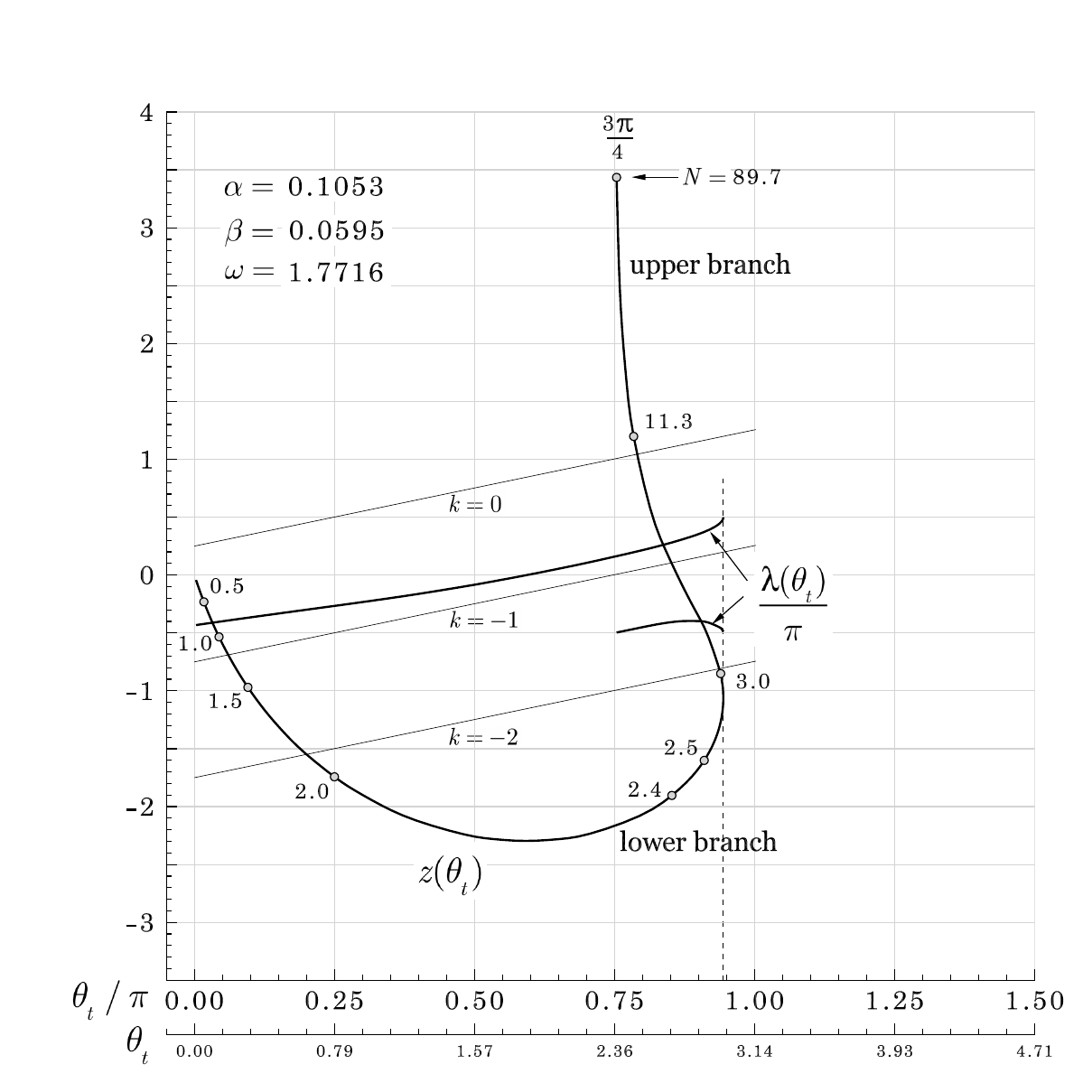}}\\  [-0.5\baselineskip]
  \caption{The curve \itm{z(\protect\thetayc_{\ssub{0.5}{t}})} and its slope \itm{\protect\lambdayc(\protect\thetayc_{\ssub{0.5}{t}})} versus \itm{\protect\thetayc_{\ssub{0.5}{t}}} relative to \protect\tcb.}\label{zt.tc2}
\end{figure}
\Rfib{zt.tc2} illustrates the situation corresponding to \tcb.
The function \itm{z(\thetay_{t})} is composed of two branches. The lower one decreases to a minimum and then increases to become vertical slightly before \itm{\thetay_{t}/\piy=0.95}; the upper branch picks up from there, increases backward going through an inflection point after which it tends asymptotically to become again vertical at \itm{\thetay_{t}/\piy=3/4}.
The branches are parameterized with the gravitational number, which increases monotonically along the curve. The latter's verticalities \itm{[\tdt{}{\!z(\thetay_{t})}{\thetay_{t}}=\pm\infty,\,\,\lambday(\thetay_{t})/\piy=\pm1/2]} do not perturb \REqd{dxidN}{dxidN.1}.
The function \itm{\lambday(\thetay_{t})} never intersects any of the straight lines of \REq{den=0}, the denominator in \REq{dxidN.1} never vanishes, and the monotonicity of the peripheral density, as well as central density by consequence, is secured in full agreement with the graph of \Rfi{xicxittc2}.

Now, suppose we reduce to 1/10 with respect to \tcb\ the fluid mass [\Reqmb{1}] of the same gas inside the spherical container and keep it at the same temperature.
The average density [\Reqmb{2}] becomes 1/10 of the original one and will pass on [\Reqsmb{31b}{31c}] the same reduction to the characteristic numbers: \itm{\alphay=0.01053\mbox{ and }\betay=0.00595}; obviously, the characteristic number $\omegay$ remains fixed to the value 1.7716 according to its definition [\REq{ab.r}].
\begin{figure}[t]
  \vspace{0.5\baselineskip}
  \gfbox{\wbox}{\includegraphics[keepaspectratio=true, trim = 6ex  2ex 0ex 8ex , clip , width=.97\columnwidth]{\gdir/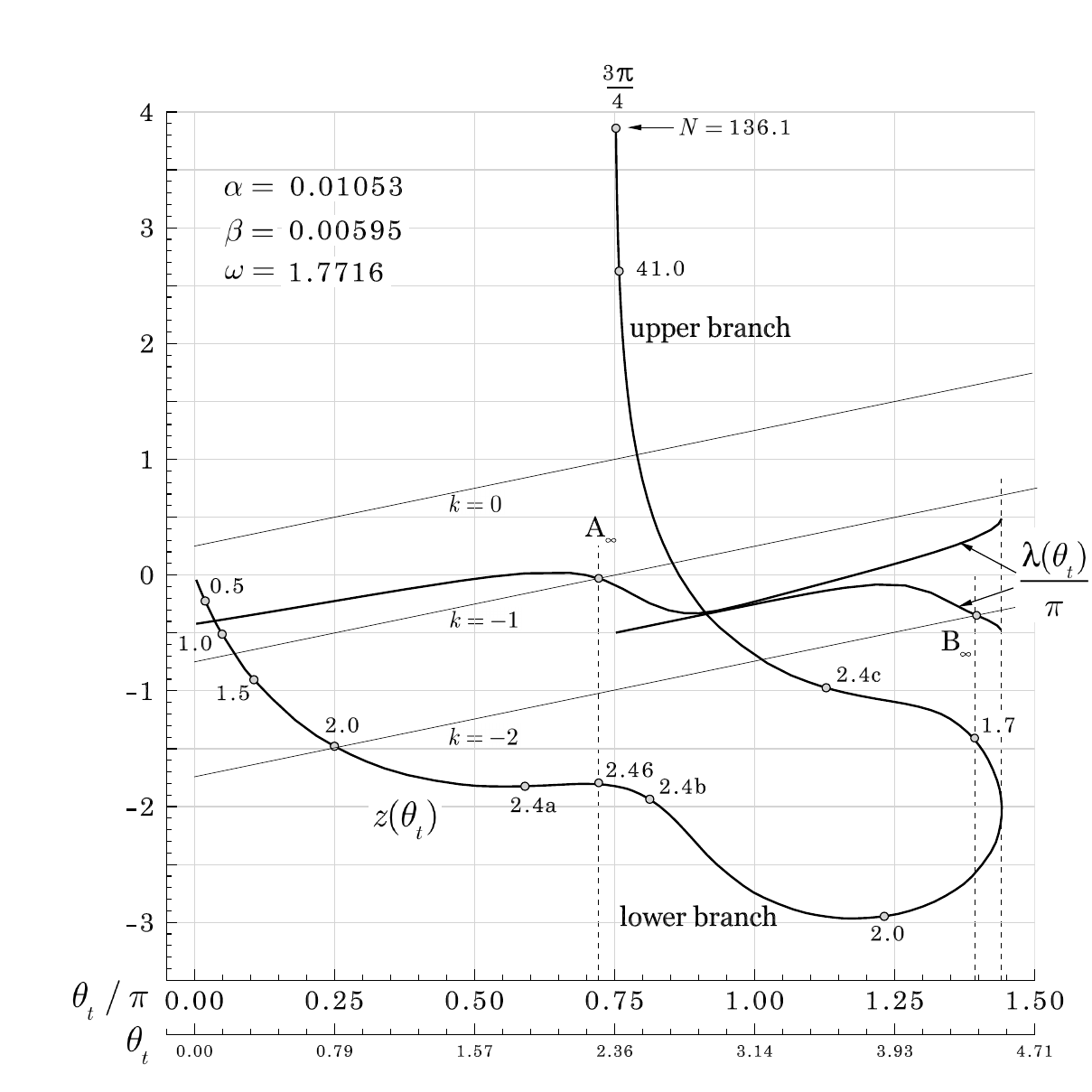}}\\  [-0.5\baselineskip]
  \caption{The curve \itm{z(\protect\thetayc_{\ssub{0.5}{t}})} and its slope \itm{\protect\lambdayc(\protect\thetayc_{\ssub{0.5}{t}})} versus \itm{\protect\thetayc_{\ssub{0.5}{t}}} relative to   \itm{\protect\omegay=1.7716}, the same of \tcb, but with \itm{\itm{\protect\betayc=0.00595}}.}\label{zt.sb}
\end{figure}
\begin{figure}[h]
  \vspace{0.5\baselineskip}
  \gfbox{\wbox}{\includegraphics[keepaspectratio=true, trim = 4ex  6ex  12ex 12ex , clip , width=.97\columnwidth]{\gdir/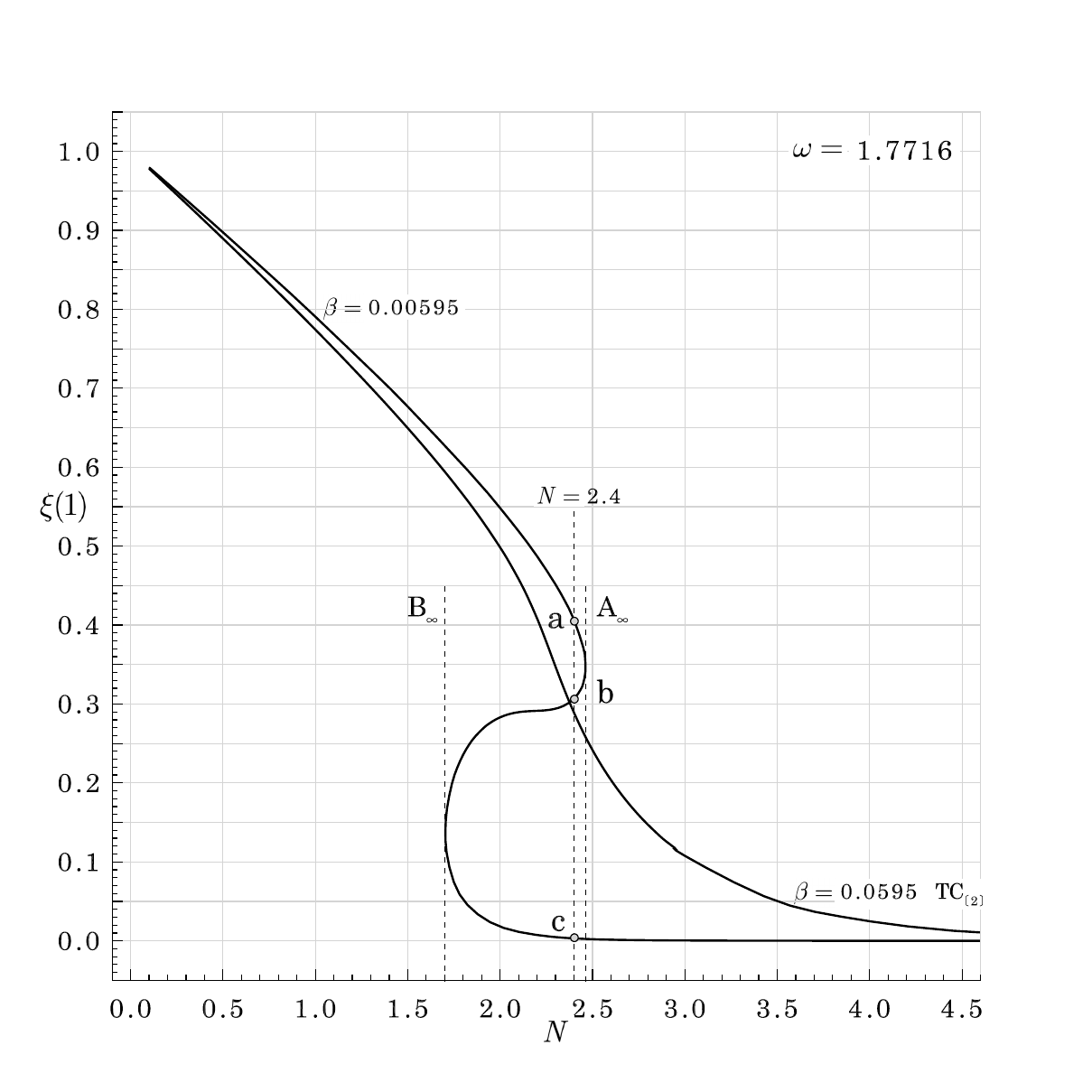}}\\  [-0.5\baselineskip]
  \caption{Effect of the characteristic number \itm{\protect\betayc} on the peripheral density's dependence on the gravitational number and appearance of multiple solutions.}\label{xit}
\end{figure}
The functions \itm{z(\thetay_{t})} and \itm{\lambday(\thetay_{t})} corresponding to this case are illustrated in \Rfi{zt.sb}.
The former is again composed of two branches, but the lower one goes through two minima and one maximum, while the upper one presents a rather flat inflection point.
These characteristics bring the latter function to intersect the straight lines \itm{k\negthickspace=\negthickspace-1, -2} of \REq{den=0} in the two points A$_{\infty}$ and B$_{\infty}$;  between them, the gravitational number decreases from 2.46 to 1.7 along the curve and, in such subinterval, triple solutions coexist.
We have purposely indicated the three solutions at \itm{N=2.4} and tagged them with a,b,c.
The triple solutions are better appreciated in the graphs of peripheral and central densities shown in \Rfid{xit}{xic}, respectively.
\begin{figure}[h]
  \vspace{0.5\baselineskip}
  \gfbox{\wbox}{\includegraphics[keepaspectratio=true, trim = 6ex  6ex  12ex 11ex , clip , width=.97\columnwidth]{\gdir/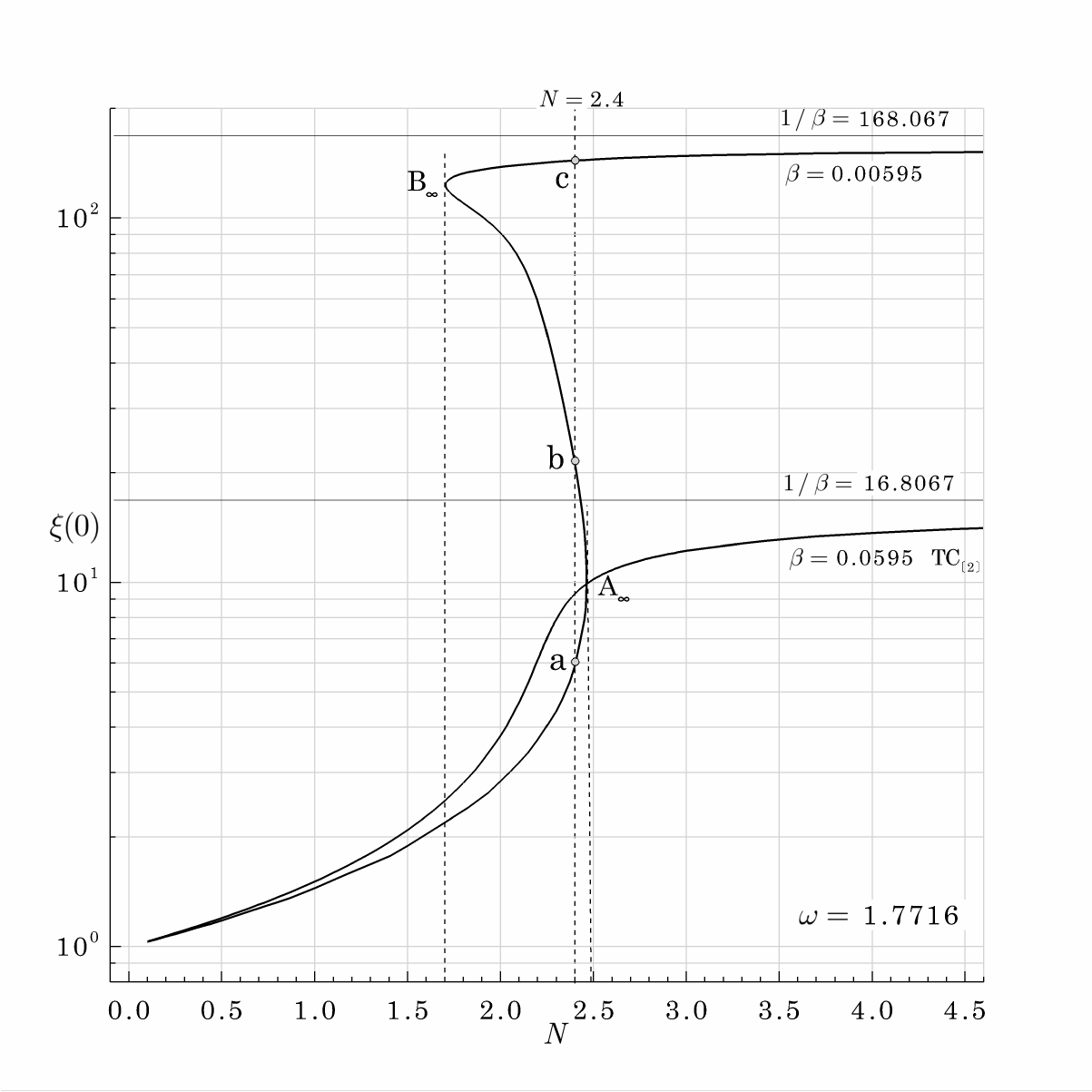}}\\  [-0.5\baselineskip]
  \caption{Effect of the characteristic number \itm{\protect\betayc} on the central density's dependence on the gravitational number and appearance of multiple solutions.}\label{xic}
\end{figure}
On the contrary of the \pg's case, there is neither spiraling peripheral density (\Rfimb{10}) nor oscillating central density (\Rfimb{11}).
The three solutions coexist because the curves become s-shaped in the subinterval of the gravitational number marked by A$_{\infty}$ and B$_{\infty}$ in \Rfi{zt.sb}.
It is also interesting to see how the solutions' coexistence is rendered in the \itm{u,v} plane.
\Rfib{uvsb2.4}, which should be compared with \Rfi{uv2.4}, clearly shows that,
\begin{figure}[h]
  \vspace{0.5\baselineskip}
  \gfbox{\wbox}{\includegraphics[keepaspectratio=true, trim = 16ex  8ex 4ex 12ex , clip , width=.97\columnwidth]{\gdir/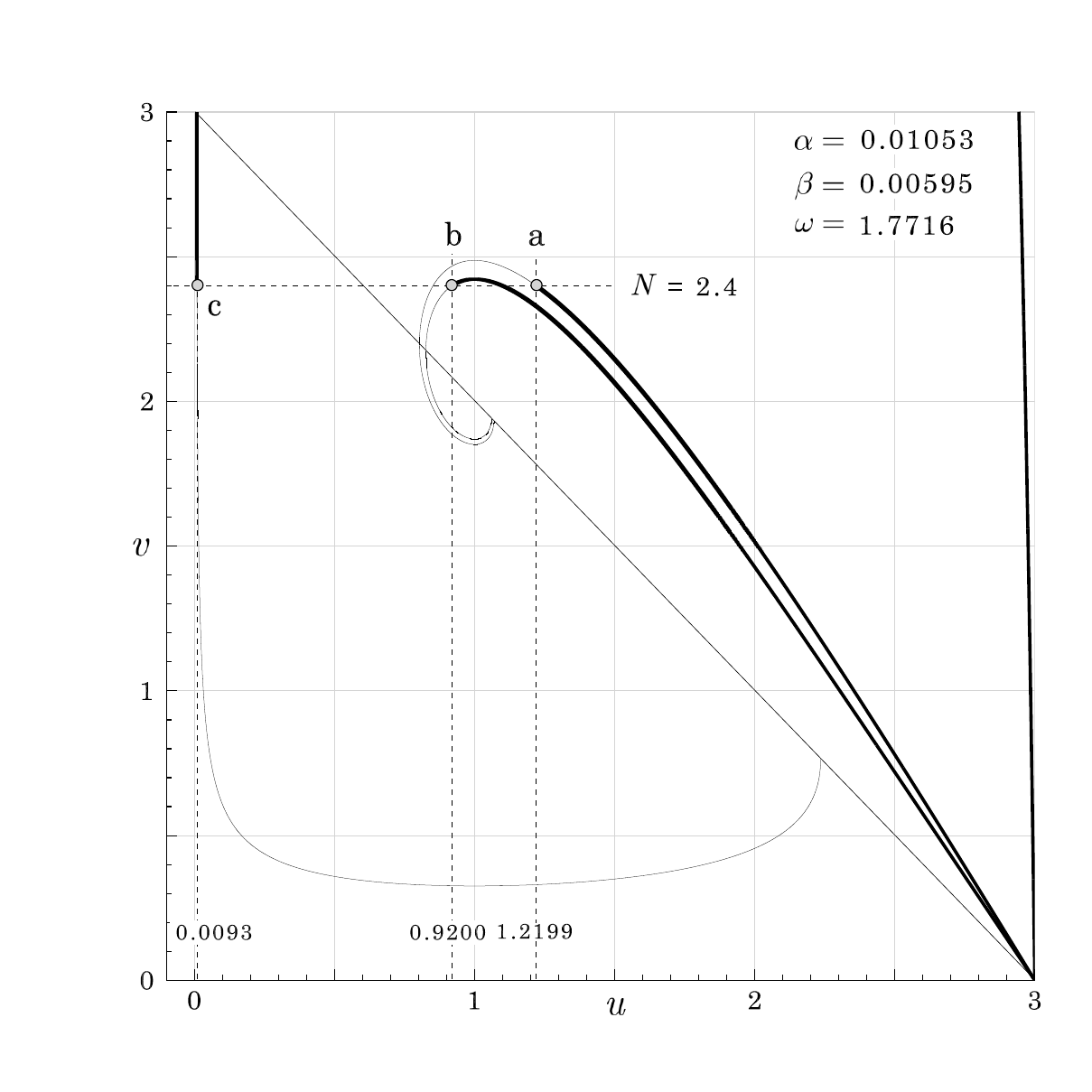}}\\  [-0.5\baselineskip]
  \caption{The three solutions at $N=2.4$ in the \itm{u,v} plane for the case of \Rfi{zt.sb}.}\label{uvsb2.4}
\end{figure}
on the contrary of what happens for the \pg, the three solutions belong to three different $v(u)$ curves.
So, multiple solutions do exist also for the \vdw\ but their genesis originates from reasons completely different from those of the \pg.

The curves belonging to \tcb\ in \Rfid{xit}{xic} have both an inflection point.
It is clear that, for a prescribed $\omegay$, there must be a value of $\betay$ between 0.00595 and 0.0595 in correspondence to which the inflection point's axis becomes vertical; that value marks a boundary point above which the solutions are unique and below which the solutions could be multiple depending on the gravitational number's value.
If $\omegay$ varies then the boundary point describes a curve in the $\omegay\,,\,\betay$ plane which separates the regions of solution's uniqueness or multiplicity.
The curve we have obtained through a series of patient and thorough calculations is illustrated in \Rfi{ombe} and echoed in \Rfi{albe} if $\alphay$ is used instead of $\omegay$; we have indicated the cases considered in this study as well as the three cases calculated by Aronson and Hansen\cite{ea1972aj} that we will reconsider in \Rse{ah}.
\begin{figure}[h]
  \vspace{0.5\baselineskip}
  \gfbox{\wbox}{\includegraphics[keepaspectratio=true, trim = 8ex  8ex 4ex 12ex , clip , width=.97\columnwidth]{\gdir/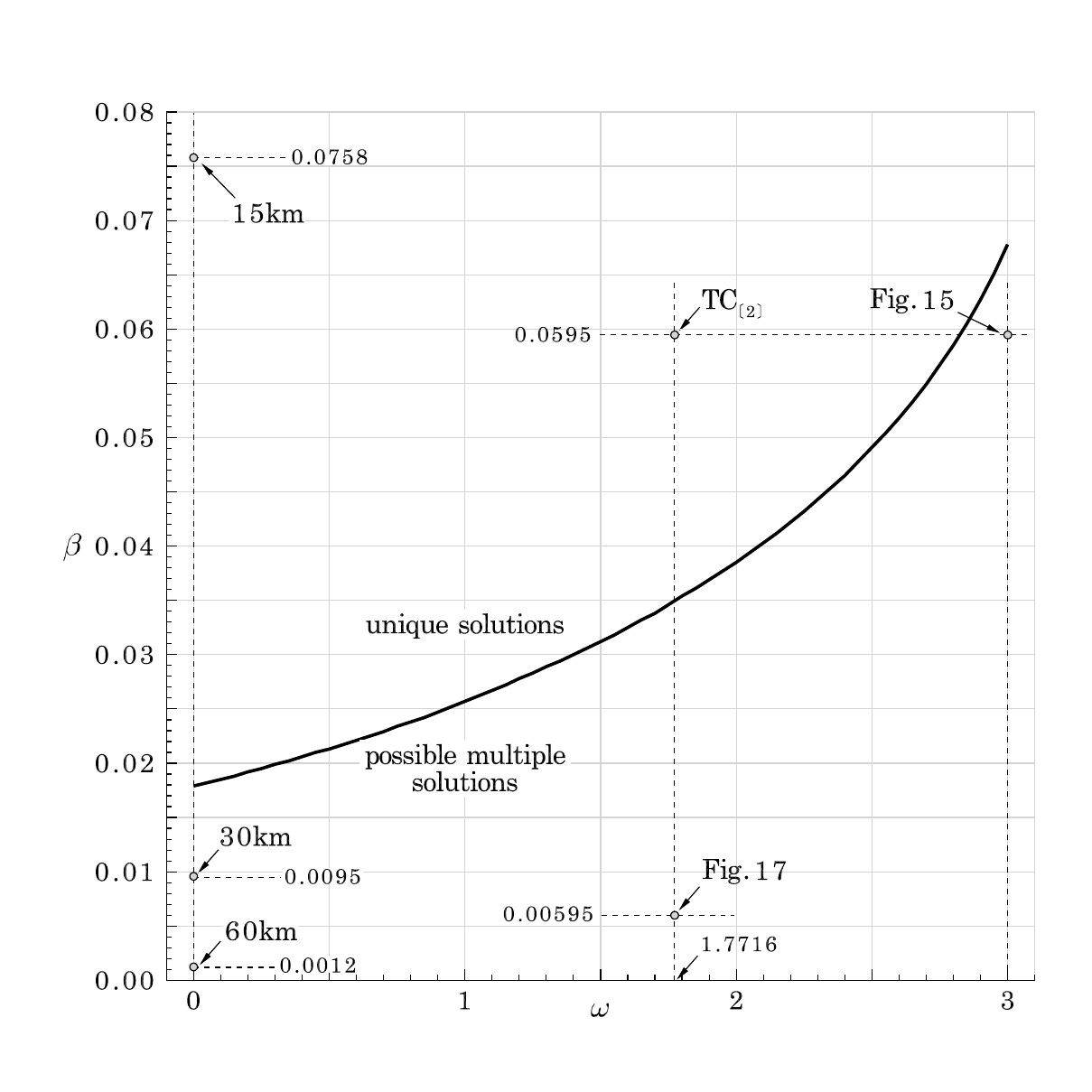}}\\  [-0.5\baselineskip]
  \caption{The \itm{\betayc-\protect\omegay} boundary separating the regions of uniqueness and possible multiplicity of the solutions; the cases dealt with in this study together with the three cases (15, 30, 60 km) investigated by Aronson and Hansen\cite{ea1972aj} \itm{(\protect\omegay=0)} are explicitly indicated.}\label{ombe}
\end{figure}
\begin{figure}[h]
  \vspace{0.5\baselineskip}
  \gfbox{\wbox}{\includegraphics[keepaspectratio=true, trim = 8ex  6ex 4ex 12ex , clip , width=.97\columnwidth]{\gdir/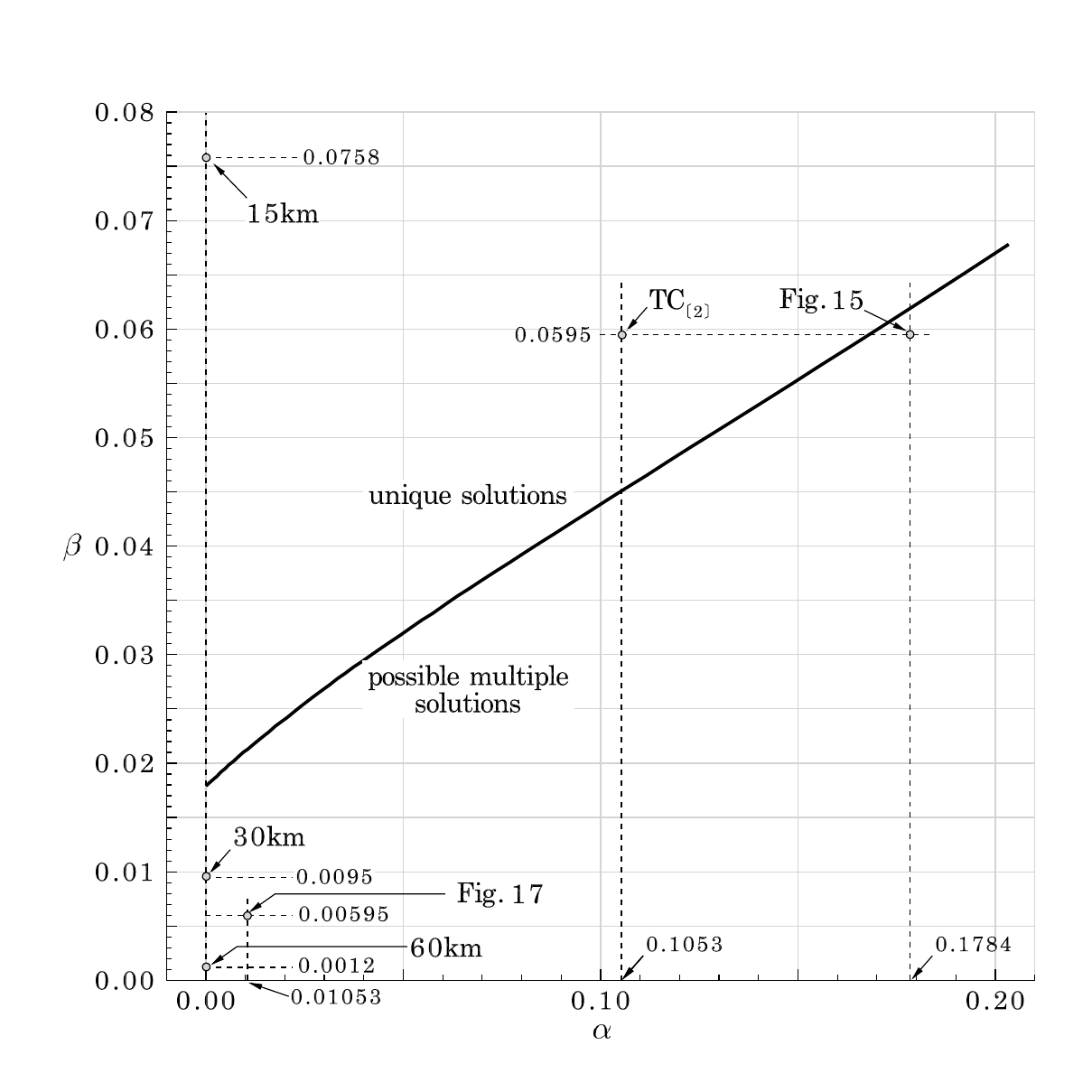}}\\  [-0.5\baselineskip]
  \caption{The  \itm{\betayc-\protect\alphayc} boundary separating the regions of uniqueness and possible multiplicity of the solutions; the cases dealt with in this study together with the three cases investigated by Aronson and Hansen\cite{ea1972aj} \itm{(\alphayc=0)} are explicitly indicated.}\label{albe}
\end{figure}

How should we interpret the existence of the boundary lines in \Rfid{ombe}{albe}?
Well, in some sense, we have already anticipated the answer in the second of the considerations we prefixed our investigation with in the second paragraph of \Rse{pr.vdw}.
We have discovered another frontier that limits \vdw's credentials as a physical model; it adds to the collection of those already known.
On the other hand, the existence of the multiple-solution region only limits but certainly does not remove the physical worthiness of \vdw\ in the unique-solution region.

\subsubsection{\anttlc Alternative attempts beyond perfect gas (reprise of \mbox{Sec.\;V B 4$_{\ssub{0.65}{\mbox{[2]}}}$}).\label{ah}}

In \Rsemb{V B 4}, we considered the cases studied by Aronson and Hansen\cite{ea1972aj} for the purpose to validate our results produced by the M$_{2}$ scheme.
We tried to reproduce the results relative to the 60 km case showed in their \mbox{Fig. 1}; we were successful with their shallow curve but we failed to obtain their steep curve.
On the contrary of what Aronson and Hansen declared in their text, we pointed out that the huge difference between the two curves could not be attributed to the tiny difference between the values (0.0820 and 0.0825 MeV$^{-1}$) of their parameter $\betay=1/kT$ because our calculations with both values basically returned the same shallow curve (\Rfimb{13}).
We were misled by the central-density value \itm{\rhoy_{c}=1.3\cdot10^{15}\mbox{ g/cm}^{3}} indicated in Aronson and Hansen's \mbox{Fig. 1} in correspondence to the steep curve because, fortuitously, that is also the central-density value for the 15 km case; so, we calculated the latter but still found a substantial divergence (\Rfimb{13}) from the steep curve.
We concluded our analysis by conceding our failure to find an explanation of the curves' divergence.

The bell rang again when we positioned Aronson and Hansen's cases on our graph of \Rfi{ombe}: the 15 km case falls in the unique-solution region but both the 30 km and 60 km cases fall in the multiple-solution region.
With the availability of our 1oM$_{2}$ scheme, we have been able to calculate all cases.
The updated \Rtamb{IV} with information relative to Aronson and Hansen's cases is given here in \Rta{ahtable}.
\begin{table}[h]
  \vspace{0.5\baselineskip}
  \gfbox{\wbox}{\includegraphics[keepaspectratio=true, trim = 0ex  0ex 0ex 0ex , clip , width=.97\columnwidth]{\gdir/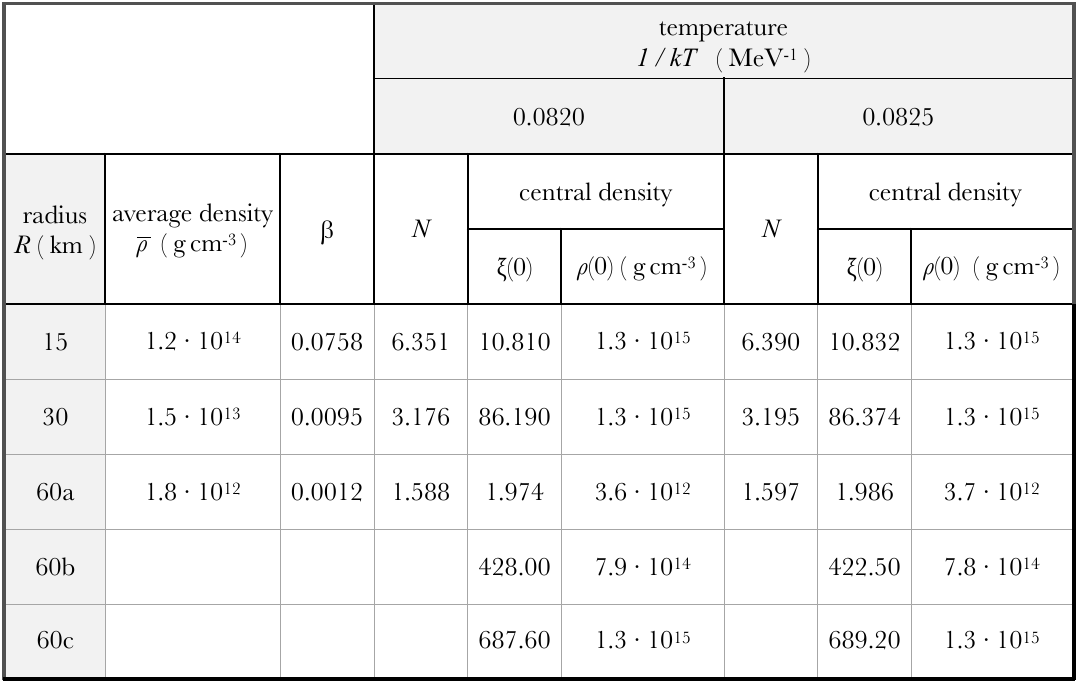}}\\  [-0.5\baselineskip]
  \caption{Update of \Rtamb{IV} with more complete information relative to Aronson and Hansen's cases;\cite{ea1972aj} we refer the reader to the original caption for details.}\label{ahtable}
\end{table}
Let us begin with the 60 km case.
The values of the gravitational number (1.588 and 1.597) fall in the triple-solution subinterval [B$_{\infty}$, A$_{\infty}$], similar to that shown in \Rfid{xit}{xic}.
With reference to Aronson and Hansen's \mbox{Fig. 1}, reconstructed and extended here in \Rfi{ah60}, the shallow curve is the a-solution and the steep curve is the c-solution with a confirmed central-density value; there also exists another solution, the b-solution purposely grayed out, whose existence seems to have escaped Aronson and Hansen's attention.
\begin{figure}[h]
  \vspace{0.5\baselineskip}
  \gfbox{\wbox}{\includegraphics[keepaspectratio=true, trim = 8ex  4ex 4ex 12ex , clip , width=.97\columnwidth]{\gdir/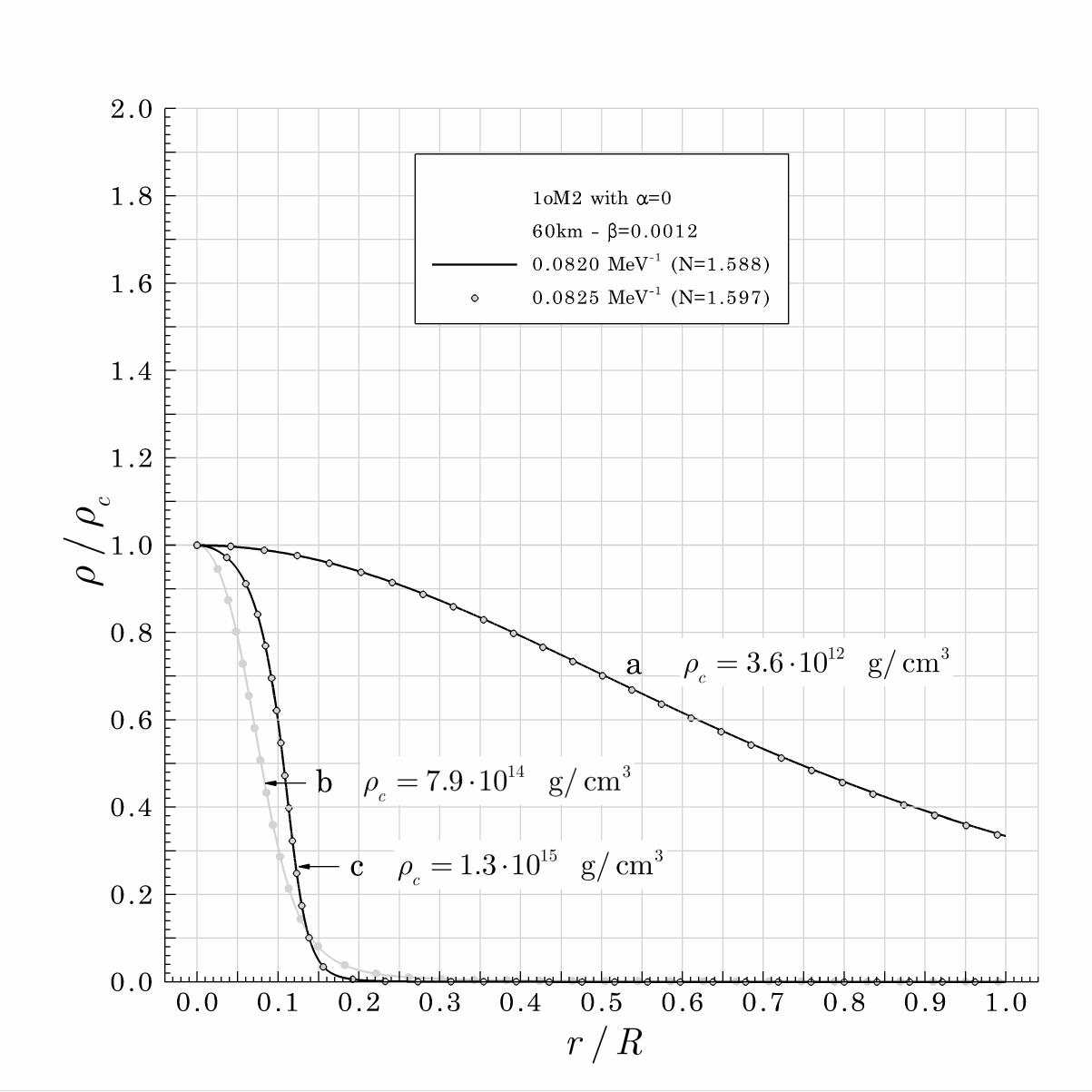}}\\  [-0.5\baselineskip]
  \caption{The three solutions of the case 60km investigated by Aronson and Hansen.\cite{ea1972aj} The b-solution is greyed out as reminder that its existence seems to have escaped those authors' attention.}\label{ah60}
\end{figure}
A triple-solution subinterval [B$_{\infty}$, A$_{\infty}$] also exists for the 30 km case, which we did not calculate in \ocite{dg2024pof}, but the values of the gravitational number (3.176 and 3.195) fall outside, to the right of A$_{\infty}$, and, therefore, the corresponding solutions turn out to be unique.
The unique solutions of the 15 km and 30 km cases are depicted in \Rfi{ah1530}; they share the same central density, again \itm{\rhoy_{c}=1.3\cdot10^{15}\mbox{ g/cm}^{3}}.
\begin{figure}[h]
  \vspace{0.5\baselineskip}
  \gfbox{\wbox}{\includegraphics[keepaspectratio=true, trim = 8ex  4ex 4ex 12ex , clip , width=.97\columnwidth]{\gdir/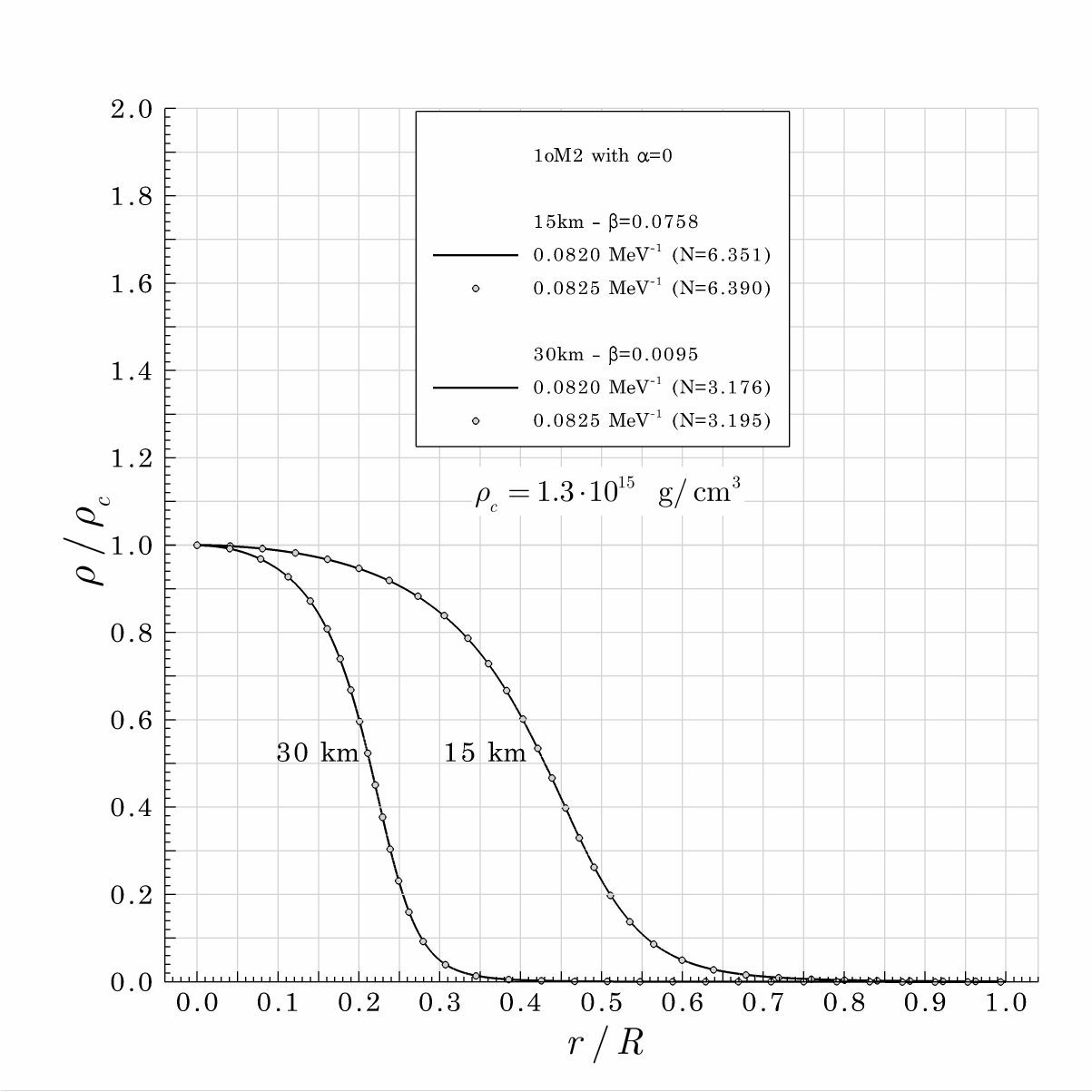}}\\  [-0.5\baselineskip]
  \caption{The single solutions of the cases 15km, 30km investigated by Aronson and Hansen.\cite{ea1972aj}}\label{ah1530}
\end{figure}
Both \Rfid{ah60}{ah1530} are reproduced in \Rfid{ah60log}{ah1530log} with the logarithmic vertical scale in order to better visualize the differences among the density profiles.
\begin{figure}[h]
  \vspace{0.5\baselineskip}
  \gfbox{\wbox}{\includegraphics[keepaspectratio=true, trim = 8ex  4ex 4ex 12ex , clip , width=.97\columnwidth]{\gdir/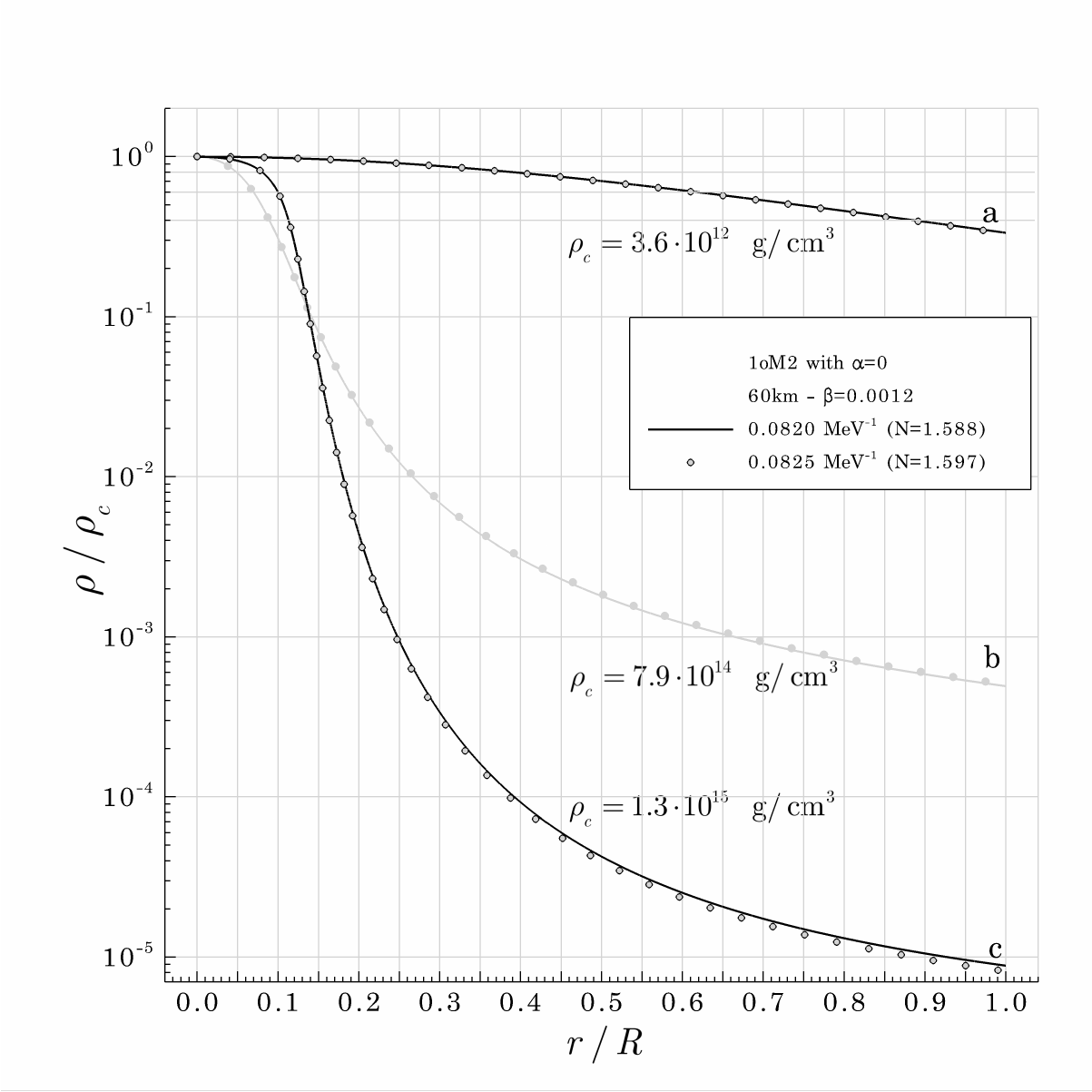}}\\  [-0.5\baselineskip]
  \caption{Same as \Rfi{ah60} but with densities in log scale to better appreciate differences.}\label{ah60log}
\end{figure}
\begin{figure}[h]
  \vspace{0.5\baselineskip}
  \gfbox{\wbox}{\includegraphics[keepaspectratio=true, trim = 8ex  4ex 4ex 12ex , clip , width=.97\columnwidth]{\gdir/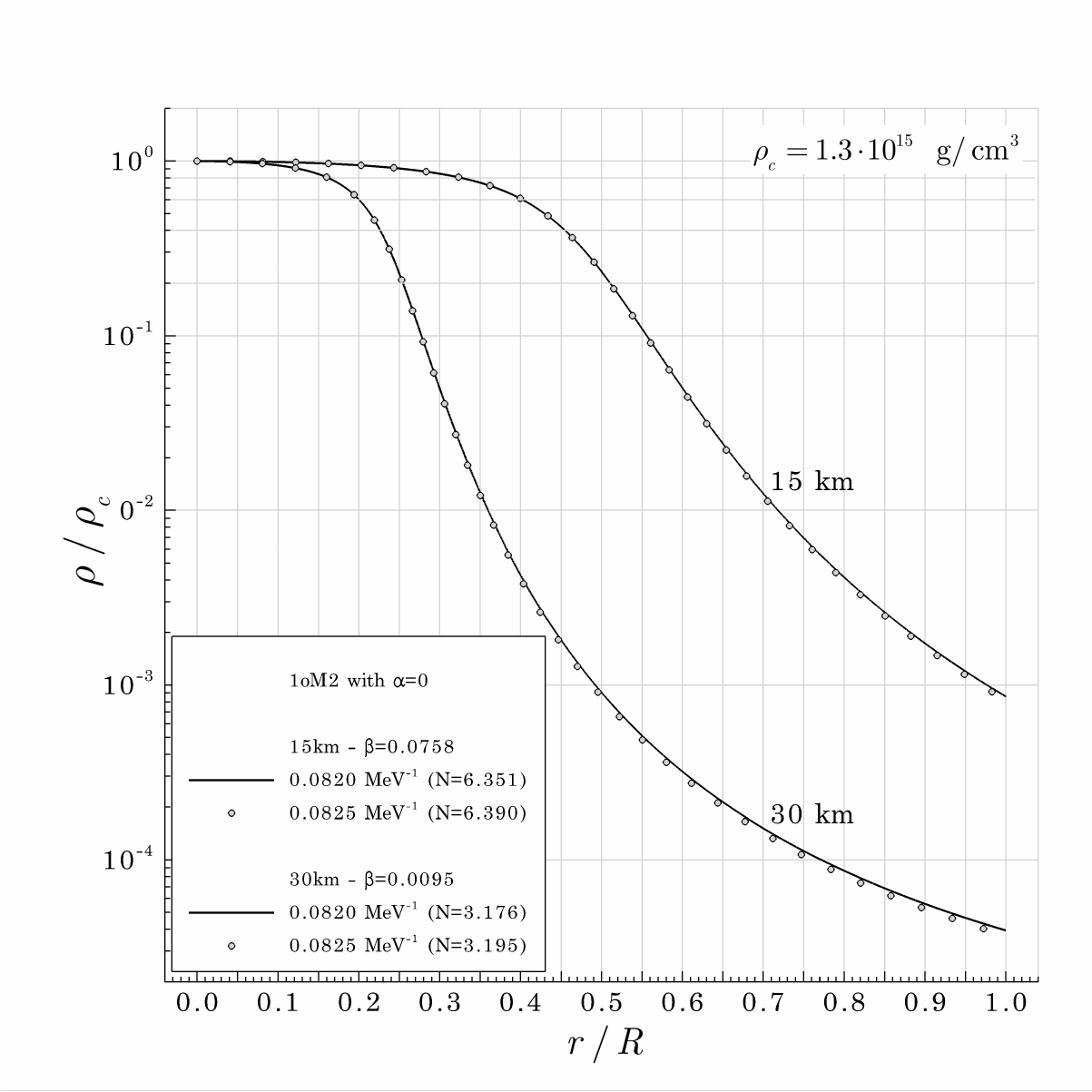}}\\  [-0.5\baselineskip]
  \caption{Same as \Rfi{ah1530} but with densities in log scale to better appreciate differences.}\label{ah1530log}
\end{figure}

\section{\protect\anttlc Conclusions}

The main motivation of this study has been the necessity to investigate the existence or absence of values of the characteristic numbers $\alphay$ and $\betay$ in correspondence to which the \pge\ appear also for the \vdw.
The intrinsic numerical limitations of the M$_{2}$ scheme, we amply utilized to obtain the majority of the results described in \ocite{dg2024pof}, with regard to such an investigation has led us to follow an alternative path based on the mathematical idea of converting the second-order differential equation [\REq{sode.ss.vdw.nd}] at the core of the M$_{2}$ scheme into a system of two equivalent first-order differential equations (\Rse{rm2s}).
The development of this conversion (\Rsed{bc}{civ}) incorporates Milne's homology invariant variables\cite{em1932mnras} adapted by Chandrasekhar\cite{sc1957,Note4} to isothermal spheres and brings smoothly to a set of four governing equations [\REq{fos.rcp}, \REq{fos.xi.u}, \REq{fos.v.u}, \REq{fos.u.rev}], the \fomt\ scheme, that is the clear and unambiguous generalization of the standalone differential equation standardly considered and discussed in the literature\cite{sc1957,tp1989ajss,tp1990pr} for the \pg\ to more general thermodynamic models, \vdw\ in particular.
We have used the \pg\ as a benchmark to test the \fomt\ (\Rse{pg}).
We believe to have given a novel contribution with our method to integrate \REq{fos.v.u.pg} (\Rse{dvdu.int}) based on the coordinate transformations [\REqq{cts}] illustrated in \Rfi{ct}; for example, one noteworthy merit of our method resides in its capability to predict (almost) analytically the bounds\cite{Note15} of the gravitational number [\REq{Ncases}] delimiting the regions of existence of the solutions.
Also, we have put the familiar results relative to the \pg\ in a more general and rational perspective that permitted us to acquire a broader and deeper understanding of the corresponding phenomenology that, in turn, has paved the way to the challenge of dealing with the \vdw\ (\Rse{vdwg}).
We have applied the \fomt\ scheme to the \vdw\ (\Rsed{vdw.h}{ge.vdw}) and have validated it by rebuilding \tcb\ (\Rse{i.hu3}) and obtaining results agreeing with those calculated with the M$_{2}$ (\Rfit{uindN2.4}{etaindN2.4}{xicxittc2}).

After having accomplished all that, somehow preliminary, work, we have proceeded to exploit (\Rse{ms}) the opportunity offered by the combination of the peripheral boundary condition [\REq{xi.is.uw}] with the parametric equations [\REq{ct.uv.N}] for the terminus $\uw$ generated by the auxiliary condition [\REq{v.is.uw}] that leads to the expression in parametric form [\REq{dxidN}] of the peripheral density's derivative with respect to the gravitational number.
This expression turns out to be the true key to unlock the way out to reach the finishing line of our investigation because it permits us to study analytically the monotonicity of peripheral density vs gravitational number, certainly another note of merit deserved by the coordinate transformations [\REqq{cts}].

The study reveals that, for a given characteristic number $\omegay$, there is a particular value of the characteristic number $\betay$ below which monotonicity breaks down, the peripheral/central densities' curves become s-shaped (\Rfid{xit}{xic}) and allow the coexistence of three solutions if the gravitational number falls in an appropriate subinterval.
However, the peripheral density does not spiral and the central density does not oscillate around \itm{N=2}.
The gravitational number is absolutely unbound; we reached an outstanding \itm{N=136} in our calculations (\Rfi{zt.sb}), a striking value that must sound unthinkable to the \pg's faithfuls.
Thus, most of the \pge\ do not afflict the \vdw.
Yes, there can be multiple solutions but their genesis is completely different from that of \pg's multiple solutions.
The curves in the \itm{\,\betay-\omegay} plane (\Rfi{ombe}) and in the \itm{\,\betay-\alphay} plane (\Rfi{albe}) marking the boundary between the regions of uniqueness and multiplicity of solution constitute another frontier, in addition to others already known from long time, which confines the applicability of the \vdw.
However, the existence of the lower region does not detract in any way whatsoever the worthiness of the \vdw\ in the upper region.

Finally, as bonus, the awareness of the existence of the uniqueness/multiplicity boundary has put us in the position (\Rse{ah}) to resolve the mystery of the missing steep curve corresponding to Aronson and Hansen's 60 km case\cite{ea1972aj} that we unsuccessfully chased in \Rsemb{V B 4}.
As compensation, this time we even found another one that had somehow escaped the attention of those authors.

\section*{\protect\anttlc Acknowledgments}
One of the authors (DG) is grateful to his colleagues R. Trasarti Battistoni and P. V\'{a}n for their systematic incitement to engage in the study presented here and for the many useful, as well as heated at times, discussions exchanged regarding the ``perfect gas and self-gravitation'' conundrum; a rare example of constructiveness produced by sometimes discordant scientific opinions.
The same author also acknowledges the valuable mathematical support of Mrs M. R. Ramirez Villar.

\vspace*{\baselineskip}
\section*{\protect\anttlc REFERENCES}


\begin{thebibliography}{32}%
\makeatletter
\providecommand \@ifxundefined [1]{%
 \@ifx{#1\undefined}
}%
\providecommand \@ifnum [1]{%
 \ifnum #1\expandafter \@firstoftwo
 \else \expandafter \@secondoftwo
 \fi
}%
\providecommand \@ifx [1]{%
 \ifx #1\expandafter \@firstoftwo
 \else \expandafter \@secondoftwo
 \fi
}%
\providecommand \natexlab [1]{#1}%
\providecommand \enquote  [1]{``#1''}%
\providecommand \bibnamefont  [1]{#1}%
\providecommand \bibfnamefont [1]{#1}%
\providecommand \citenamefont [1]{#1}%
\providecommand \href@noop [0]{\@secondoftwo}%
\providecommand \href [0]{\begingroup \@sanitize@url \@href}%
\providecommand \@href[1]{\@@startlink{#1}\@@href}%
\providecommand \@@href[1]{\endgroup#1\@@endlink}%
\providecommand \@sanitize@url [0]{\catcode `\\12\catcode `\$12\catcode
  `\&12\catcode `\#12\catcode `\^12\catcode `\_12\catcode `\%12\relax}%
\providecommand \@@startlink[1]{}%
\providecommand \@@endlink[0]{}%
\providecommand \url  [0]{\begingroup\@sanitize@url \@url }%
\providecommand \@url [1]{\endgroup\@href {#1}{\urlprefix }}%
\providecommand \urlprefix  [0]{URL }%
\providecommand \Eprint [0]{\href }%
\providecommand \doibase [0]{http://dx.doi.org/}%
\providecommand \selectlanguage [0]{\@gobble}%
\providecommand \bibinfo  [0]{\@secondoftwo}%
\providecommand \bibfield  [0]{\@secondoftwo}%
\providecommand \translation [1]{[#1]}%
\providecommand \BibitemOpen [0]{}%
\providecommand \bibitemStop [0]{}%
\providecommand \bibitemNoStop [0]{.\EOS\space}%
\providecommand \EOS [0]{\spacefactor3000\relax}%
\providecommand \BibitemShut  [1]{\csname bibitem#1\endcsname}%
\let\auto@bib@innerbib\@empty
\bibitem [{\citenamefont {Giordano}\ \emph {et~al.}(2019)\citenamefont
  {Giordano}, \citenamefont {Amodio}, \citenamefont {Iavernaro}, \citenamefont
  {Labianca}, \citenamefont {Lazzo}, \citenamefont {Mazzia},\ and\
  \citenamefont {Pisani}}]{dg2019ejmb}%
  \BibitemOpen
  \bibfield  {author} {\bibinfo {author} {\bibfnamefont {D.}~\bibnamefont
  {Giordano}}, \bibinfo {author} {\bibfnamefont {P.}~\bibnamefont {Amodio}},
  \bibinfo {author} {\bibfnamefont {F.}~\bibnamefont {Iavernaro}}, \bibinfo
  {author} {\bibfnamefont {A.}~\bibnamefont {Labianca}}, \bibinfo {author}
  {\bibfnamefont {M.}~\bibnamefont {Lazzo}}, \bibinfo {author} {\bibfnamefont
  {F.}~\bibnamefont {Mazzia}}, \ and\ \bibinfo {author} {\bibfnamefont
  {L.}~\bibnamefont {Pisani}},\ }\href@noop {} {\bibfield  {journal} {\bibinfo
  {journal} {European Journal of Mechanics / B Fluids}\ }\textbf {\bibinfo
  {volume} {78}},\ \bibinfo {pages} {66} (\bibinfo {year} {2019})}\BibitemShut
  {NoStop}%
\bibitem [{\citenamefont {Giordano}\ \emph {et~al.}(2024)\citenamefont
  {Giordano}, \citenamefont {Amodio}, \citenamefont {Iavernaro}, \citenamefont
  {Mazzia}, \citenamefont {V\'{a}n},\ and\ \citenamefont
  {Sz\"{u}cs}}]{dg2024pof}%
  \BibitemOpen
  \bibfield  {author} {\bibinfo {author} {\bibfnamefont {D.}~\bibnamefont
  {Giordano}}, \bibinfo {author} {\bibfnamefont {P.}~\bibnamefont {Amodio}},
  \bibinfo {author} {\bibfnamefont {F.}~\bibnamefont {Iavernaro}}, \bibinfo
  {author} {\bibfnamefont {F.}~\bibnamefont {Mazzia}}, \bibinfo {author}
  {\bibfnamefont {P.}~\bibnamefont {V\'{a}n}}, \ and\ \bibinfo {author}
  {\bibfnamefont {M.}~\bibnamefont {Sz\"{u}cs}},\ }\href@noop {} {\bibfield
  {journal} {\bibinfo  {journal} {Physics of Fluids}\ }\textbf {\bibinfo
  {volume} {36}},\ \bibinfo {pages} {056127 1} (\bibinfo {year}
  {2024})}\BibitemShut {NoStop}%
\bibitem [{Note1()}]{Note1}%
  \BibitemOpen
  \bibinfo {note} {Admittedly, we did not prove it.}\BibitemShut {Stop}%
\bibitem [{\citenamefont {Chandrasekhar}(1967)}]{sc1957}%
  \BibitemOpen
  \bibfield  {author} {\bibinfo {author} {\bibfnamefont {S.}~\bibnamefont
  {Chandrasekhar}},\ }\href@noop {} {\emph {\bibinfo {title} {An introduction
  to the study of stellar structure}}}\ (\bibinfo  {publisher} {Dover},\
  \bibinfo {address} {New York NY},\ \bibinfo {year} {1967})\BibitemShut
  {NoStop}%
\bibitem [{\citenamefont {Padmanabhan}(1989)}]{tp1989ajss}%
  \BibitemOpen
  \bibfield  {author} {\bibinfo {author} {\bibfnamefont {T.}~\bibnamefont
  {Padmanabhan}},\ }\href@noop {} {\bibfield  {journal} {\bibinfo  {journal}
  {The Astrophysical Journal Supplement Series}\ }\textbf {\bibinfo {volume}
  {71}},\ \bibinfo {pages} {651} (\bibinfo {year} {1989})}\BibitemShut
  {NoStop}%
\bibitem [{\citenamefont {Padmanabhan}(1990)}]{tp1990pr}%
  \BibitemOpen
  \bibfield  {author} {\bibinfo {author} {\bibfnamefont {T.}~\bibnamefont
  {Padmanabhan}},\ }\href@noop {} {\bibfield  {journal} {\bibinfo  {journal}
  {Physics Reports}\ }\textbf {\bibinfo {volume} {188}},\ \bibinfo {pages}
  {285} (\bibinfo {year} {1990})}\BibitemShut {NoStop}%
\bibitem [{\citenamefont {Horedt}(2004)}]{gh2004}%
  \BibitemOpen
  \bibfield  {author} {\bibinfo {author} {\bibfnamefont {G.~P.}\ \bibnamefont
  {Horedt}},\ }\href@noop {} {\emph {\bibinfo {title} {Polytropes. Applications
  in astrophysics and related fields.}}},\ \bibinfo {series} {Astrophysics and
  Space Science Library}, Vol.\ \bibinfo {volume} {306}\ (\bibinfo  {publisher}
  {Kluwer Academic Publishers},\ \bibinfo {address} {Dordrecht, The
  Netherlands},\ \bibinfo {year} {2004})\BibitemShut {NoStop}%
\bibitem [{\citenamefont {Milne}(1932)}]{em1932mnras}%
  \BibitemOpen
  \bibfield  {author} {\bibinfo {author} {\bibfnamefont {E.~A.}\ \bibnamefont
  {Milne}},\ }\href@noop {} {\bibfield  {journal} {\bibinfo  {journal} {Monthly
  Notices of the Royal Astronomical Society}\ }\textbf {\bibinfo {volume}
  {92}},\ \bibinfo {pages} {610} (\bibinfo {year} {1932})}\BibitemShut
  {NoStop}%
\bibitem [{Note2()}]{Note2}%
  \BibitemOpen
  \bibinfo {note} {More precisely, item 3 of Sec. VI at page 182.}\BibitemShut
  {Stop}%
\bibitem [{Note3()}]{Note3}%
  \BibitemOpen
  \bibinfo {note} {For specific details, we refer the reader to Eq. (453) and
  Fig. 20 in \protect \mbox {Sec. 27} at \protect \mbox {page 168} of \protect
  \mbox {Ref.~\protect \rev@citealpnum {sc1957}} or Eq. (18) and Fig. 1 in
  \protect \mbox {Sec. II} at \protect \mbox {page 651} of \protect \mbox
  {Ref.~\protect \rev@citealpnum {tp1989ajss}} or Eq. (4.26) and Fig. 4.1 in
  \protect \mbox {Sec. 4.3} at \protect \mbox {page 313} of \protect \mbox
  {Ref.~\protect \rev@citealpnum {tp1990pr}}}\BibitemShut {NoStop}%
\bibitem [{Note4()}]{Note4}%
  \BibitemOpen
  \bibinfo {note} {Perhaps we should better say Chandrasekhar's. As a matter of
  fact, Milne did not consider isothermal spheres in \protect \mbox
  {Ref.~\protect \rev@citealpnum {em1932mnras}}; he introduced the \protect
  \mbox {$u,v$} variables [Eqs. (42) at page 622] for a non-isothermal sphere
  composed by a zone of perfect-gas polytrope of index \protect \mbox {$n=3$}
  in contact with a zone of degenerate-gas polytrope of index \protect \mbox
  {$n=3/2$}. To the best of the literature we have consulted, it seems to us
  that Chandrasekhar was the first to introduce the \protect \mbox {$u,v$}
  variables for isothermal spheres of perfect gas in the first edition of
  \protect \mbox {Ref.~\protect \rev@citealpnum {sc1957}} that appeared in
  1939; our conjecture seems to be corroborated by the brief note\cite
  {sc1949aj} that Chandrasekhar and Wares published in the Astrophysical
  Journal in 1949 \protect \textbf {}and by Horedt's citation in the beginning
  of Sec. 2.2.2 at page 34 of \protect \mbox {Ref.~\protect \rev@citealpnum
  {gh2004}}. Besides, we reassure the reader that we have thoroughly
  cross-checked our definitions [\protect \mbox {Eqs.~(\ref {v.is})} and (\ref
  {u.is})] with those of Chandrasekhar [Eqs. (400) at page 160 and the
  auxiliary Eqs. (373) at page 155 of \protect \mbox {Ref.~\protect
  \rev@citealpnum {sc1957}}] through a patient exercise of notation conversion
  and definition verification; of course, Chandrasekhar's definitions apply
  only to the PG-m.}\BibitemShut {Stop}%
\bibitem [{Note5()}]{Note5}%
  \BibitemOpen
  \bibinfo {note} {The attentive reader may have noticed that \protect \mbox
  {Eq.~(\ref {v-dv.is.0.c})} (top) follows also as direct consequence of
  \protect \mbox {Eq.~(\ref {dv.is.tmp1})} but this fits the logic in which
  \protect \mbox {Eq.~(\ref {bc.rcp.r=0.nd})} is pre-assigned. Here we are
  following the reverse logical path: \protect \mbox {Eq.~(\ref {v.is.0})} is
  pre-assigned with the purpose in mind to prove \protect \mbox {Eq.~(\ref
  {bc.rcp.r=0.nd})}; therefore, \protect \mbox {Eq.~(\ref {dv.is.tmp1})} must
  be looked at posteriorly as a sort of consistency indicator in connection
  with \protect \mbox {Eq.~(\ref {v-dv.is.0.c})} (top).}\BibitemShut {Stop}%
\bibitem [{Note6()}]{Note6}%
  \BibitemOpen
  \bibinfo {note} {This would be the case, for example, for the thermodynamic
  model proposed by Stahl et al.\cite {bs1995pss} that we considered in
  \protect \mbox {Sec.\protect \tmspace +\thickmuskip {.2777em}V B
  4}$_{\scalebox {0.65}{$\protect \mbox {[\protect \rev@citealpnum
  {dg2024pof}]}$}}$; see \protect \mbox {Eq.\protect \tmspace +\thickmuskip
  {.2777em}(77a)}$_{\scalebox {0.65}{$\protect \mbox {[\protect \rev@citealpnum
  {dg2024pof}]}$}}$.}\BibitemShut {Stop}%
\bibitem [{Note7()}]{Note7}%
  \BibitemOpen
  \bibinfo {note} {For the case of the PG-m, see Chandrasekhar's derivation
  that leads to his Eq. (466) at page 169 of \protect \mbox {Ref.~\protect
  \rev@citealpnum {sc1957}}. Padmanabhan also mentions the same result at page
  653 of \protect \mbox {Ref.~\protect \rev@citealpnum {tp1989ajss}}, just
  below Eq. (18), and at page 314 of \protect \mbox {Ref.~\protect
  \rev@citealpnum {tp1990pr}}, just below Eq. (4.26), but without giving a
  derivation.}\BibitemShut {Stop}%
\bibitem [{Note8()}]{Note8}%
  \BibitemOpen
  \bibinfo {note} {We conjecture that this attitude hinges on historical
  reasons tracing back to the pioneers in astronomy and astrophysics. See
  \protect \mbox {\protect \mbox {Refs.\protect \tmspace +\thickmuskip
  {.2777em}45-52}$_{\scalebox {0.65}{$\protect \mbox {[\protect \rev@citealpnum
  {dg2019ejmb}]}$}}$}.}\BibitemShut {Stop}%
\bibitem [{Note9()}]{Note9}%
  \BibitemOpen
  \bibinfo {note} {We belong to this camp.}\BibitemShut {Stop}%
\bibitem [{Note10()}]{Note10}%
  \BibitemOpen
  \bibinfo {note} {The physical fallacy hidden behind this move is tremendously
  brought to light by the outstanding analysis carried out by Saslaw\cite
  {wc1968mnras} in his first paper of the series addressing
  gravithermodynamics. In Sec. 3, he considered a vdWG-m\ as an example of
  ``imperfect self-gravitating gas''. In Sec. 3.2, he evaluated the van der
  Waals' constant $a$, notoriously connected to the molecular forces, and
  showed beyond any doubt in \protect \mbox {Eq. (5)} that $a$ is linearly
  proportional to $G$; thus, the removal of molecular gravitational forces
  \protect \mbox {$(a=0)$} called for by the PG-m\ requires {a fortiori} the
  inescapable vanishing of the gravitational constant \protect \mbox {$(G=0)$}.
  There is no way around it.}\BibitemShut {Stop}%
\bibitem [{Note11()}]{Note11}%
  \BibitemOpen
  \bibinfo {note} {One of those strokes of magic that mathematics is
  surprisingly capable of sometimes.}\BibitemShut {Stop}%
\bibitem [{Note12()}]{Note12}%
  \BibitemOpen
  \bibinfo {note} {Padmanabhan\cite {tp1989ajss,tp1990pr} used the total energy
  as selector but the two approaches are consistent. To prove the equivalence,
  we start from \protect \mbox {Eq.\protect \tmspace +\thickmuskip
  {.2777em}(121)}$_{\scalebox {0.65}{$\protect \mbox {[\protect \rev@citealpnum
  {dg2019ejmb}]}$}}$, substitute $\lambda $ for its left-hand side in
  accordance to Padmanabhan's notation, $u_{t}/3$ for $\protect \xiyf
  (1,N)\equiv \protect \xiyf (u_{t})$ [\protect \mbox {Eq.~(\ref {xi.is.uw})}],
  and $v(u_{t})$ for $N$ [\protect \mbox {Eq.~(\ref {v.is.uw})}]. Thus, the
  total energy reads $\lambda = \left (u_{t}- {3}/{2}\right )/v(u_{t})$ that
  compares with \protect \mbox {Eq. (22)} of \protect \mbox {Ref.~\protect
  \rev@citealpnum {tp1989ajss}} or \protect \mbox {Eq. (4.29)} of \protect
  \mbox {Ref.~\protect \rev@citealpnum {tp1990pr}}. The latter expression can
  be rewritten as $v(u_{t}) = \left (u_{t}- {3}/{2}\right )/\lambda $ which
  reveals the terminus $u_{t}$ being fixed by the intersection between the
  universal solution $v(u)$ and the straight line $ v = \left (u -
  {3}/{2}\right )/\lambda $ shown in Fig. 2 of \protect \mbox {Ref.~\protect
  \rev@citealpnum {tp1989ajss}} or Fig. 4.2 of \protect \mbox {Ref.~\protect
  \rev@citealpnum {tp1990pr}}. However, it must be kept in mind that multiple
  solutions with same total energy (Padmanabhan's selector) correspond to
  different gravitational numbers and that multiple solutions with same
  gravitational number (our selector) correspond to different total
  energies.}\BibitemShut {Stop}%
\bibitem [{Note13()}]{Note13}%
  \BibitemOpen
  \bibinfo {note} {See for example \protect \mbox {\protect \mbox
  {Figs.\protect \tmspace +\thickmuskip {.2777em}7}$_{\scalebox
  {0.65}{$\protect \mbox {[\protect \rev@citealpnum {dg2019ejmb}]}$}}$\protect
  \,and\protect \tmspace +\thickmuskip {.2777em}8$_{\scalebox {0.65}{$\protect
  \mbox {[\protect \rev@citealpnum {dg2019ejmb}]}$}}$} which illustrate the two
  solutions corresponding to \protect \mbox {$N=2.4$}.}\BibitemShut {Stop}%
\bibitem [{\citenamefont {Chandrasekhar}\ and\ \citenamefont
  {Wares}(1949)}]{sc1949aj}%
  \BibitemOpen
  \bibfield  {author} {\bibinfo {author} {\bibfnamefont {S.}~\bibnamefont
  {Chandrasekhar}}\ and\ \bibinfo {author} {\bibfnamefont {G.~W.}\ \bibnamefont
  {Wares}},\ }\href@noop {} {\bibfield  {journal} {\bibinfo  {journal} {The
  Astrophysical Journal}\ }\textbf {\bibinfo {volume} {109}},\ \bibinfo {pages}
  {551} (\bibinfo {year} {1949})}\BibitemShut {NoStop}%
\bibitem [{Note14()}]{Note14}%
  \BibitemOpen
  \bibinfo {note} {\label {en.hints}\textcolor {black}{Some hints for the
  interested reader. Assume \protect \mbox {$\protect \thetayf \ll 1$} and
  linearize numerator and denominator of \protect \mbox {Eq.~(\ref
  {fos.v.u.pg.zt})}, including the exponentials; then passing the resulting
  expression to the limit \protect \mbox {$\protect \thetayf \rightarrow 0$}
  yields a quadratic equation for \protect \mbox {$z_{\theta }(0)$}.
  Compatibility with \protect \mbox {Eq.~(\ref {fos.xi.u.3.f})} selects the
  applicable root.}}\BibitemShut {Stop}%
\bibitem [{Note15()}]{Note15}%
  \BibitemOpen
  \bibinfo {note} {We consider the possibility to calculate these endpoints as
  another mark of superiority of the 1oM$_{2}$\ scheme with respect to the
  M$_{2}$ scheme. In the latter scheme, we basically treated the isothermal
  Lane-Emden equation as a boundary-value problem in which the gravitational
  number is prescribed [\protect \mbox {Eqs.\protect \tmspace +\thickmuskip
  {.2777em}(42)}$_{\scalebox {0.65}{$\protect \mbox {[\protect \rev@citealpnum
  {dg2019ejmb}]}$}}$]; in that calculational context, upper and lower bounds of
  the gravitational number can only be obtained through a strategy of tedious
  trial and error to gain accurate decimal digits which require repetitive and
  time consuming running of the numerical algorithm, the tediousness being
  particularly acute for the upper bound. In connection with \protect \mbox
  {Eq.~(\ref {ct.uv})} (bottom), Darwin's statement ``... I am unable to find
  any analytical relationship by which the minimum value of \protect \mbox
  {$\protect \frac {1}{3}\protect \betayf ^{2}$} can be deduced.'' at page 19
  of \protect \mbox {Ref.~\protect \rev@citealpnum {gd1889ptrs}}, comes to mind
  again; in Darwin's notation, \protect \mbox {$\protect \frac {1}{3}\protect
  \betayf ^{2}$} corresponds to our \protect \mbox {$1/N$}. We quoted it in
  \protect \mbox {Sec.\protect \tmspace +\thickmuskip {.2777em}4}$_{\scalebox
  {0.65}{$\protect \mbox {[\protect \rev@citealpnum {dg2019ejmb}]}$}}$, at the
  bottom of the right column of page 85, where we also conceded the same
  incapability. We are pleased to have moved forward to a position in which we
  are not obliged to admit the same defeat this time.}\BibitemShut {Stop}%
\bibitem [{Note16()}]{Note16}%
  \BibitemOpen
  \bibinfo {note} {For example, the vdWG-m\ cannot describe water vapor because
  the latter's compressibility ratio is $\sim 0.23$.}\BibitemShut {Stop}%
\bibitem [{Note17()}]{Note17}%
  \BibitemOpen
  \bibinfo {note} {It is easy to verify that setting \protect \mbox {$\protect
  \mathfrak {h}=1$} in \protect \mbox {Eq.~(\ref {fos.w.z.vdw.z})} retrieves
  the differential equation of the PG-m\ [\protect \mbox {Eq.~(\ref
  {fos.v.u.pg.zt})}] and decouples it from \protect \mbox {Eq.~(\ref
  {fos.w.z.vdw.w})}.}\BibitemShut {Stop}%
\bibitem [{Note18()}]{Note18}%
  \BibitemOpen
  \bibinfo {note} {Another quick verification: setting \protect \mbox
  {$\protect \mathfrak {h}(0)=1$} in \protect \mbox {Eq.~(\ref
  {fos.w.z.vdw.z.0})} retrieves \protect \mbox {Eq.~(\ref
  {fos.v.u.pg.zt.0})}.}\BibitemShut {Stop}%
\bibitem [{Note19()}]{Note19}%
  \BibitemOpen
  \bibinfo {note} {We wish to stress that this is not the same as imposing the
  central density as boundary condition; we have already expressed our opinion
  about this matter in \protect \mbox {Refs.~\protect \rev@citealpnum
  {dg2019ejmb,dg2024pof}}.}\BibitemShut {Stop}%
\bibitem [{Note20()}]{Note20}%
  \BibitemOpen
  \bibinfo {note} {Paragraph just above \protect \mbox {Fig.\protect \tmspace
  +\thickmuskip {.2777em}10}$_{\scalebox {0.65}{$\protect \mbox {[\protect
  \rev@citealpnum {dg2024pof}]}$}}$ at page 11.}\BibitemShut {Stop}%
\bibitem [{\citenamefont {Aronson}\ and\ \citenamefont
  {Hansen}(1972)}]{ea1972aj}%
  \BibitemOpen
  \bibfield  {author} {\bibinfo {author} {\bibfnamefont {E.}~\bibnamefont
  {Aronson}}\ and\ \bibinfo {author} {\bibfnamefont {C.}~\bibnamefont
  {Hansen}},\ }\href@noop {} {\bibfield  {journal} {\bibinfo  {journal} {The
  Astrophysical Journal}\ }\textbf {\bibinfo {volume} {177}},\ \bibinfo {pages}
  {145} (\bibinfo {year} {1972})}\BibitemShut {NoStop}%
\bibitem [{\citenamefont {Stahl}, \citenamefont {Kiessling},\ and\
  \citenamefont {Schindler}(1995)}]{bs1995pss}%
  \BibitemOpen
  \bibfield  {author} {\bibinfo {author} {\bibfnamefont {B.}~\bibnamefont
  {Stahl}}, \bibinfo {author} {\bibfnamefont {M.}~\bibnamefont {Kiessling}}, \
  and\ \bibinfo {author} {\bibfnamefont {K.}~\bibnamefont {Schindler}},\
  }\href@noop {} {\bibfield  {journal} {\bibinfo  {journal} {Planetary and
  Space Science}\ }\textbf {\bibinfo {volume} {43}},\ \bibinfo {pages} {271}
  (\bibinfo {year} {1995})}\BibitemShut {NoStop}%
\bibitem [{\citenamefont {Saslaw}(1968)}]{wc1968mnras}%
  \BibitemOpen
  \bibfield  {author} {\bibinfo {author} {\bibfnamefont {W.~C.}\ \bibnamefont
  {Saslaw}},\ }\href@noop {} {\bibfield  {journal} {\bibinfo  {journal}
  {Monthly Notices of the Royal Astronomical Society}\ }\textbf {\bibinfo
  {volume} {141}},\ \bibinfo {pages} {1} (\bibinfo {year} {1968})}\BibitemShut
  {NoStop}%
\bibitem [{\citenamefont {Darwin}(1889)}]{gd1889ptrs}%
  \BibitemOpen
  \bibfield  {author} {\bibinfo {author} {\bibfnamefont {G.~H.}\ \bibnamefont
  {Darwin}},\ }\href@noop {} {\bibfield  {journal} {\bibinfo  {journal}
  {Philosophical Transactions of the Royal Society of London A}\ }\textbf
  {\bibinfo {volume} {180}},\ \bibinfo {pages} {1} (\bibinfo {year}
  {1889})}\BibitemShut {NoStop}%
\end{thebibliography}
%

\end{document}